\documentclass[moor,nonblindrev]{informs3} 

\usepackage{stmaryrd}
\usepackage[dvipsnames]{xcolor}
\usepackage{tikz}
\usepackage{dsfont}
\usetikzlibrary{arrows,positioning}
\usepackage{caption}
\usepackage{subcaption}
\usepackage{xspace}
\usepackage{cases}
\usepackage{algorithm}
\usepackage{algpseudocode}
\usetikzlibrary{patterns}
\usepackage{mathtools}

\DeclareSymbolFont{yhlargesymbols}{OMX}{yhex}{m}{n}
\DeclareMathAccent{\wideparen}{\mathord}{yhlargesymbols}{"F3}

\newif\ifarXiv

\arXivtrue 

\ifarXiv
\OneAndAHalfSpacedXI

\else
\OneAndAHalfSpacedXI
\fi

\usepackage{natbib}

 \NatBibNumeric
 \def\bibfont{\small}%
 \bibpunct[, ]{[}{]}{,}{n}{}{,}%

\usepackage[colorlinks=true,breaklinks=true,bookmarks=true,urlcolor=blue,
     citecolor=blue,linkcolor=blue,bookmarksopen=false,draft=false]{hyperref}

\def\EMAIL#1{\href{mailto:#1}{#1}}
\def\URL#1{\href{#1}{#1}}         

\TheoremsNumberedThrough     

\EquationsNumberedThrough    

\MANUSCRIPTNO{MOR-2018-305} 

{\theoremstyle{THkey}}
{\theoremstyle{EXkey}}
\begin{document}

\newcommand{\supp}{\operatorname{supp}}
\newcommand{\Val}{\operatorname{F}}
\newcommand{\Costop}{\operatorname{C}}
\newcommand{\Costflow}{\operatorname{T}}
\newcommand{\flow}{f}
\newcommand{\att}{S}
\newcommand{\defender}{\textbf{P1}\xspace}
\newcommand{\attacker}{\textbf{P2}\xspace}
\newcommand{\nedges}{N}
\newcommand{\ndef}{n_1}
\newcommand{\natt}{n}
\newcommand{\aref}[1]{\text{(\hyperref[#1]{A\ref{#1}})}}
\newcommand{\bref}[1]{\text{(\hyperref[#1]{B\ref{#1}})}}
\newcommand{\lref}[1]{(\hyperref[#1]{\text{L}\ref{#1}})}
\newcommand{\pref}[1]{\hyperref[#1]{Proof}}
\newcommand{\sref}[1]{\hyperref[#1]{\ref{#1}}}
\newcommand{\rd}{b_1}
\newcommand{\ra}{b_2}
\newcommand{\fmax}{\Val^{max}}
\newcommand{\tmin}{\Costflow^{min}}
\newcommand{\mfs}{\Omega_{max}}
\newcommand{\mcmfs}{\Omega^*}
\newcommand{\set}{\mathcal{P}}
\newcommand{\sets}{\mathcal{S}}
\newcommand{\sett}{\mathcal{T}}
\newcommand{\edges}{\mathcal{E}}
\newcommand{\nodes}{\mathcal{V}}
\newcommand{\nash}{\Sigma}

\newcommand{\Value}[1]{\Val\left(#1\right)}
\newcommand{\Eff}[2]{\Val\left({#1}^{#2}\right)}
\newcommand{\Effp}[2]{\Val\left(\left(#1\right)^{#2}\right)}
\newcommand{\Loss}[2]{\Val\left(#1 - {#1}^{#2}\right)}
\newcommand{\Lossp}[2]{\Val\left(#1 - (#1)^{#2}\right)}
\newcommand{\Cost}[1]{\Costop\left(#1\right)}
\newcommand{\Costf}[1]{\Costflow\left(#1\right)}
\newcommand{\Exponebis}[2]{\mathbb{E}_{#1}\left[#2\right]}
\newcommand{\Exptwobis}[2]{\mathbb{E}_{#1}\left[#2\right]}
\newcommand{\Exps}[3]{\mathbb{E}_{(#1,#2)}\left[#3\right]}
\newcommand{\Expone}[2][\empty]{\mathbb{E}_{{\sigma}^{#1}}\left[#2 \right]}
\newcommand{\Exptwo}[2][\empty]{\mathbb{E}_{{\sigma}^{#1}}\left[#2 \right]}
\newcommand{\Expboth}[3]{\mathbb{E}_{\sigma^{#1}}\left[#3\right]}
\newcommand{\tb}[1]{\textcolor{blue}{#1}}
\newcommand{\tr}[1]{\textcolor{red}{#1}}
\newcommand{\tg}[1]{\textcolor{Green!80!black}{#1}}
\newcommand{\tp}[1]{\textcolor{violet!100!black}{#1}}
\newcommand{\Circ}{(\hyperlink{(P)}{$\mathcal{P}$})\xspace}

\newcommand{\pathtight}[1]{\overline{\Chains}^{#1}}
\newcommand{\pathloose}[1]{\widehat{\Chains}^{#1}}
\newcommand{\component}[1]{X^{#1}}
\newcommand{\weight}[1]{w^{#1}}
\newcommand{\poset}[1]{P^{#1}}
\newcommand{\setmin}[1]{S^{#1}}
\newcommand{\maxweight}[1]{W^{#1}}
\newcommand{\pathprob}[2]{\pi_{#1}^{#2}}
\newcommand{\edgeprob}[2]{\rho_{#1}^{#2}}
\newcommand{\edgeweight}[1]{\beta_{#1}}
\newcommand{\combiprob}[2]{\delta_{#1}^{#2}}
\newcommand{\offset}[1]{\widetilde{\alpha}^{#1}}
\newcommand{\stepmax}{n^*}
\newcommand{\paths}[1]{\Lambda^{#1}}
\newcommand{\comparability}[1]{\mathcal{H}^{#1}}
\newcommand{\tree}[1]{\mathcal{T}^{#1}}
\newcommand{\chain}{C}
\newcommand{\Chains}{\mathcal{C}}
\newcommand{\ground}{X}
\newcommand{\mat}{M}
\newcommand{\allpaths}[1]{\Chains^{#1}}
\newcommand{\restrict}[1]{\mkern 2mu \vrule height 2ex\mkern3mu #1}
\newcommand{\Cross}{$\mathbin{\tikz [x=1.4ex,y=1.4ex,line width=.2ex, red] \draw (0,0) -- (1,1) (0,1) -- (1,0);}$}%

\newcommand{\FPOC}{\text{$($\hyperlink{DDD}{\ensuremath{\mathcal{D}}}$)$}\xspace}
\newcommand{\OPOC}{\text{$($\hyperlink{QQQ}{\ensuremath{\mathcal{Q}}}$)$}\xspace}
\newcommand{\OPOCt}{\text{$($\hyperlink{QQQ}{\ensuremath{\mathcal{Q}}}$)$}\xspace}

\newcommand{\primal}{\text{$($\hyperlink{LPs}{\ensuremath{\mathcal{M}_P}}$)$}\xspace}
\newcommand{\dual}{\text{$($\hyperlink{LPs}{\ensuremath{\mathcal{M}_D}}$)$}\xspace}
\newcommand{\original}{\text{$($\hyperlink{MCCP}{\ensuremath{\mathcal{M}}}$)$}\xspace}

\newcommand{\polyp}{\text{$($\hyperlink{Polyp}{\ensuremath{\mathcal{M}_P^\prime}}$)$}\xspace}
\newcommand{\polyd}{\text{$($\hyperlink{Polyd}{\ensuremath{\mathcal{M}_D^\prime}}$)$}\xspace}
\newcommand{\polyo}{\text{$($\hyperlink{Polyp}{\ensuremath{\mathcal{M}^\prime}}$)$}\xspace}

\newcommand{\combi}{(\hyperlink{Combi}{\ensuremath{\mathcal{T}}})\xspace}

\newcommand{\StateNew}[1]{\algrenewcommand{\alglinenumber}[1]{\footnotesize A##1:}\State #1}
\newcommand{\StateNewtwo}[1]{\algrenewcommand{\alglinenumber}[1]{\footnotesize B##1:}\State #1}

\newcommand{\rev}[1]{{\color{black} #1}}
\newcommand{\revm}[1]{{\color{black} #1}}
\newcommand{\rem}[1]{}
\newcommand{\arcs}{A}


 \RUNAUTHOR{Dahan, Amin, and Jaillet}

\RUNTITLE{Probability Distributions on Partially Ordered Sets and Network Interdiction Games}
\TITLE{Probability Distributions on Partially Ordered Sets and Network Interdiction Games}

\ARTICLEAUTHORS{%
\AUTHOR{Mathieu Dahan}
\AFF{School of Industrial and Systems Engineering, 
Georgia Institute of Technology, 
Atlanta, Georgia 30332, \EMAIL{mathieu.dahan@isye.gatech.edu} \URL{}}
\AUTHOR{Saurabh Amin}
\AFF{Laboratory for Information and Decision Systems, Massachusetts Institute of Technology, 
Cambridge, Massachusetts 02139, \EMAIL{amins@mit.edu} \URL{}}
\AUTHOR{Patrick Jaillet}
\AFF{Department of Electrical Engineering and Computer Science, 
Laboratory for Information and Decision Systems,\\ and
Operations Research Center, 
Massachusetts Institute of Technology,
Cambridge, Massachusetts 02139, \EMAIL{jaillet@mit.edu} \URL{}}
} 

\ABSTRACT{%
\rev{This article poses the following problem:} Does there exist a probability distribution over subsets of a finite partially ordered set (poset), such that a set of constraints involving marginal probabilities of the poset's elements and maximal chains is satisfied? We present a combinatorial algorithm to positively resolve this question. \rev{The algorithm can be implemented in polynomial time in the special case where maximal chain probabilities are affine functions of their elements.} \rev{This existence problem is relevant for the equilibrium characterization of a generic strategic} interdiction game on a capacitated flow network. The game involves a routing entity that sends its flow through the network while facing path transportation costs, and an interdictor who simultaneously interdicts one or more edges while facing edge interdiction costs. \rem{The first (resp. second) player seeks to maximize the value of effective (resp. interdicted) flow net the total transportation (resp. interdiction) cost.}Using our existence result on posets and strict complementary slackness in linear programming, we show that the \rev{Nash equilibria} of this game can be \rev{fully} described using primal and dual solutions of a minimum-cost circulation problem. Our analysis provides a new characterization of the critical components \rev{in the interdiction game}. \rev{It also leads to a polynomial-time approach for equilibrium computation.}


}%





\maketitle
%




\ifarXiv
\vspace{-0.6cm}

\else

\fi

\section{Introduction.}\label{intro}


%

In this article, we study the problem of showing the existence of a probability distribution over a partially ordered set (or poset) that satisfies a set of constraints involving marginal probabilities of the poset's elements and maximal chains. This problem is \rev{essential for} the equilibrium  analysis \rev{and computation} of a generic network \rev{interdiction} game, in which a strategic interdictor seeks to disrupt the flow of a routing entity. In particular, our existence result on posets enables us to show that the equilibrium structure of the game can be described using primal and dual solutions of a minimum-cost circulation problem. 

\subsection{Probability distributions over posets.}

For a given finite nonempty poset, we consider a problem in which each element is associated with a value between 0 and 1; additionally, each maximal chain has a value at most 1. We want to determine if there exists a probability distribution over the subsets of the poset such that: $(i)$ The probability \rev{that} each element of the poset is in a subset is \emph{equal to} its corresponding value; and $(ii)$  the probability  \rev{that} each maximal chain of the poset intersects with a subset is \emph{as large as} its corresponding value. This problem, denoted \FPOC, is equivalent to resolving the feasibility of a polyhedral set. However, geometric ideas -- such as the ones involving the use of Farkas' lemma or Carath{\'e}odory's theorem -- cannot be applied to solve this problem, because they do not capture the  structure of posets. We positively resolve problem \FPOC under two conditions that are naturally satisfied for \rev{typical situations}:

%
\begin{enumerate}
\item The value of each maximal chain is no more than the sum of the values of its elements. 

\item The values of the maximal chains  satisfy a conservation law: 
\rev{For any decomposition of two intersecting maximal chains, the sum of the corresponding maximal chain values is constant.}

\end{enumerate}
%
      
Under these two conditions, we prove the feasibility of problem \FPOC (Theorem~\ref{thm:finally}). First, we show that solving \FPOC is equivalent to proving that the optimal value of an exponential-size linear optimization problem, denoted \OPOC, is no more than 1 (Proposition~\ref{Equivalence}). Then, to optimally solve \OPOC, we design a combinatorial algorithm (Algorithm~\ref{ALG3}) that exploits the relation between the values associated with the poset's elements and maximal chains. Each iteration of the algorithm involves constructing a subposet, selecting its set of minimal elements, and assigning a specific weight to it.
Importantly, in the design of the algorithm, we \rem{need to }ensure that the conservation law satisfied by the values associated with the maximal chains of the poset is preserved after each iteration. This design feature enables us to obtain a relation between maximal chains after each iteration, which leads to optimality guarantee of the algorithm (Propositions~\ref{Never Negative}~-~\ref{Termination}). We show that the optimal value of \OPOC \rem{can be computed in closed form: it }is equal to the largest value associated with an element or maximal chain of the poset, \rev{and} is no more than 1 (Theorem~\ref{optimal value}).

%
%
%
%
%

\rev{In the special case where the value of each maximal chain is an affine function of the constituting elements, we refine our combinatorial algorithm to efficiently solve \OPOC (Proposition~\ref{Same}). Our polynomial algorithm (Algorithm~\ref{PolyALG}) relies on subroutines based on the shortest path algorithm in directed acyclic graphs, and does not require the enumeration of maximal chains.}

 
Next, we show that the feasibility of problem \FPOC on posets is crucial for the equilibrium analysis of a class of two-player interdiction games on flow networks. 

\subsection{Network \rev{interdiction} games.}
We  model a network \rev{interdiction} game between player 1 (routing entity) that sends its flow through the network while facing heterogeneous path transportation costs; and player 2 (interdictor) who simultaneously chooses an interdiction plan comprised of one or more edges. Player 1 (resp. player 2) seeks to maximize the value of effective (resp. interdicted) flow net the transportation (resp. interdiction) cost. We adopt mixed strategy Nash equilibria as the solution concept of this game.

Our \rev{interdiction} game is general in that it models heterogeneous costs of transportation and interdiction. It models the strategic situation in which player 1 is an operator who wants to route flow (e.g. water, oil, or gas) through pipelines, while player 2 is an attacker who targets multiple pipes in order to steal or disrupt the flow. \rev{Another relevant} setting is the one where player 1 is a malicious entity composed of routers who carry illegal (or dangerous) goods through a transportation network (i.e., roads, rivers, etc.), and player 2 is a security agency that dispatches interdictors to intercept malicious routers and prevent the illegal goods from crossing the network. In both these settings, mixed strategies can be viewed as the players introducing randomization in implementing their respective actions. For instance, player 1's mixed strategy  models a randomized choice of paths for routing its flow of goods through the network, while player 2's mixed strategy indicates a randomized dispatch of interdictors to disrupt or intercept the flow. 


The existing literature in network interdiction \rev{and robust flow problems} has dealt with this type of problems in a sequential (Stackelberg) setting (see \citet{DBLP:reference/complexity/AvenhausC09}, \citet{Ball:1989:FMV:2309676.2309706},  \citet{doi:10.1287/mnsc.21.5.531}, \citet{Wollmer1964}).  Typically, these problems are solved using integer programming techniques, and are staple for designing system interdiction and defense (see \citet{doi:10.1002/net.10001},  \citet{BertsimasNS13}, \citet{Morton_98}, \citet{4753111}, \citet{NET:NET21561}, \citet{Wood19931}). However, these models do not capture the situations in which the interdictor is capable of simultaneously interdicting multiple edges, possibly in a randomized manner.  
\rev{Our model is closely tied to the randomized network interdiction problem considered by \citet{BertsimasNO13}, in which} the interdictor first randomly interdicts a fixed number of edges, and then the operator routes a feasible flow in the network. The interdictor's goal is to minimize the largest amount of flow that reaches the destination node. Although this model is equivalent to a simultaneous game, our model \rev{differs} in that we do not impose any restriction on the number of edges that can be simultaneously interdicted. Additionally, we account for transportation and interdiction costs faced by the players. 



Our work is also motivated by previous problems studied in network security games (e.g. \citet{Baykal-GursoyDPG14}, \citet{10.1007/978-3-642-35582-0_20}, \citet{Szeto}). However, the available results in this line of work are for simpler cases, and do not apply to our model.
%
%
%
%
%
Related to our work are the network security games proposed by  \citet{Washburn1995} and Gueye and Marbukh \cite{10.1007/978-3-642-34266-0_11}. \citet{Washburn1995}\rem{, the authors} consider a simultaneous game where an evader chooses one source-destination path and the interdictor inspects one edge. \rem{In this model, }The interdictor's (resp. evader's) objective is to maximize (resp. minimize) the probability \rev{that} the evader is detected by the interdictor. Gueye and Marbukh  \cite{10.1007/978-3-642-34266-0_11} model an operator who routes a feasible flow in the network, and an attacker who disrupts one edge. The attacker's (resp. operator's) goal is to maximize (minimize) the amount of lost flow. The attacker \rev{also} faces a cost of attack.
In contrast, our model allows the interdictor to inspect multiple edges simultaneously, and \rev{also} accounts for the transportation cost faced by the routing entity. 

The generality of our model renders known methods for analyzing security games inapplicable to our game. Indeed, prior work has considered solution approaches based on max-flows and min-cuts, and used these objects as metrics of criticality for network components (see  \citet{Assadi:2014:MVP:2685231.2685250}, \citet{6062676},  \citet{10.1007/978-3-642-35582-0_20}). However, these objects cannot be applied to describe the critical network components in our game due to the heterogeneity of path interdiction probabilities resulting from the transportation costs.
%
%
%
%
A related issue is that  computing a Nash equilibrium of our game is \rev{challenging} because of the large size of the players' action sets. Indeed, player 1 (resp. player 2) chooses a probability distribution over an infinite number of feasible flows (resp. exponential number of subsets of edges). Therefore, well-known algorithms for computing (approximate) Nash equilibria are practically inapplicable for this setting (see Lipton et al. \cite{Lipton:2003:PLG:779928.779933}, McMahan et al. \cite{Mcmahan03planningin}, \rev{and \citet{10.1007/978-3-540-77105-0_9}}). Guo et al. \cite{Guo:2016:OII:3060832.3060972} developed a column and constraint generation algorithm to approximately solve their network security game. However, it cannot be applied to our model due to the transportation and interdiction costs that we consider.



Instead, we propose an approach for \rev{solving} our game based on a minimum-cost circulation problem, which we denote \original, and our existence problem on posets \FPOC. The main \rev{findings are the following:}


\begin{enumerate}
\item \rev{\rev{\emph{Every}} Nash equilibrium of the game can be described using primal and dual optimal solutions of \original (\rev{Theorem}~\ref{One_NE_general}). \rev{Specifically,} the expected flow of an equilibrium routing strategy for player 1 is an optimal flow of \original. Furthermore, equilibrium interdiction strategies for player 2 are such that the marginal interdiction probabilities of the network edges and source-destination paths can be expressed using the optimal dual solutions and the properties of the network.
%
%
In fact, these equilibrium conditions rely on our results on posets (Theorems~\ref{thm:finally} and \ref{optimal value}) for the existence problem \FPOC.
%
%
%
The players' payoffs in equilibrium can be expressed in terms of the optimal solutions of \original, and are independent of the chosen path decomposition of player 1's strategy. \citet{BertsimasNS13} showed that such property does not necessarily hold in path-based formulations of the Robust Maximum Flow Problem (RMFP) with multiple interdictions.}

\item \rev{Our solution approach shows that Nash equilibria of the game can be computed in polynomial time: The first step consists of solving the minimum-cost circulation problem \original using known algorithms (see \citet{Karmarkar1984} and \citet{Orlin1993}). The optimal flow is shown to be an equilibrium routing strategy for player 1. Using the optimal dual solution, the second step of our approach consists of running our polynomial algorithm on posets (Algorithm~\ref{PolyALG}) to construct an equilibrium interdiction strategy for player 2 that satisfies marginal interdiction probabilities. This result contrasts with the \emph{NP}-hardness of the RMFP (\citet{DISSER202018}).}

%
%
%



\item \rev{The critical components in the network can be computed from a primal-dual pair of solutions of \original that satisfy strict complementary slackness.} Specifically, the primal (resp. dual) solution provides the paths (resp. edges) that are chosen (resp. interdicted) in at least one Nash equilibrium of the game (\rev{Proposition}~\ref{Nec_conds}). This result generalizes the classical min-cut-based metrics of network criticality previously studied in the network interdiction literature (see   \citet{ASSIMAKOPOULOS1987413}, McMasters et al. \cite{doi:10.1002/nav.3800170302},   \citet{Washburn1995},  \citet{Wood19931}). Indeed, we show that in our more general setting, multiple edges in a source-destination path may be interdicted in equilibrium, and cannot be represented with a single cut of the network. \rem{We address this issue by computing the dual solutions of $(\mathcal{M})$, and by constructing an equilibrium interdiction strategy using our \rev{polynomial} algorithm (Algorithm~\ref{PolyALG}) for posets.[May be removed (since it is repetitive)]}
%
%
%
%

\end{enumerate}

The rest of the paper is organized as follows: In Section~\ref{sec:problem}, we pose our existence problem on posets, and introduce our main feasibility result. 
\rev{Section~\ref{Big Proof} presents and analyzes a combinatorial algorithm for solving the existence problem. A polynomial implementation of the algorithm is described in Section~\ref{sec:refinement} when the maximal chain values are affine.}
\rev{Applications} of our \rev{results on posets} are then demonstrated in Section~\ref{sec:games}, where we study our \rev{strategic} network \rev{interdiction} game. Lastly, we provide some concluding remarks in Section~\ref{sec:conclusion}.

 
\section{\rev{Probability distributions on posets.}}\label{sec:problem}

In this section, we first recall some standard definitions in order theory. We then pose our problem of proving the existence of probability distributions over partially ordered sets, and introduce our main result about its feasibility. 



\subsection{\rev{Preliminaries.}}\label{sec:order_theory}
A finite \emph{partially ordered set} or \emph{poset} $P$ is a pair $(\ground,\preceq)$, where $\ground$ is a finite set and $\preceq$ is a partial order on $X$, i.e., $\preceq$ is a binary relation on $\ground$ satisfying:
\begin{itemize}
\item[--] Reflexivity: \rev{For all} $x \in \ground, \ x \preceq x$ in $P$.
\item[--] Antisymmetry: \rev{For all} $(x,y) \in \ground^2$, if $x \preceq y$ in $P$ and $y \preceq x$ in $P$, then $x = y$. 
\item[--] Transitivity: \rev{For all} $(x,y,z) \in \ground^3$, if $x \preceq y$ in $P$ and $y \preceq z$ in $P$, then $x \preceq z$ in $P$.
\end{itemize}

Given $(x,y) \in \ground^2$, we denote $x \prec y$ in $P$ if $x \preceq y$ in $P$ and $x \neq y$. We say that $x$ and $y$ are \emph{comparable} in $P$ if either $x \prec y$ in $P$ or $y \prec x$ in $P$. On the other hand, $x$ and $y$ are \emph{incomparable} in $P$ if neither $x \prec y$ in $P$\rev{,} nor $y \prec x$ in $P$. We say that $x$ is \emph{covered} in $P$ by $y$, denoted $x \prec: y$ in $P$, if $x \prec y$ in $P$ and there does not exist $z \in \ground$ such that $x \prec z$ in $P$ and $z \prec y$ in $P$. When there is no confusion regarding the poset, we abbreviate $x \preceq y$ in $P$ by writing $x \preceq y$, etc.

Let $Y$ be a nonempty subset of $X$, and let $\preceq_{\restrict{Y}}$ denote the restriction of $\preceq$ to $Y$. Then, $\preceq_{\restrict{Y}}$ is a partial order on $Y$, and $(Y,\preceq_{\restrict{Y}})$ is a \emph{subposet} of $P$. A poset $P = (\ground,\preceq)$ is called a \emph{chain} (resp. \emph{antichain}) if every distinct pair of elements in $\ground$ is comparable (resp. incomparable) in $P$.  Given a poset $P = (X,\preceq)$, a nonempty subset $Y \subseteq \ground$ is a \emph{chain} (resp. an \emph{antichain}) in $P$ if the subposet $(Y,\preceq_{\restrict{Y}})$ is a chain (resp. an antichain). A single element of $\ground$ is both a chain and an antichain.


Given a poset $P = (X,\preceq)$, an element $x \in \ground$ is a \emph{minimal} element (resp. \emph{maximal} element) if there are no elements $y \in \ground$ such that $y \prec x$ (resp. $x \prec y$).  Note that any chain has a unique minimal and maximal element. A chain $C \subseteq \ground$ (resp. antichain $A \subseteq \ground$) is \emph{maximal} in $P$ if there are no other chains $C^\prime$ (resp. antichains $A^\prime$) in $P$ that contain $C$ (resp. $A$). Let $\mathcal{C}$ and $\mathcal{A}$ respectively denote the set of maximal chains and antichains in $P$. A maximal chain $C \in \mathcal{C}$ of size $n$ can be represented as $C = \{x_1,\dots,x_n\}$ where \rev{for all} $k \in \llbracket 1,n-1\rrbracket, \ x_k \prec:x_{k+1}$. We state the following property:

\begin{lemma}\label{Minimal Elements}
Given a finite nonempty poset $P$, the set of minimal elements of $P$ is an antichain of $P$, and intersects with every maximal chain of $P$.
\end{lemma}

%
%

\emph{Proof in Appendix \ref{sec:additional}.}

Given a poset $P = (X,\preceq)$, we consider its \rev{directed} \emph{cover} graph, denoted $H_P = (X,E_P)$. $H_P$ is a \rev{directed acyclic} graph whose set of \rev{nodes} is $\ground$, and whose set of edges is given by $E_P \coloneqq \{(x,y) \in \ground^2 \ | \ x \prec: y\}$. When $H_P$ is represented such that for all \revm{$(x,y) \in X^2$ with $x \prec:y$}, the vertical coordinate of the \rev{node} corresponding to $y$ is higher than the vertical coordinate of the \rev{node} corresponding to $x$, the resulting diagram is called a \emph{Hasse diagram} of $P$. 

We now introduce the notion of subposet generated by a subset of maximal chains. 
Given a poset $P = (\ground,\preceq)$, let $\ground^\prime \subseteq \ground$ be a subset of elements, let $\Chains^\prime \subseteq \Chains$ be a subset of maximal chains of $P$, and consider the binary relation $\preceq_{\Chains^\prime}$ defined by\rev{: for all} $(x,y) \in {\ground^\prime}^2, \ x \preceq_{\Chains^\prime} y \rev{\text{ if and only if }} (x = y) \text{ or }(\rev{\text{there exists }} C \in \Chains^\prime $ such that $x,y \in C \text{ and } x \prec y).$ Furthermore, we \rev{assume} that if $C^1 = \{x_{-k},\dots,x_{-1},x^*,x_1,\dots,x_n\}$ and $C^2 = \{y_{-l},\dots,y_{-1},x^*,y_1,\dots,y_m\}$ are in $\Chains^\prime$ and intersect in $x^* \in \ground^\prime$, then $\Chains^\prime$ also contains $C_1^2 = \{x_{-k},\dots,x_{-1},x^*,y_1,\dots,y_m\}$ and $C_2^1 =\{y_{-l},\dots,y_{-1},x^*,x_1,\dots,x_n\}$. In other words, $\Chains^\prime$ preserves the decomposition of  maximal chains intersecting in $\ground^\prime$. Then, the following lemma \rev{holds}:



\begin{lemma}\label{new poset general}
Consider the poset $P = (X,\preceq)$, a subset $\ground^\prime \subseteq \ground$, and a subset $\Chains^\prime \subseteq \Chains$ that preserves the decomposition of maximal chains intersecting in  $\ground^\prime$.  Then, $P^\prime = (\ground^\prime,\preceq_{\Chains^\prime})$ is also a poset. 
Furthermore, for any maximal chain $C$ of $P^\prime$ of size at least two, there exists a maximal chain $C^\prime$ in $\Chains^\prime$ such that $C = C^\prime \cap \ground^\prime$.
\end{lemma}

\emph{Proof in Appendix~\ref{sec:additional}.}


The subposet $P^\prime = (\ground^\prime,\preceq_{\Chains^\prime})$ of $P$ in Lemma~\ref{new poset general} satisfies the property that if two elements in $\ground^\prime$ are comparable in $P$, and belong to a same maximal chain $C \in \Chains^\prime$, then they are also comparable in $P^\prime$. Graphically, this is equivalent to removing the edges from the Hasse diagram $H_P$ if their two end nodes do not belong to a same maximal chain $C \in \Chains^\prime$.

\begin{example}
Consider the poset $P$ represented by the Hasse Diagram $H_P$ in Figure~\ref{fig:def}.
\begin{figure}[htbp]
 \centering
        \begin{tikzpicture}[->,>=stealth',shorten >=0pt,auto,x=1.6cm, y=2cm,
  thick,main node/.style={circle,draw},main node2/.style={circle,draw,inner sep = 0.06cm},flow_a/.style ={blue!100}]
\tikzstyle{edge} = [draw,thick,->]
\tikzstyle{cut} = [draw,very thick,-]
\tikzstyle{flow} = [draw,line width = 1pt,->,blue!100]
\small
	\node[main node] (1) at (0,0) {$1$};
	\node[main node] (2) at (0.5,0.5) {$3$};
	\node[main node] (3) at (0,1) {$4$};
	\node[main node] (4) at (1,0) {$2$};
	\node[main node] (5) at (1,1) {$5$};
	\node[main node] (6) at (1,1.6) {$6$};
	
	\path[edge]
	(1) edge (2)
	(2) edge  (3)
	(4) edge (2)
	(2) edge (5)
	(5) edge (6);

\node[main node] (10) at (0+5,0) {$1$};
	\node[main node] (20) at (0.5+5,0.5) {$3$};
	\node[main node] (30) at (-1+5,0) {$4$};
	\node[main node] (40) at (1+5,0) {$2$};
	\node[main node] (60) at (0.5+5,1.1) {$6$};
	
	\path[edge]
	(10) edge (20)
	(20) edge (60)
	(40) edge (20);

\normalsize
\end{tikzpicture}
    \caption{On the left is represented a Hasse diagram of a poset $P$. On the right is represented a Hasse diagram of the subposet $P^\prime = (\ground^\prime,\preceq_{\Chains^\prime})$ of $P$, where $\ground^\prime = \{1,2,3,4,6\} $ and $\Chains^\prime = \{\{1,3,5,6\},\{2,3,5,6\}\}$.}
    \label{fig:def}
\end{figure}
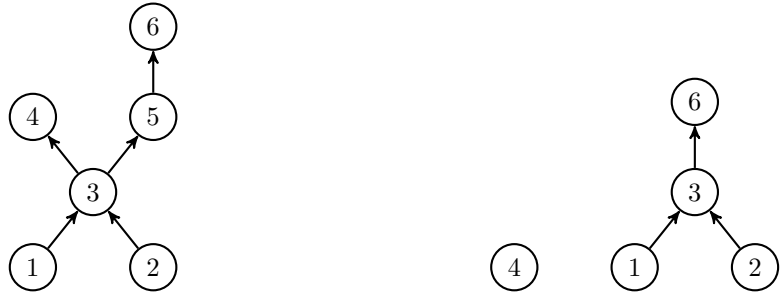

We observe that $1 \prec 4$, $2 \prec: 3$; $1$ and $3$ are comparable, but $4$ and $6$ are incomparable; $\{2,4\}$ is a chain in $P$, but is not maximal since it is contained in the maximal chain $\{2,3,4\}$. Similarly, $\{4\}$ is an antichain in $P$, but is not maximal since it is contained in the maximal antichain $\{4,5\}$. The set of maximal chains and antichains of $P$ are given by $\mathcal{C} = \{\{1,3,4\},\{2,3,5,6\},\{1,3,5,6\},\{2,3,4\}\}$ and $\mathcal{A} = \{\{1,2\},\{3\},\{4,5\},\{4,6\}\}$, respectively. The set of minimal elements of $P$ is given by $\{1,2\}$, and intersects with every maximal chain in $\mathcal{C}$. Finally, $P^\prime = (\ground^\prime,\preceq_{\Chains^\prime})$, where $\ground^\prime = \{1,2,3,4,6\} $ and $\Chains^\prime = \{\{1,3,5,6\},\{2,3,5,6\}\}$, is a poset \rev{as} illustrated in Figure~\ref{fig:def}.
\hfill $\triangle$
\end{example}

\subsection{\rev{Problem formulation and main result.}}\label{sec:Setting}

Consider a finite nonempty poset $P = (\ground,\preceq)$. Let $\mathcal{P}\coloneqq2^\ground$ denote the power set of $\ground$, and let $\Delta(\mathcal{P}) \coloneqq \{\sigma \in \revm{\mathbb{R}_{\geq0}^{\mathcal{P}}} \ | \ \sum_{\att \in \mathcal{P}} \sigma_\att = 1\}$ denote the set of probability distributions over $\mathcal{P}$. We are concerned with the setting where each element $x \in X$ is associated with a value  $\rho_x \in [0,1]$, and each maximal chain $C \in \Chains$ has a value $\pi_C \leq 1$. Our problem is to determine if there exists a probability distribution $\sigma \in \Delta(\mathcal{P})$ such that for every element $x \in \ground$, the probability that $x$ is in a subset $S \in \mathcal{P}$ is equal to $\rho_x$; and for every maximal chain $C \in \mathcal{C}$, the probability that $C$ intersects with a subset $S \in \mathcal{P}$ is at least $\pi_C$. That is,\hypertarget{DDD}{}
%
\begin{subnumcases}{(\mathcal{D}): \quad \exists \, \sigma \in \revm{\mathbb{R}_{\geq0}^{\mathcal{P}}} \ \text{ such that }\ }
 \displaystyle \ \, \sum_{\{\att \in \mathcal{P}\, | \, x \in \att\}}\sigma_{\att}  = \rho_{x}, & $\forall x \in X,$ \label{Equal}\\
  \displaystyle \sum_{\{\att \in \mathcal{P} \, | \,\att \cap C \neq \emptyset\}} \sigma_{\att}\geq \pi_C, & $\forall C \in \mathcal{C},$ \label{Inequal}\\
\quad  \ \ \, \sum_{S \in \mathcal{P}} \sigma_S = 1. &    \label{Total}
\end{subnumcases}


For the case in which $\pi_C \leq 0$ for all maximal chains $C \in \mathcal{C}$, constraints \eqref{Inequal} can be removed, and the feasibility of \FPOC follows from Carath\'eodory's theorem. However, no known results can be applied to the general case. Note that although \eqref{Equal}-\eqref{Total} form a polyhedral set, Farkas' lemma cannot be directly used to evaluate its feasibility. Instead, in this article, we study the feasibility of \FPOC using order-theoretic properties of the problem. We assume two natural conditions on $\rho= (\rho_x)_{x \in \ground}$ and $\pi= (\pi_C)_{C \in \Chains}$, which we introduce next.

%
%
%

Firstly, for feasibility of \FPOC, $\rho$ and $\pi$ must \rev{necessarily} satisfy the following inequality:
%
\begin{align}
\forall C \in \mathcal{C}, \quad \sum_{x \in C} \rho_x \geq \pi_C.\label{Nec Cond}
\end{align}
Indeed, if \FPOC is feasible, then for $\sigma \in \revm{\mathbb{R}_{\geq0}^{\mathcal{P}}}$ satisfying \eqref{Equal}-\eqref{Total}, the following holds:
\begin{align*}
\forall C \in \mathcal{C}, \ \sum_{x \in C}\rho_x \overset{\eqref{Equal}}{=} \sum_{x \in C}\sum_{\{\att \in \mathcal{P}\, | \, x \in \att\}}\sigma_{\att} = \sum_{\att \in \mathcal{P}}\sigma_{\att}\sum_{x \in C}\mathds{1}_{\{x \in \att\}} = \sum_{\att \in \mathcal{P}}\sigma_{\att}|S \cap C| \geq \sum_{\{\att \in \mathcal{P} \, | \, \att \cap C \neq \emptyset\}}\sigma_{\att} \overset{\eqref{Inequal}}{\geq} \pi_C.
\end{align*}

%
That is, the necessity of \eqref{Nec Cond} follows from the fact that for any probability distribution over $\mathcal{P}$, and any subset of elements $C \subseteq X$, the probability that $C$ intersects with a subset $S \in \mathcal{P}$ is upper bounded by the sum of the probabilities with which each element in $C$ is in a subset $S \in \mathcal{P}$.
%

\begin{figure}[ht]
 \centering
        \begin{tikzpicture}[->,>=stealth',shorten >=0pt,auto,x=1.6cm, y=2cm,
  thick,main node/.style={circle,draw},main node2/.style={circle,draw,inner sep = 0.06cm},flow_a/.style ={blue!100}]
\tikzstyle{edge} = [draw,thick,->]
\tikzstyle{cut} = [draw,very thick,-]
\tikzstyle{flow} = [draw,line width = 1pt,->,blue!100]
\small

	\node[main node] (6) at (2.5,0) {1};
	\node[main node] (7) at (3,0.5) {3};
	\node[main node] (8) at (2.5,1) {4};

	\node[main node] (12) at (4+0.5,0.5) {3};
	\node[main node] (13) at (4.5+0.5,0) {2};
	\node[main node] (14) at (4.5+0.5,1) {5};
	\node[main node] (14b) at (4.5+0.5,1.6) {6};

	\node[main node] (15) at (5.5+1,0) {1};
	\node[main node] (16) at (6+1,0.5) {3};
	\node[main node] (17) at (6.5+1,1) {5};
	\node[main node] (17b) at (6.5+1,1.6) {6};

	\node[main node] (18) at (8+1.5,0.5) {3};
	\node[main node] (19) at (7.5+1.5,1) {4};
	\node[main node] (20) at (8.5+1.5,0) {2};
	
%
	

	\path[edge]
	(6) edge (7)
	(7) edge  (8)
	(13) edge (12)
	(12) edge (14)
	(15) edge (16)
	(16) edge (17)
	(18) edge (19)
	(20) edge (18)
	(14) edge (14b)
	(17) edge (17b);
	
%
\node (110) at (3.1,1.05) {$C^1$};
\node (110) at (3.9+0.5,1.05) {$C^2$};
\node (110) at (5.9+1.0,1.02) {$C_1^2$};
\node (110) at (8.1+01.5,1.02) {$C_2^1$};

\normalsize
\end{tikzpicture}
    \caption{Four maximal chains of the poset shown in Figure~\ref{fig:def}.}
    \label{small cross}
\end{figure}
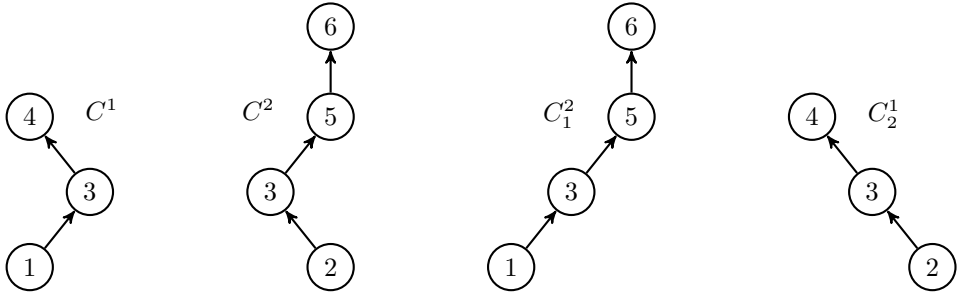

Secondly, we \rev{assume} that $\pi$ satisfies a specific condition for each pair of maximal chains that intersect each other. Consider any pair of maximal chains $C^1$ and $C^2$ of $P$, with $C^1 \cap C^2 \neq \emptyset$. Let $x^* \in C^1 \cap C^2$, and let us rewrite $C^1 = \{x_{-k},\dots,x_{-1},x^*,x_{1},\dots,x_{n}\}$ and $C^2 = \{y_{-l},\dots,y_{-1},x^*,y_{1},\dots,y_{m}\}$. Then, $P$ also contains two maximal chains $C_1^2 = \{x_{-k},\dots,x_{-1},x^*,y_{1},\dots,y_{m}\}$ and $C_2^1 = \{y_{-l},\dots,y_{-1},x^*,x_{1},\dots,x_{n}\}$ that satisfy $C^1 \cup C^2 = C_1^2 \cup C_2^1$; see Figure~\ref{small cross} for an illustration. \rev{We require $\pi$ to satisfy} the following condition:
%
\begin{align}
\pi_{C^1} + \pi_{C^2} = \pi_{C_1^2} + \pi_{C_2^1}. \label{Conservation}
\end{align}
\rev{Essentially}, \eqref{Conservation} can be viewed as a \emph{conservation law} on the maximal chains in $\Chains$.
 %
%
%

We now present our main result regarding the feasibility of \FPOC, under conditions \eqref{Nec Cond} and \eqref{Conservation}.

\begin{theorem}\label{thm:finally}
The problem \FPOC is feasible for any finite nonempty poset $(X,\preceq)$, with parameters $\rho = (\rho_x) \in [0,1]^{X}$ and $\pi = (\pi_C) \in (-\infty,1]^{\mathcal{C}}$ that satisfy~\eqref{Nec Cond} and \eqref{Conservation}.  
\end{theorem}

This result plays a crucial role in solving a \rev{two-player interdiction game on a flow network (Section~\ref{sec:games}).} The game involves a ``router'' who sends a flow of goods to maximize her value of flow crossing the network while facing transportation costs, and an ``interdictor'' who inspects one or more network edges to maximize the value of interdicted flow while facing interdiction costs. Our \rev{equilibrium} analysis in Section~\ref{sec:games} shows that \rev{interdiction strategies in Nash equilibria interdict each edge $x$ with a probability $\rho_x$, and interdict each path $C$ with a probability at least $\pi_C$}. Essentially, for this game, $(\rho_x)$ and $(\pi_C)$ are governed by network properties, such as edge transportation and interdiction costs, and naturally satisfy \eqref{Nec Cond} and \eqref{Conservation}.
When the network is a directed acyclic graph, a partial order can be defined on the set of edges, such that the set of maximal chains is exactly the set of source-destination paths of the network. Thus, showing the existence of interdiction strategies satisfying the above-mentioned \rev{equilibrium conditions} is an instantiation of the problem \FPOC. \rev{In fact, Theorem~\ref{thm:finally} is useful for deriving several properties satisfied by the equilibrium strategies of this network interdiction game.}

\rem{
Importantly, note that \FPOC may not be feasible if $P$ is not a poset. Let us consider the following example:  $X = \{1,2,3\}$, $\mathcal{C} = \{\{1,2\},\{1,3\},\{2,3\}\}$, $\rho_x = 0.5$, $\forall x \in X$, and $\pi_C = 0.5$, $\forall C \in \mathcal{C}$. There is no poset that has $\Chains$ as its set of maximal chains. If $\sigma \in \mathbb{R}_+^{\mathcal{P}}$ satisfies \eqref{Equal} and \eqref{Inequal}, then necessarily, $\sigma_{\{x\}} = 0.5, \ \forall x \in X$. However, this implies that $\sum_{S \in \mathcal{P}}\sigma_S \geq 1.5 >1$, which renders \FPOC infeasible for this example. }

 \rev{It is important to note that \FPOC may not be feasible if the conservation law \eqref{Conservation} is not satisfied, as illustrated in the following counterexample:
  
\begin{example}
Let $P$ be the poset represented by the Hasse Diagram in Figure~\ref{Counter_Poset}.
 \begin{figure}[ht]
 \centering
        \begin{tikzpicture}[->,>=stealth',shorten >=0pt,auto,x=1.4cm, y=1.3cm,
  thick,main node/.style={circle,draw},main node2/.style={circle,draw,inner sep = 0.06cm},flow_a/.style ={blue!100}]
\tikzstyle{edge} = [draw,thick,->]
\tikzstyle{cut} = [draw,very thick,-]
\tikzstyle{flow} = [draw,line width = 1pt,->,blue!100]
\small

\def \Spp{2}
\def \Ofp{3}
\def \lep{1}
	\node[main node] (1) at (0,0) {$1$};
	\node[main node] (3) at (0,1) {$3$};
	\node[main node] (2) at (1,0) {$2$};
	\node[main node] (4) at (1,1) {$4$};
	\node[main node] (5) at (0,2) {$5$};
	\node[main node] (6) at (1,2) {$6$};
	
	\path[edge]
	(1) edge (4)
	(1) edge  (3)
	(2) edge (4)
	(4) edge (5)
	(3) edge (5)
	(4) edge (6);

\draw[dashed,-] (\Ofp/2 + 1/2,-0.5) -- (\Ofp/2 + 1/2,2.5);

\normalsize
	\node (100) at (-0.55+0+\Ofp,1) {$C^1$};
	
	\node (100) at (-0.55+\lep/2+\Ofp+\Spp,1) {$C^2$};
	\node (100) at (-0.55+\lep/2+\Ofp+2*\Spp,1) {$C^3$};
	\node (100) at (-0.55+\lep/2+\Ofp+3*\Spp,1) {$C^4$};
	\node (100) at (-0.55+0+\Ofp+4*\Spp,1) {$C^5$};

	\small
	
	\node[main node] (11) at (0+\Ofp,0) {$1$};
	\node[main node] (13) at (0+\Ofp,1) {$3$};
	\node[main node] (15) at (0+\Ofp,2) {$5$};

	\path[edge]
	(11) edge (13)
	(13) edge (15);

	\node[main node] (21) at (0-\lep/2+\Ofp+\Spp,0) {$1$};
	\node[main node] (24) at (\lep/2+\Ofp+\Spp,1) {$4$};
	\node[main node] (25) at (-\lep/2+\Ofp+\Spp,2) {$5$};
	
	\path[edge]
	(21) edge (24)
	(24) edge (25);
	
	\node[main node] (31) at (-\lep/2+\Ofp+2*\Spp,0) {$1$};
	\node[main node] (34) at (\lep/2+\Ofp+2*\Spp,1) {$4$};
	\node[main node] (36) at (\lep/2+\Ofp+2*\Spp,2) {$6$};	
	
	\path[edge]
	(31) edge (34)
	(34) edge (36);

	\node[main node] (42) at (\lep/2+\Ofp+3*\Spp,0) {$2$};
	\node[main node] (44) at (\lep/2+\Ofp+3*\Spp,1) {$4$};
	\node[main node] (45) at (-\lep/2+\Ofp+3*\Spp,2) {$5$};
	
	\path[edge]
	(42) edge (44)
	(44) edge (45);

	\node[main node] (52) at (0+\Ofp+4*\Spp,0) {$2$};
	\node[main node] (54) at (0+\Ofp+4*\Spp,1) {$4$};
	\node[main node] (56) at (0+\Ofp+4*\Spp,2) {$6$};	
	
	\path[edge]
	(52) edge (54)
	(54) edge (56);
	
\normalsize
\end{tikzpicture}
    \caption{\rev{Hasse diagram of a poset $P$ (left), and its five maximal chains (right).}}
    \label{Counter_Poset}
\end{figure}
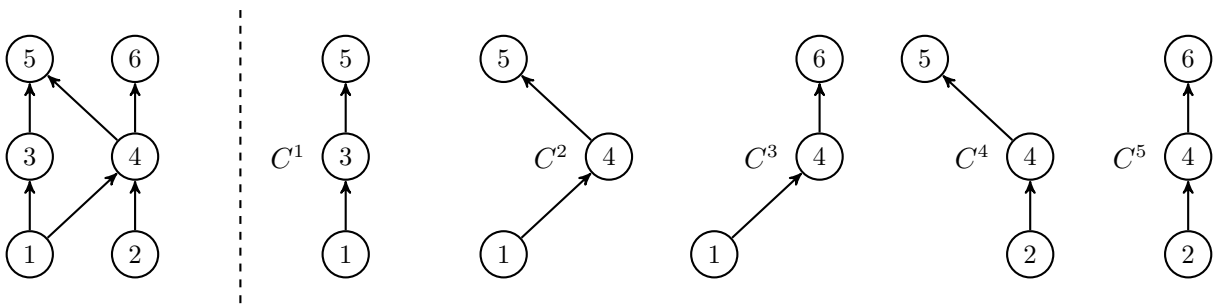

 In this poset, the maximal chains are $C^1=\{1,3,5\}$, $C^2 = \{1,4,5\}$, $C^3 = \{1,4,6\}$, $C^4 = \{2,4,5\}$, $C^5 = \{2,4,6\}$. Consider the following values: $\rho_x = 0.4$ for $x \in \{1,4,5\}$, and $\rho_x = 0$ for $x \in \{2,3,6\}$; $\pi_{C^5} = 0.4$ and $\pi_C = 0.8$ for $C \in \Chains\backslash\{C^5\}$. We note that \eqref{Nec Cond} is satisfied. However, $\pi_{C^2} + \pi_{C^5} = 1.2 \neq 1.6 = \pi_{C^3} + \pi_{C^4}$, which violates \eqref{Conservation}.
%
%
If $\sigma \in \revm{\mathbb{R}_{\geq0}^{\mathcal{P}}}$ satisfies \eqref{Equal} and \eqref{Inequal}, then necessarily, $\sigma_{\{x\}} = 0.4$ for all $x \in \{1,4,5\}$, which violates \eqref{Total}. Thus, problem \FPOC is infeasible for this example. \hfill $\triangle$

\end{example}}

\rem{Thus, in proving Theorem~\ref{thm:finally}, we \rev{assume} that the problem \FPOC is defined for a poset.} \rem{Henceforth, in proving Theorem~\ref{thm:finally}, we consider that conditions \eqref{Nec Cond} and \eqref{Conservation} are satisfied.} 

Next, we show that \FPOC is feasible if and only if the optimal value of a linear program is no more than 1.





\subsection{Equivalent optimization problem.}\label{sec:equivalents}
\rem{Consider the problem \FPOC for a given poset $P = (X,\preceq)$, and vectors $\rho \in [0,1]^{X}$ and $\pi \in ]-\infty,1]^{\mathcal{C}}$ satisfying~\eqref{Nec Cond} and \eqref{Conservation}. }
We observe that when $\sum_{x \in X} \rho_{x} \leq1$, a trivial solution for \FPOC is given by: $\widetilde{\sigma}_{\{x\}}= \rho_x$ \rev{for all} $x \in \ground$, and $\widetilde{\sigma}_{\emptyset} = 1 - \sum_{x \in \ground}\rho_x$. The vector $\widetilde{\sigma}$ so constructed indeed represents a probability distribution over $\mathcal{P}$, and satisfies constraints \eqref{Equal}. Furthermore, for each maximal chain $C \in \Chains$, $ \sum_{\{\att \in \mathcal{P} \, | \,\att \cap C \neq \emptyset\}} \widetilde{\sigma}_{\att} = \sum_{x \in C} \rho_x \overset{\eqref{Nec Cond}}{\geq} \pi_C$. Therefore, $\widetilde{\sigma}$ is a feasible solution of \FPOC.  However, in general, $\sum_{x \in X} \rho_{x}$ may be larger than $1$, which prevents the aforementioned construction of $\widetilde{\sigma}$  from being a probability distribution. Thus, to construct a feasible solution of \FPOC, \rev{some probability must be assigned} to subsets of elements of size larger than 1. This is governed by the following quantity\rev{, defined for each maximal chain $C \in \Chains$}:
%
%
%
%
%
%
\begin{align}
\delta_\chain \coloneqq \sum_{x \in \chain} \rho_{x} - \pi_\chain. \label{delta_C}
\end{align}

The role of $\delta = (\delta_C)_{C \in \Chains}$ \rev{in} assigning probabilities to subsets of elements \rev{can be better understood by considering} the following optimization problem:\hypertarget{QQQ}{}
%
%
%
%
%
%
%
%
%
%
%
\ifarXiv
\begin{alignat}{3}
(\mathcal{Q}): \quad \  &\text{minimize}  \quad \quad&& \displaystyle\sum_{\att \in \mathcal{P}} \sigma_\att \nonumber\\[0.1cm]
&\text{subject to}  \quad \quad && \displaystyle\sum_{\mathclap{\{\att \in \mathcal{P}\, | \, x \in \att\}}}\sigma_{\att} = \rho_{x},  && \quad \forall x \in X \label{Equal_2}\\[0.1cm]
& && \displaystyle\sum_{\mathclap{\{\att \in \mathcal{P} \, | \,|\att \cap \chain| \geq 2\}}} \sigma_{\att} (|\att \cap \chain|-1) \leq \delta_\chain,  && \quad \forall \chain \in \Chains \label{small combi}\\[0.1cm]
& &&  \sigma_{\att} \geq0, && \quad \forall \att \in \mathcal{P}.\nonumber
\end{alignat}

\fi


Problems \OPOC and \FPOC are related in that the set of constraints \eqref{Equal}-\eqref{Inequal} is equivalent to the set of constraints \eqref{Equal_2}-\eqref{small combi}; see the proof of Proposition~\ref{Equivalence} below. Furthermore, the objective function in \OPOC is analogous to the constraint \eqref{Total} in \FPOC.  The feasibility of \OPOC is straightforward (for example, $\widetilde{\sigma}$ constructed above is a feasible solution); however,  a feasible solution of \OPOC may not be a probability distribution.

Given a maximal chain $C \in \Chains$, constraint \eqref{small combi} bounds the total amount of probability that can be assigned to subsets that contain more than one element in $C$. One can see that for a subset $S\in \mathcal{P}$ such that $|S\cap C| \leq 1$, the probability $\sigma_S$ assigned to $S$ does not influence constraint \eqref{small combi}.  However, the more elements from $C$ a subset $\att$ contains, the smaller the probability that can be assigned to $\att$, due to scaling by the factor $(|\att \cap C| - 1)$. Thus, $\delta$ determines the amount of probability that can be assigned to larger subsets.


Let $z^*_{\OPOCt}$ denote the optimal value of \OPOC. \rev{Then, the following result holds}:


\begin{proposition}\label{Equivalence}
\FPOC is feasible if and only if $z^*_{\OPOCt} \leq 1$.

\end{proposition}
\proof{Proof of Proposition~\ref{Equivalence}.}

First, let us show that the set of constraints \eqref{Equal}-\eqref{Inequal} is equivalent to the set of constraints \eqref{Equal_2}-\eqref{small combi}. Let $\sigma \in \revm{\mathbb{R}_{\geq0}^{\mathcal{P}}}$ that satisfies $\sum_{\{\att \in \mathcal{P}\, | \, x \in \att\}}\sigma_{\att}  = \rho_{x}$ \rev{for all} $x \in X$. For every maximal chain $\chain \in \Chains$, the following equality \rev{holds}:
\begin{align}
\sum_{x \in \chain} \rho_{x} = \sum_{x \in \chain}\sum_{\{\att \in \mathcal{P}\, | \, x \in \att\}}\sigma_{\att} = \sum_{\att \in \mathcal{P}}\sigma_{\att}\sum_{x \in \chain}\mathds{1}_{\{x \in \att\}} =  \sum_{\{\att \in \mathcal{P} \, | \,\att \cap \chain \neq \emptyset\}} \sigma_{\att} |\att \cap \chain|.\label{int10}
\end{align}

Therefore, for every maximal chain $\chain \in \Chains$, \rev{the following equivalence is satisfied}:
\begin{align}
\sum_{\{\att \in \mathcal{P} \, | \,\att \cap \chain \neq \emptyset\}} \sigma_{\att}\geq \pi_\chain  \overset{\eqref{delta_C},\eqref{int10}}{\Longleftrightarrow} \delta_\chain \geq  \sum_{\{\att \in \mathcal{P} \, | \,\att \cap \chain \neq \emptyset\}}\sigma_{\att} (|\att \cap \chain|-1)=  \sum_{\{\att \in \mathcal{P} \, | \, |\att \cap \chain| \geq 2 \}} \sigma_{\att} (|\att \cap \chain|-1).\label{Eq_inequality}
\end{align}



Now, let us show that \FPOC is feasible if and only if the optimal value of \OPOC satisfies $z^*_{\OPOCt} \leq 1$. 
\begin{itemize}
\item[--] If \rev{there exists} $\sigma \in \revm{\mathbb{R}_{\geq0}^{\mathcal{P}}}$ that satisfies \eqref{Equal}-\eqref{Total}, then we showed that \rem{$\sigma$ also satisfies \eqref{Equal_2}-\eqref{small combi}. Therefore,}$\sigma$ is a feasible solution of \OPOC. Furthermore, the objective value of $\sigma$ is equal to 1, which implies that $z^*_{\OPOCt} \leq 1$. 

\item[--] If $z^*_{\OPOCt} \leq 1$, let $\sigma^*$ be an optimal solution of \OPOC. Necessarily, $\sigma^*_\emptyset = 0$, 
\rev{and the vector $\sigma \in \mathbb{R}^{\mathcal{P}}$ defined as follows is feasible for \FPOC: $\sigma_{\att} = \sigma^*_\att$, \rev{for every} $\att \in \mathcal{P}\backslash\emptyset$, and $\sigma_\emptyset = 1 - z^*_{\OPOCt}.$}
\hfill
\Halmos
\end{itemize}

\endproof

Therefore, proving Theorem~\ref{thm:finally} is equivalent to showing that $z^*_{\OPOCt} \leq 1$. In fact, we show a stronger result, which will be \rev{crucial for solving our network interdiction game} in Section~\ref{sec:games}:
\begin{theorem}\label{optimal value}
$z^*_{\OPOCt} = \max\{\max\{\rho_x, \ x \in \ground\},\max\{\pi_C, \ \chain \in \Chains\}\}$.
\end{theorem}
%
%

%



It is easy to see that $z^*_{\OPOCt} \geq \max\{\max\{\rho_x, \ x \in \ground\},\max\{\pi_C, \ \chain \in \Chains\}\}$. Indeed, any feasible solution $\sigma \in \revm{\mathbb{R}_{\geq0}^{\mathcal{P}}}$ of \OPOC satisfies $\sum_{\att \in \mathcal{P}}\sigma_\att \geq \sum_{\{\att \in \mathcal{P} \, | \, x \in \att \}}\sigma_\att = \rho_x$ \rev{for every} $x \in \ground$, and $\sum_{\att \in \mathcal{P}}\sigma_\att \geq \sum_{\{\att \in \mathcal{P} \, | \, \att \cap \chain \neq \emptyset\}}\sigma_\att \overset{\eqref{Eq_inequality}}{\geq} \pi_\chain$ \rev{for every} $\chain \in \Chains$. To show the reverse inequality, we \rev{must} prove that there exists a feasible solution of \OPOC with objective value equal to $\max\{\max\{\rho_x, \ x \in \ground\},\max\{\pi_C, \ \chain \in \Chains\}\}$. This is the focus of the next section.

%




%


\section{\color{black}Constructive proof of Theorem~\ref{optimal value}.} \label{Big Proof}
\rev{Essentially,} we design a combinatorial algorithm to compute a feasible solution of \OPOC with objective value exactly equal to $\max\{\max\{\rho_x, \ x \in \ground\},\max\{\pi_C, \ \chain \in \Chains\}\}$. Recall from Section~\ref{sec:equivalents} that such a feasible solution is optimal for \OPOC, and can be used to construct a feasible solution of \FPOC; see the proof of Proposition~\ref{Equivalence}.

%



%
%

Before formally introducing our algorithm, we discuss the main ideas behind its design. In each  iteration, the algorithm selects a subset of elements, and assigns a positive weight to it. Let us discuss the execution of the first iteration of the algorithm.
  
Firstly, we \rem{need to }determine the collection of subsets that can be assigned a positive weight without violating any of the constraints in the problem \OPOC. Essentially, this is dictated by the maximal chains $C \in \Chains$ for which $\delta_C = 0$.  Indeed, for any $C \in \Chains$ with $\delta_C = 0$, the following equivalence \rev{holds}: $\sum_{\{\att \in \mathcal{P} \, | \, |\att \cap \chain| \geq 2 \}} \underset{\geq 0}{\underbrace{\sigma_{\att}}} \underset{>0}{\underbrace{(|\att \cap \chain|-1)}} \leq 0$ \rev{if and only if} $\sigma_{\att} = 0$ \rev{for all} $\att \in \mathcal{P}$ such that $|\att \cap \chain| \geq 2.$
\rev{Therefore, our algorithm must select a subset of elements $S \in \mathcal{P}$ that intersects every maximal chain $C \in \Chains$ for which $\delta_C = 0$ in at most one element.}

To precisely characterize this collection of subsets, we consider the notion of subposet generated by a subset of maximal chains, introduced in Section~\ref{sec:order_theory}. In particular, by considering $\Chains^\prime$  the set of maximal chains $C \in \Chains$ such that $\delta_C = 0$, and $X^\prime$ the subset of elements $x \in \ground$ such that $\rho_x >0$, we can show (in Proposition~\ref{Never Negative} below) that the condition stated in  Lemma~\ref{new poset general} is satisfied, and $P^\prime = (\ground^\prime,\preceq_{\Chains^\prime})$ is a poset. Interestingly, \rem{we can then deduce that}the subsets of elements that \rev{the algorithm} can select from at that iteration are the antichains of $P^\prime$. In any poset, a chain and an antichain intersect in at most one element. By definition of $\preceq_{\Chains^\prime}$, this implies that $|S \cap C| \leq 1$ for every antichain $S \subseteq X^\prime$ of $P^\prime$ and every maximal chain $C \in \Chains$ of $P$ such that $\delta_C = 0$.

Now, we need to determine which antichain of $P^\prime$ to select. Let $S^\prime \subseteq \ground^\prime$ denote the subset of elements selected by the algorithm in the first iteration. Recall that an optimal solution of \OPOC satisfies constraints \eqref{Equal}-\eqref{Inequal} with the least total amount of weight assigned to subsets of elements of $X$. Thus, it is desirable that the weight assigned to $S^\prime$ in this iteration contribute towards satisfying all constraints \eqref{Inequal}. To capture this requirement, our algorithm selects $S^\prime$  as the set of minimal elements of $P^\prime$. The selected $S^\prime$ is an antichain of $P^\prime$, intersects with every maximal chain of $P$, and enables us to prove the optimality of the algorithm.

Secondly, we discuss how to determine the maximum amount of weight $w^\prime$ that can be assigned to $S^\prime$ in the first iteration, without violating any of the constraints \eqref{Equal_2} and \eqref{small combi}. This is governed by the remaining chains $C \in \Chains$ for which $\delta_C >0$ and the elements constituting $S^\prime$. If $w^\prime$ is larger than $\frac{\delta_C}{|S^\prime \cap C| - 1}$ for $C \in \mathcal{C}$ such that $|S^\prime \cap C| \geq 2$, then the corresponding constraint \eqref{small combi} will be violated. Similarly, $w^\prime$ cannot be larger than \rev{any} $\rho_x$, $x \in S^\prime$.
%
Thus, the weight \rev{to} assign to $S^\prime$ is:
 $$w^\prime = \min\left\{\min\left\{\rho_x, \ x \in S^\prime\right\},\min\left\{\frac{\delta_C}{|S^\prime \cap C| - 1}, \ C \in \Chains \ | \  \delta_C > 0 \text{ and } |S^\prime \cap C| \geq 2\right\}\right\}.$$ 
%
%
%

At the end of the iteration, \rev{the algorithm} updates the vectors $\rho$ and $\delta$, as well as the sets of elements $\ground^\prime$ and maximal chains $\Chains^\prime$ to consider in subsequent iterations. In particular, we will show that some maximal chains need to be removed in order to preserve the conservation law at each iteration. The algorithm terminates when there are no more elements $x \in \ground$ with positive $\rho_x$. \rem{This completes the discussion of the main points that we \rev{must} account for in designing the algorithm. }We are now in the position to formally present Algorithm~\ref{ALG3}.

%
\begin{algorithm}
\caption{\textbf{: Optimal solution of $(\mathcal{Q})$}}\label{ALG3}

    \hspace*{\algorithmicindent} \textbf{Input}: Finite nonempty poset $P = (\ground,\preceq)$, and vectors $\rho \in \revm{\mathbb{R}_{\geq0}^{\ground}}$,  $\delta \in \revm{\mathbb{R}_{\geq0}^{\Chains}}$.\\
    \hspace*{\algorithmicindent} \textbf{Output}: Vector $\sigma \in \revm{\mathbb{R}_{\geq0}^{\mathcal{P}}}$.
\begin{algorithmic}[1]



\StateNew{$\allpaths{1} \gets \Chains, \quad \quad  \edgeprob{x}{1} \gets \rho_{x}, \ \forall x \in X$, \quad \quad  
$\combiprob{\chain}{1} \gets \delta_\chain, \ \forall \chain \in \allpaths{1}$}\label{Initiate}
\State{$\component{1} \gets \{x \in X \ | \ \edgeprob{x}{1} >0\}$, \quad \quad  $\pathtight{1} \gets  \{\chain \in \allpaths{1} \ | \ \combiprob{\chain}{1} = 0\}$, \quad \quad  $\pathloose{1} \gets \{\chain \in \allpaths{1} \ | \ \combiprob{\chain}{1} > 0\}$}


\State{$k \gets 1$}
\While{$\component{k} \neq \emptyset$}

\State{Construct the poset $\poset{k} = (\component{k},\preceq_{\pathtight{k}})$}\label{alg:Subposet}

\State{\revm{Select} $\setmin{k}$ the set of minimal elements of $\poset{k}$}\label{set_min_elem}
\State{$\weight{k} = \min\{\min\{\edgeprob{x}{k}, \ x \in \setmin{k}\},\min\{\frac{\combiprob{\chain}{k}}{|\setmin{k} \cap \chain| - 1}, \ \chain \in \pathloose{k} \ | \ |\setmin{k} \cap \chain| \geq 2\}\}$,  \ and \ $\sigma_{\setmin{k}} \gets \weight{k}$}\label{max_weight}


\State{$\edgeprob{x}{k+1} \gets \edgeprob{x}{k} - \weight{k} \mathds{1}_{\{x \in \setmin{k}\}}, \ \forall x \in \ground$, \ and \ $\combiprob{\chain}{k+1}  \gets \combiprob{\chain}{k} - \weight{k}(|\setmin{k} \cap \chain|-1)\mathds{1}_{\{|\setmin{k} \cap C| \geq 2\}}, \ \forall \chain \in \Chains$} \label{update_vectors}

\State{$\allpaths{k+1} \gets \{C \in \allpaths{k} \ | \ \text{the minimal element of $C\cap \component{k}$ in $P$ is in } \setmin{k}\}$} \label{hardest_update}

\State{$\component{k+1} \gets \{x \in \component{k} \ | \ \edgeprob{x}{k+1} >0\}$,   $\pathtight{k+1} \gets  \{\chain \in \allpaths{k+1} \ | \ \combiprob{\chain}{k+1} = 0\}$,   $\pathloose{k+1} \gets \{\chain \in \allpaths{k+1} \ | \ \combiprob{\chain}{k+1} > 0\}$}  \label{update_sets}

\State{$k \gets k+1$}

\EndWhile

\end{algorithmic}
\end{algorithm}

We illustrate Algorithm~\ref{ALG3} with an example in Appendix~\ref{app:example}.

Let $\stepmax$ denote the number of iterations of Algorithm~\ref{ALG3}. Since \rev{it has not yet been shown to terminate}, we suppose that $\stepmax \in \mathbb{N} \cup \{+\infty\}$. For every maximal chain $C \in \Chains$, let us define the sequence $(\pi^k_C)_{k \in \llbracket 1,\stepmax +1\rrbracket}$ induced by  Algorithm~\ref{ALG3} as follows: 
\begin{align}
\pi^1_C \rev{\ \coloneqq \ } \pi_C, \text{ and for every } k \in \llbracket 1, \stepmax \rrbracket, \ \pathprob{C}{k+1}  \rev{\ \coloneqq\ } \pathprob{\chain}{k} - \weight{k} \mathds{1}_{\{\setmin{k} \cap \chain \neq \emptyset\}}. \label{update_pi}
\end{align} 
Given $k \in \llbracket 1,\stepmax +1\rrbracket$, $\pi_C^k$ (resp. $\edgeprob{x}{k}$) represents the remaining value associated with the maximal chain $C \in \Chains$ (resp. the element $x \in \ground$) after the first $k-1$ iterations of the algorithm.
%
%
 For convenience, we let $\component{0} \gets \ground$.

We now proceed with proving Theorem~\ref{optimal value}. Our proof consists of three main parts: 
\begin{enumerate}
\item[\textbf{Part 1:}] Algorithm~\ref{ALG3} is well-defined (Proposition~\ref{Never Negative}). 

\item[\textbf{Part 2:}] It terminates and outputs a feasible solution of \OPOC (Proposition~\ref{Terminates}).

\item[\textbf{Part 3:}] It assigns a total weight $\sum_{k=1}^{\stepmax} \weight{k}$ equal to $\max\{\max\{\rho_x, \ x \in \ground\},\max\{\pi_C, \ \chain \in \Chains\}\}$ at termination (Proposition~\ref{Termination}). 

\end{enumerate}

\subsection*{Part 1: Well-definedness  of Algorithm~\ref{ALG3}.}
To show that Algorithm~\ref{ALG3} is well-defined, we need to ensure that at each iteration $k \in \llbracket 1,\stepmax \rrbracket$ of the algorithm, $P^k$ is a poset. 
 Lemma~\ref{new poset general} can be applied to show this, provided that we are able to prove that $\pathtight{k}$ preserves the decomposition of  maximal chains intersecting in $\component{k}$. This property, and some associated results, are stated below:

\begin{proposition}\label{Never Negative}
Each iteration of Algorithm~\ref{ALG3} is well-defined. In particular, for every $k\in \llbracket 1, \stepmax +1\rrbracket$, the following hold:
\begin{enumerate}
\item[(i)] For every maximal chain $C \in \Chains$, $\combiprob{\chain}{k}$ determines the remaining weight that can be assigned to subsets that intersect $C$ at more than one element:
\begin{align}
&\forall \chain \in \Chains, \quad \combiprob{\chain}{k} = \sum_{x \in \chain} \edgeprob{x}{k} - \pathprob{\chain}{k},\label{Relation_k}\\
& \forall \chain \in \allpaths{k}, \quad \combiprob{\chain}{k} \geq 0. \label{Inequality_k}
\end{align}

\item[(ii)] $\allpaths{k}$ preserves the decomposition of maximal chains intersecting in $\component{k-1}$:
\begin{align*}
\forall (C^1,C^2) \in \Chains^2 \ | \ C^1 \cap C^2 \cap \component{k-1} \neq \emptyset, \  (C^1,C^2) \in (\allpaths{k})^2 \Longrightarrow (C_1^2,C_2^1) \in (\allpaths{k})^2. 
\end{align*}

\item[(iii)] $\pi^k$ satisfies the conservation law on the maximal chains of $\allpaths{k}$ that intersect in $\component{k-1}$:
\begin{align}
\forall (C^1,C^2) \in (\allpaths{k})^2 \ | \ C^1 \cap C^2 \cap \component{k-1}\neq \emptyset, \quad \pathprob{C^1}{k} + \pathprob{C^2}{k} = \pathprob{C_1^2}{k} + \pathprob{C_2^1}{k}. \label{Conservation_k}
\end{align}

\item[(iv)] $P^k = (\component{k}, \preceq_{\pathtight{k}})$ is a poset.

\end{enumerate}

%



\end{proposition}

\proof{Proof of Proposition~\ref{Never Negative}.}

We show $(i)-(iv)$ by induction. 

First, consider $k=1$. Since $\allpaths{1} = \Chains$, $\rho^1 = \rho$, $\pi^1 = \pi$, and $\delta^1 = \delta$, then $(i)$ follows from \eqref{Nec Cond} and \eqref{delta_C}. Since $\component{0} = \ground$ and $\allpaths{1} = \Chains$\rem{ contains all maximal chains}, then $(ii)$ is automatically satisfied. $(iii)$ is a direct consequence of \eqref{Conservation}.

Now we apply Lemma~\ref{new poset general} to show $(iv)$, i.e., $P^1 = (\component{1},\preceq_{\pathtight{1}})$ is a poset. Specifically, we show that $\pathtight{1}$ preserves the decomposition of maximal chains intersecting in $\component{1}$. Consider $C^1,C^2 \in \pathtight{1}$ such that $C^1 \cap C^2 \cap \component{1}\neq \emptyset$, and let us consider the other two maximal chains $C_1^2$ and $C_2^1$, which we know from $(ii)$ are in $\allpaths{1}$, since $\component{1} \subseteq \component{0}$. \rem{We need to show that they are also in $\pathtight{1}$. }Let $x^* \in C^1 \cap C^2 \cap \component{1}$, and let us rewrite $C^1 = \{x_{-k},\dots,x_{-1},x_0 = x^*,x_{1},\dots,x_{n}\}$ and $C^2 = \{y_{-l},\dots,y_{-1},y_0 = x^*,y_{1},\dots,y_{m}\}$. Then, $C_1^2 = \{x_{-k},\dots,x_{-1},x^*,y_{1},\dots,y_{m}\}$ and $C_2^1 = \{y_{-l},\dots,y_{-1},x^*,x_{1},\dots,x_{n}\}$. 
\rev{From} $(i)-(iii)$, since $C^1,C^2 \in \pathtight{1}$; the conservation law  is satisfied by $\pi^1$ on the maximal chains in $\allpaths{1}$ intersecting in $\component{0}$; $C_1^2, C_2^1 \in \allpaths{1}$; and  since $\delta^1 \geq 0$ on $\allpaths{1}$:
\begin{align*}
\sum_{i=-k}^{n} \edgeprob{x_i}{1} +  \sum_{j=-l}^{m} \edgeprob{y_{j}}{1}  =\pathprob{C^1}{1} + \pathprob{C^2}{1} = \pathprob{C_1^2}{1} + \pathprob{C_2^1}{1} =  \sum_{x \in C_1^2} \edgeprob{x}{1} + \sum_{x \in C_2^1} \edgeprob{x}{1}   - \combiprob{C_1^2}{1} - \combiprob{C_2^1}{1} \leq \sum_{i=-k}^{n} \edgeprob{x_i}{1} +  \sum_{j=-l}^{m} \edgeprob{y_{j}}{1}.
\end{align*}
Therefore, $\combiprob{C_1^2}{1} = \combiprob{C_2^1}{1} = 0$, and $C_1^2, C_2^1 \in \pathtight{1}$. From Lemma~\ref{new poset general}, $P^1 = (\component{1},\preceq_{\pathtight{1}})$ is a poset. 

We now assume that $(i)-(iv)$ hold for $k \in \llbracket 1, \stepmax \rrbracket$, and show that they also hold for $k+1$:


\begin{enumerate}
\item[$(i)$] Since $P^k$ is a poset, the $k-$th iteration of the algorithm is well-defined, and we can consider the set $\setmin{k}$ and the weight $\weight{k}$ at that iteration.  Then, for every $C \in \Chains$,  \aref{update_vectors} and \eqref{update_pi} give us:
\begin{align*}
\sum_{x \in \chain} \edgeprob{x}{k+1} - \pathprob{\chain}{k+1} &= \sum_{x \in \chain} \edgeprob{x}{k} - \pathprob{\chain}{k}  -  \weight{k} |\setmin{k} \cap \chain| +  \weight{k} \mathds{1}_{\{\setmin{k} \cap \chain \neq \emptyset\}}= \combiprob{\chain}{k} - \weight{k}(|\setmin{k} \cap \chain| - 1)\mathds{1}_{\{\setmin{k} \cap \chain \neq \emptyset\}} = \combiprob{\chain}{k+1}.
\end{align*}

Now, consider a maximal chain $\chain \in \allpaths{k}$. Since $\delta^k \geq 0$ on $\allpaths{k}$, then $\allpaths{k} = \pathtight{k} \cup \pathloose{k}$ (from \aref{update_sets}). 
\begin{enumerate}
\item If $\chain \in \pathtight{k}$, then by definition of $\preceq_{\pathtight{k}}$, $\chain \cap \component{k}$ is a chain in $P^k$. 
From Lemma~\ref{Minimal Elements}, \rem{we know that }$\setmin{k}$ is an antichain of $P^k$. Therefore, $|\setmin{k} \cap (\chain \cap \component{k})| \leq 1$. Since $\setmin{k} \subseteq \component{k}$, we obtain that $|\setmin{k} \cap \chain| = |(\setmin{k} \cap \component{k}) \cap \chain| = |\setmin{k} \cap (\chain \cap \component{k})| \leq 1$. Thus, $\combiprob{\chain}{k+1} \overset{\aref{update_vectors}}{=} \combiprob{\chain}{k} - \weight{k}(|\setmin{k} \cap \chain| - 1)\mathds{1}_{\{|\setmin{k} \cap \chain| \geq 2\}} =  \combiprob{\chain}{k} = 0$. 

\item If $\chain \in \pathloose{k}$, then by definition of $\weight{k}$:\rem{, we have} $\combiprob{\chain}{k+1} \overset{\aref{update_vectors}}{=} \combiprob{\chain}{k} - \weight{k} (|\setmin{k} \cap \chain| - 1)\mathds{1}_{\{|\setmin{k} \cap \chain| \geq 2\}} \overset{\aref{max_weight}}{\geq} 0$.


\end{enumerate}

In summary, for all $C \in \allpaths{k}, \ \combiprob{C}{k+1}\geq 0$. Since $\allpaths{k+1} \overset{\aref{hardest_update}}{\subseteq} \allpaths{k}$, then for all $C \in \allpaths{k+1}, \ \combiprob{C}{k+1}\geq 0$.



%


\item[$(ii)$] Consider $C^1,C^2 \in \allpaths{k+1} \subseteq \allpaths{k}$ such that $C^1 \cap C^2 \cap \component{k} \neq \emptyset$, and let $C_1^2$ and $C_2^1$ be the other two maximal chains such that $C_1^2 \cup C_2^1 = C^1 \cup C^2$. Since $\component{k} \overset{\aref{update_sets}}{\subseteq} \component{k-1}$, then $C^1 \cap C^2 \cap \component{k-1} \neq \emptyset$. Therefore, by inductive hypothesis,  $C_1^2$, $C_2^1  \in \allpaths{k}$ as well. Let $x_1$ (resp. $y_1$) denote the minimal element of the chain $C^1 \cap \component{k}$ (resp.  $C^2 \cap \component{k}$) in $P$. 
Since $C^1$, $C^2 \in \allpaths{k+1}$, then  $(x_1,y_1) \overset{\aref{hardest_update}}{\in} (\setmin{k})^2$. Let $x^* \in \component{k}$ denote an intersecting \revm{element} of $C^1$ and $C^2$. Since $C^1 \cap \component{k}$ is a chain in $P$, contains $x^*$, and whose minimal element is $x_1$, then necessarily $x_1 \preceq x^*$. Similarly, we obtain that $y_1 \preceq x^*$. 
Therefore, the minimal element of $C_1^2\cap \component{k}$ (resp. $C_2^1\cap \component{k}$) \revm{in $P$} is $\revm{x_1  \in S^k}$ (resp. $\revm{y_1 \in S^k}$).
Thus, $C_1^2, C_2^1 \in \allpaths{k+1}$, and $\allpaths{k+1}$ preserves the decomposition of maximal chains of $P$ intersecting in $\component{k}$. 

\item[$(iii)$] Now, given $C^1$, $C^2$ in $\allpaths{k+1}$ that intersect in $\component{k}$, we just proved that $C_1^2$ and $C_2^1$ are in $\allpaths{k+1}$ as well. Therefore, \rev{for all} $C \in \{C^1,C^2,C_1^2,C_2^1\}$, \rem{we have }$\pathprob{\chain}{k+1} \overset{\eqref{update_pi}}{=} \pathprob{\chain}{k} - \weight{k}$ (since $\setmin{k} \cap \chain \neq \emptyset$).
%
By inductive hypothesis, since $\allpaths{k+1} \subseteq \allpaths{k}$ and $\component{k+1} \subseteq \component{k}$, $\pi^k$ satisfies the conservation law between $C^1$, $C^2$, $C_1^2$, and $C_2^1$.
Thus, we conclude that $\pathprob{C^1}{k+1} +  \pathprob{C^2}{k+1} = \pathprob{C^1}{k} +  \pathprob{C^2}{k} - 2\weight{k}=  \pathprob{C_1^2}{k} + \pathprob{C_2^1}{k} - 2 \weight{k}=\pathprob{C_1^2}{k+1} + \pathprob{C_2^1}{k+1}$.

\item[$(iv)$] This is a consequence of $(i)-(iii)$; the proof is analogous to the one derived for \rev{$k=1$}. 
\end{enumerate}

Therefore, we conclude by induction that $(i)-(iv)$ hold for every $k \in \llbracket 1,\stepmax +1\rrbracket$.
\hfill
\Halmos
\endproof

The proof of Proposition~\ref{Never Negative} highlights the importance of our construction of $\Chains^{k+1}$ for $k \in \llbracket 1,\stepmax \rrbracket$ as given in \aref{hardest_update}. This step of the algorithm ensures that $\allpaths{k+1}$ preserves the decomposition of maximal chains intersecting in $X^k$. It also ensures that each maximal chain in $\allpaths{k+1}$ intersects $\setmin{k}$. A direct consequence is that  $\pi^{k+1}$ satisfies the conservation law on the maximal chains of $\allpaths{k+1}$ that intersect in $X^k$. \rev{This implies that} $\pathtight{k+1}$ preserves the decomposition of maximal chains intersecting in $\component{k+1}$, \rev{and} $P^{k+1}$ is a poset (Lemma~\ref{new poset general}). The issue however is that some maximal chains in $\allpaths{k}$ may be removed when constructing $\allpaths{k+1}$, and we must ensure that the corresponding constraints \eqref{small combi} will still be satisfied by the output of the algorithm. This is the focus of the next part.
%


\subsection*{Part 2: Feasibility of Algorithm~\ref{ALG3}'s output.} \rem{Now that we have shown the algorithm to be well-defined, }The second main part of the proof of Theorem~\ref{optimal value} is to show that the algorithm terminates, and outputs a feasible solution of \OPOC. 
Showing that the algorithm terminates is based on the fact that there is a finite number of elements and maximal chains.
\rem{To show the feasibility of the solution generated by the algorithm, we need to verify that constraints \eqref{Equal_2} and \eqref{small combi} are satisfied. }From \aref{update_sets}, we deduce that constraints \eqref{Equal_2} are automatically satisfied at termination, since  an element $x \in \ground$ is removed whenever the remaining value $\rho_x^k$ is 0. Similarly, from Proposition~\ref{Never Negative}, we obtain that constraints \eqref{small combi} are satisfied for all maximal chains in $\allpaths{\stepmax +1}$, i.e., the maximal chains that are not removed by the algorithm. \rev{For the remaining maximal chains}
%
%
$C \in \Chains\backslash \allpaths{\stepmax+1}$, we create a finite sequence of ``dominating'' maximal chains, and show that constraint \eqref{small combi} being satisfied for the last maximal chain of the sequence implies that it is also satisfied for the initial maximal chain $C$. To carry out this argument, we essentially  need the following lemma:
 
%

\begin{lemma}\label{Dominates}
Consider $C^{(1)} \in \Chains$, and suppose that $C^{(1)} \in \allpaths{k_1}\backslash\allpaths{k_1+1}$ and $C^{(1)} \cap \component{k_1} \neq \emptyset$ \rev{for some $k_1 \in \llbracket 1,\stepmax \rrbracket$}. Then,  \rev{there exists} $C^{(2)} \in \allpaths{k_1+1}$ such that  $\combiprob{C^{(1)}}{k_1} \geq \combiprob{C^{(2)}}{k_1}$ and $C^{(2)} \cap \component{k_1} \supseteq C^{(1)} \cap \component{k_1}$. 

\end{lemma}

\proof{Proof of Lemma~\ref{Dominates}.}

Consider $C^{(1)} \in \Chains$, and suppose that \rev{there exists} $k_1 \in \llbracket 1,\stepmax \rrbracket$ such that $C^{(1)} \in \allpaths{k_1}\backslash\allpaths{k_1+1}$ and $\chain^{(1)} \cap \component{k_1} \neq \emptyset$. This case arises when the minimal element of $C^{(1)} \cap \component{k_1}$ in $P$ is not a minimal element of $\poset{k_1}$. Then, \rev{there is} a chain in $\poset{k_1}$ whose maximal element is the minimal element of $C^{(1)} \cap \component{k_1}$ in $P$, and whose minimal element is a minimal element of $\poset{k_1}$. \rev{By} definition of $\poset{k_1}$, this chain is contained in a maximal chain in $\pathtight{k_1}$ \rev{(Lemma~\ref{new poset general})}. We then exploit $(i)-(iii)$ in Proposition~\ref{Never Negative} to show that there exists a maximal chain in $\allpaths{k_1+1}$ \rev{satisfying} the desired properties.

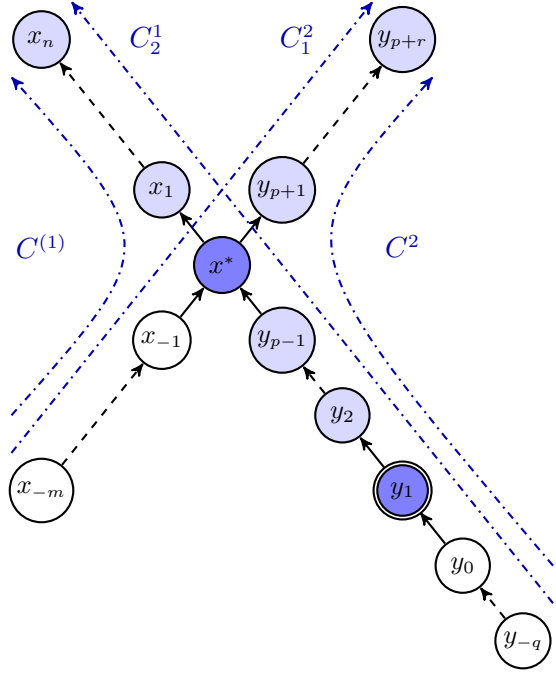
\begin{figure}[ht]
 \centering
        \begin{tikzpicture}[->,>=stealth',shorten >=0pt,auto,x=1.6cm, y=2cm,
  thick,main node/.style={circle,draw},main node2/.style={circle,draw,inner sep = 0.06cm},flow_a/.style ={blue!100}]
\tikzstyle{edge} = [draw,thick,->]
\tikzstyle{cut} = [draw,very thick,-]
\tikzstyle{flow} = [draw,line width = 1pt,->,blue!100]
\small
	\node[main node2,fill=blue!15] (1) at (0,0) {$y_{p-1}$};
	\node[main node,fill=blue!50] (2) at (-0.5,0.5) {$x^*$};
	\node[main node2,fill=blue!15] (3) at (-0,1) {$y_{p+1}$};
	\node[main node2] (4) at (-1,0) {$x_{-1}$};
	\node[main node,fill=blue!15] (5) at (-1,1) {$x_{1}$};
	\node[main node2] (6) at (-2,-1) {$x_{-m}$};
	
	\node[main node,fill=blue!15] (7) at (-2,2) {$x_{n}$};
	
	\node[main node,fill=blue!50,double] (8) at (1,-1) {$y_{1}$};
	\node[main node] (9) at (1.5,-1.5) {$y_{0}$};
	\node[main node2] (10) at (2,-2) {$y_{-q}$};
	\node[main node,fill=blue!15] (11) at (0.5,-0.5) {$y_{2}$};
	\node[main node2,fill=blue!15] (12) at (1,2) {$y_{p+r}$};
	
	
	\path[edge]
	(1) edge (2)
	(2) edge  (3)
	(4) edge (2)
	(2) edge (5)
	(6) edge[dashed] (4)
	(5) edge[dashed] (7)
	(9) edge (8)
	(10) edge[dashed] (9)
	(8) edge (11)
	(11) edge[dashed] (1)
	(3) edge[dashed] (12);

	
\draw[blue!70!black, dashdotted] 
    (-2.25,-0.5) .. controls (-1.00,0.625) .. (-2.25,1.75);
    
    \draw[blue!70!black, dashdotted] 
    (-2.25,-0.75) -- (0.75,2.25);

    \draw[blue!70!black, dashdotted] 
    (2.25,-1.75) -- (-1.75,2.25);

   \draw[blue!70!black, dashdotted] 
    (2.25,-1.5) .. controls (-0.00,0.625) .. (1.25,1.75);
    
    \normalsize
    \node[blue!70!black] (44) at (-2,0.625) {$C^{(1)}$};
    \node[blue!70!black] (44) at (1,0.625) {$C^2$};
    
    \node[blue!70!black] (44) at (0.125,2.00) {$C_1^2$};
    \node[blue!70!black] (44) at (-1.125,2.00) {$C_2^1$};

%

\normalsize
\end{tikzpicture}
    \caption{Illustration of $C^{(1)}$, $C^2$, $C_1^2$, and $C_2^1$. In dark blue are the elements in $\component{k_1}$, in light blue are the elements that may or may not be in $\component{k_1}$, and in white are the elements that are not in $\component{k_1}$. The ``double'' node $y_1$ is in $\setmin{k_1}$.}
    \label{proof3}
\end{figure}

Formally, let $x^*$ denote the minimal element of $C^{(1)} \cap \component{k_1}$  in $P$. Since $C^{(1)} \notin \allpaths{k_1+1}$, then $x^* \notin \setmin{k_1}$\rem{, i.e., $x^*$ is not a minimal element of $\poset{k_1}$}. Let $C^\prime \subseteq \component{k_1}$ denote a maximal chain of $P^{k_1}$ that contains $x^*$. From Lemma~\ref{Minimal Elements}, the minimal element of $C^\prime$ in $P^{k_1}$, which we denote $y_1$, is a minimal element of $P^{k_1}$. Therefore $y_1 \in \setmin{k}$ and $y_1 \neq x^*$.
%
Thus, $|C^\prime|\geq 2$, and there exists a maximal chain $C^2 \in \pathtight{k_1}$ such that $C^\prime = C^2 \cap \component{k_1}$ (Lemma~\ref{new poset general}). Since $C^{(1)} \cap C^2 \cap \component{k_1-1} \supseteq \{x^*\} \neq \emptyset$, let us consider the other two maximal chains $C_1^2, C_2^1 \in \Chains$ such that $C_1^2 \cup C_2^1 = C^{(1)} \cup C^2$. Since $C^{(1)}$ and $C^2$ are in $\allpaths{k_1}$, then from Proposition~\ref{Never Negative},  $C_1^2$ and $C_2^1$ are in $\allpaths{k_1}$ as well. Let us rewrite  $C^{(1)} = \{x_{-m},\dots,x_{0}=x^*,\dots,x_{n}\}$,   $C^2 = \{y_{-q},\dots,y_{0},y_1,\dots,y_p=x^*,\dots,y_{p+r}\}$, $C_1^2 = \{x_{-m},\dots,x_{-1},y_{p},\dots,y_{p+r}\}$, and $C_2^1 = \{y_{-q},\dots,y_p,x_1,\dots,x_n\}$; they are illustrated in Figure~\ref{proof3}.

Since $x^*$ is the minimal element of $C^{(1)} \cap \component{k_1}$ in $P$, then \rev{for all} $i \in \llbracket -m,-1\rrbracket$, \rem{$x_i \notin \component{k_1}$ and }$\edgeprob{x_i}{k_1} = 0$. Since $C^2 \in \pathtight{k_1}$ and $C_1^2 \in \allpaths{k_1}$, and from the conservation law between $C^{(1)}$, $C^2$, $C_1^2$ and $C_2^1$, we obtain:
%
%
%
\begin{align}
\pathprob{C_2^1}{k_1} - \pathprob{C^{(1)}}{k_1}\overset{\eqref{Conservation_k}}{=}  \pathprob{C^2}{k_1} - \pathprob{C_1^2}{k_1}\overset{\eqref{Relation_k}}{=}
\sum_{j=-q}^{p+r} \edgeprob{y_j}{k_1}  - \underset{=0}{\underbrace{\combiprob{C^2}{k_1}}} -
\sum_{i=-m}^{-1} \underset{=0}{\underbrace{\edgeprob{x_i}{k_1}}} - \sum_{j=p}^{p+r} \edgeprob{y_j}{k_1} + \underset{\geq 0}{\underbrace{\combiprob{C_1^2}{k_1}}} 
\overset{\eqref{Inequality_k}}{\geq}  \sum_{j=-q}^{p-1} \edgeprob{y_j}{k_1}. \label{almossst}
\end{align}

This implies that:
\begin{align*}
\combiprob{C^{(1)}}{k_1} & \overset{\eqref{Relation_k}}{=}  \sum_{i=0}^n \edgeprob{x_i}{k_1} - \pathprob{C^{(1)}}{k_1} + \sum_{j=-q}^{p-1} \edgeprob{y_j}{k_1} - \sum_{j=-q}^{p-1} \edgeprob{y_j}{k_1}  \overset{\eqref{Relation_k}}{=} \combiprob{C_2^1}{k_1} + \pathprob{C_2^1}{k_1} - \pathprob{C^{(1)}}{k_1} - \sum_{j=-q}^{p-1} \edgeprob{y_j}{k_1} \overset{\eqref{almossst}}{\geq}\combiprob{C_2^1}{k_1}. 
\end{align*}

Since $y_1$ is the minimal element of $C^2 \cap \component{k_1}$ in $P^{k_1}$, it is also the minimal element of $C^2 \cap \component{k_1}$ in $P$.
\rev{Therefore,} $y_1$ is the minimal element of $C_2^1 \cap \component{k_1}$ in $P$. \rev{Since $y_1 \in S^{k_1}$, then $C_2^1 \in \allpaths{k_1+1}$.}


Finally, since \rev{for all} $i \in \llbracket -m,-1\rrbracket$, $x_i \notin \component{k_1}$, then $C_2^1 \cap \component{k_1} \supseteq \{x^*,x_1,\dots,x_n\} \cap \component{k_1} =  C^{(1)} \cap \component{k_1}$, as illustrated in Figure~\ref{proof3}.
%
In conclusion, given $C^{(1)} \in \allpaths{k_1}\backslash\allpaths{k_1+1}$ such that $C^{(1)} \cap \component{k_1} \neq \emptyset$, \rev{there exists} $C^{(2)}\coloneqq C_2^1  \in \allpaths{k_1+1}$ such that $\combiprob{C^{(1)}}{k_1} \geq \combiprob{C^{(2)}}{k_1}$ and $C^{(2)}\cap \component{k_1} \supseteq C^{(1)} \cap \component{k_1}$.
\hfill \Halmos
\endproof

%

As shown in the next proposition, one of \rev{Lemma~\ref{Dominates}'s implications} is that if a maximal chain $C^{(1)}$ is removed after the $k_1-$th iteration of the algorithm, then there exists another maximal chain $C^{(2)}$ \rev{that} dominates $C^{(1)}$ in that if the output of the algorithm satisfies constraint~\eqref{small combi} for $C^{(2)}$, then it also satisfies that constraint for $C^{(1)}$. Additionally, it is guaranteed that $C^{(2)}$ is not removed before the $k_1+1-$th iteration of the algorithm. We now show the \rev{feasibility of Algorithm~\ref{ALG3}'s output:} 

\begin{proposition}\label{Terminates}
Algorithm~\ref{ALG3} terminates, and outputs a feasible solution of \OPOC.
\end{proposition}

\proof{Proof of Proposition~\ref{Terminates}.}
We recall that the algorithm terminates after iteration $\stepmax$ if $\component{\stepmax +1} = \emptyset$.  First, note that $\component{1} \subseteq \ground$ and \rev{for all} $k \in \llbracket 1,\stepmax \rrbracket, \ \component{k+1} \overset{\aref{update_sets}}{\subseteq} \component{k}$. Additionally,  $\pathloose{1} \subseteq \Chains$, and from \aref{update_vectors}, $\pathloose{k+1} \subseteq \pathloose{k}$ \rev{for every} $k \in \llbracket 1,\stepmax \rrbracket$.
Now, consider $k \in \llbracket 1,\stepmax\rrbracket$, and the weight $\weight{k}$ chosen by the algorithm at iteration $k$. From~\aref{max_weight}, \rev{there exists} $x \in \component{k}$ such that $\weight{k} = \edgeprob{x}{k}$, or \rev{there exists} $\chain \in \pathloose{k}$ such that $\weight{k} = \frac{\combiprob{\chain}{k}}{|\setmin{k} \cap \chain| - 1}$. In the first case, $x \notin \component{k+1}$, so $\component{k+1} \subsetneq \component{k}$. \rev{In the second case, either $C \notin \allpaths{k+1}$, or $C \in \allpaths{k+1}$ with $\combiprob{C}{k+1} = 0$; this implies that $C \notin \pathloose{k+1}$ and $\pathloose{k+1} \subsetneq \pathloose{k}$.}

Thus, \rev{for every} $k \in \llbracket 1,\stepmax \rrbracket$, $|\component{k+1} \times \pathloose{k+1}| < |\component{k} \times \pathloose{k}|$. Since $|\component{1} \times \pathloose{1}| \in \mathbb{N}$, if $\stepmax$ were equal to $+\infty$, we would obtain an infinite decreasing sequence of natural integers. Therefore, we conclude that $\stepmax \in \mathbb{N}$, i.e., the algorithm terminates. At termination, $\component{\stepmax +1} = \emptyset$.

Next, we show that the output $\sigma \in \revm{\mathbb{R}_{\geq0}^{\mathcal{P}}}$ of the algorithm is a feasible solution of \OPOC. First, the equality constraints \eqref{Equal_2} are trivially satisfied:
\begin{align*}
\forall x \in \ground, \ \rho_{x} \overset{\aref{Initiate}}{=} \edgeprob{x}{1} \overset{\aref{update_vectors}}{=} \underset{=0}{\underbrace{\edgeprob{x}{\stepmax+1}}} + \sum_{k=1}^{\stepmax} \weight{k} \mathds{1}_{\{x \in \setmin{k}\}} \overset{\aref{max_weight}}{=} \sum_{k=1}^{\stepmax} \sigma_{\setmin{k}} \mathds{1}_{\{x \in \setmin{k}\}} = \sum_{\{\att \in \mathcal{P} \, | \, x \in \att\}}\sigma_{\att}.
\end{align*}

Regarding constraints \eqref{small combi}, we first show the following equality:
%
\begin{align*}
\forall C \in \Chains, \ \combiprob{C}{\stepmax +1} &\overset{\aref{update_vectors}}{=}  \combiprob{\chain}{1} - \sum_{k=1}^{\stepmax}\weight{k}(|\setmin{k}\cap \chain | - 1)\mathds{1}_{\{|\setmin{k} \cap \chain| \geq 2\}} \overset{\aref{Initiate},\aref{max_weight}}{=} \delta_C - \sum_{\{\att \in \mathcal{P} \, | \, |\att \cap \chain| \geq 2\}}\sigma_{\att}(|\att \cap \chain| - 1).
\end{align*}
Therefore, constraints \eqref{small combi} are satisfied if and only if \rev{for every} $C \in \Chains, \ \combiprob{C}{\stepmax +1} \geq 0$.

From Proposition~\ref{Never Negative}, we know that \rev{for all} $C \in \allpaths{\stepmax +1}$, $\combiprob{C}{\stepmax +1} \geq 0$.
Now, consider $C^{(1)} \in \Chains$, and suppose that \rev{there exists} $k_1 \in \llbracket 1,\stepmax \rrbracket$ such that $C^{(1)} \in \allpaths{k_1}\backslash\allpaths{k_1+1}$.  If $C^{(1)} \cap \component{k_1} = \emptyset$, then \rev{for every} $l \in \llbracket k_1, \stepmax \rrbracket, \ |\setmin{l} \cap C^{(1)}| = 0$ since $\setmin{l} \overset{\aref{set_min_elem}}{\subseteq} \component{l}$ and $\component{l} \overset{\aref{update_sets}}{\subseteq} \component{k_1}$. Therefore, since $C^{(1)} \in \allpaths{k_1}$, we have $\combiprob{C^{(1)}}{\stepmax +1} \overset{\aref{update_vectors}}{=} \combiprob{C^{(1)}}{k_1} - \sum_{l=k_1}^{\stepmax}\weight{l}(| \setmin{l} \cap \chain^{(1)}| - 1)\mathds{1}_{\{|\setmin{l} \cap \chain^{(1)}| \geq 2\}} =  \combiprob{C^{(1)}}{k_1}\overset{\eqref{Inequality_k}}{\geq} 0.
$

If $C^{(1)} \cap \component{k_1} \neq \emptyset$, then  \rev{there exists} $C^{(2)} \in \allpaths{k_1+1}$ such that $ \combiprob{C^{(1)}}{k_1} \geq \combiprob{C^{(2)}}{k_1}$ and $C^{(2)} \cap \component{k_1} \supseteq C^{(1)} \cap \component{k_1}$ (Lemma~\ref{Dominates}). \rev{For any  $i \in \llbracket k_1,\stepmax\rrbracket$, $\setmin{i}\cap C^{(2)} \supseteq \setmin{i} \cap C^{(1)}$ since $\setmin{i}  \overset{\aref{set_min_elem},\aref{update_sets}}{\subseteq} \component{k_1}$.} \rev{Then,} we obtain:
\begin{align}
\forall l \in \llbracket k_1,\stepmax +1\rrbracket, \ \combiprob{C^{(1)}}{l} &\overset{\aref{update_vectors}}{=} \combiprob{C^{(1)}}{k_1} - \sum_{i=k_1}^{l-1} \weight{i} (|\setmin{i} \cap C^{(1)}| -1)\mathds{1}_{\{|\setmin{i} \cap C^{(1)}| \geq 2\}} \nonumber\\
& \geq \combiprob{C^{(2)}}{k_1} - \sum_{i=k_1}^{l-1} \weight{i} (|\setmin{i} \cap C^{(2)}| -1)\mathds{1}_{\{|\setmin{i} \cap C^{(2)}| \geq 2\}} \overset{\aref{update_vectors}}{=} \combiprob{C^{(2)}}{l}. \label{dominated}
\end{align}
In particular, $\combiprob{C^{(1)}}{\stepmax +1} \geq \combiprob{C^{(2)}}{\stepmax +1}$.

\rem{We note that $C^{(2)} \in \allpaths{k_1+1}$, and two cases can arise: 
\begin{enumerate}
\item[1)] $C^{(2)} \in \allpaths{\stepmax +1}$. In this case, $\combiprob{C^{(2)}}{\stepmax +1} \geq 0$ (Proposition~\ref{Never Negative}).
%
\item[2)] \rev{There exists} $k_2 \in \llbracket k_1+1, \stepmax \rrbracket$ such that $C^{(2)} \in \allpaths{k_2}\backslash\allpaths{k_2+1}$.  Then we reiterate the same argument:
\begin{enumerate}
\item If $C^{(2)} \cap \component{k_2} = \emptyset$, then $\combiprob{C^{(2)}}{\stepmax+1} = \combiprob{C^{(2)}}{k_2} \overset{\eqref{Inequality_k}}{\geq} 0$.

\item If $C^{(2)} \cap \component{k_2} \neq \emptyset$, then there exists $C^{(3)} \in \allpaths{k_2+1}$ such that $\combiprob{C^{(2)}}{k_2} \geq \combiprob{C^{(3)}}{k_2}$ and $C^{(3)} \cap \component{k_2} \supseteq C^{(2)} \cap \component{k_2}$ (Lemma~\ref{Dominates}). Analogous calculations to \eqref{dominated} show that $\combiprob{C^{(2)}}{\stepmax +1}\geq \combiprob{C^{(3)}}{\stepmax +1}$.
\end{enumerate}
\end{enumerate}}

By induction, we construct a sequence of maximal chains $(C^{(s)})$, a sequence of increasing integers $(k_s)$, and a termination point $s^* \in \mathbb{N}$, such that \rev{for all} $s \in \llbracket 1,s^*-1\rrbracket$, $C^{(s)} \in \allpaths{k_s}\backslash\allpaths{k_s+1}$, $\combiprob{C^{(s)}}{\stepmax+1} \geq \combiprob{C^{(s+1)}}{\stepmax+1}$, and $\combiprob{C^{(s^*)}}{\stepmax +1} \geq 0$. 
Note that $s^*$ exists since $k_s \leq \stepmax +1$. \rev{This implies} that $\combiprob{C^{(1)}}{\stepmax +1} \geq \cdots \geq \combiprob{C^{(s^*)}}{\stepmax +1} \geq 0$.

Thus, \rev{for every} $C \in \Chains, \ \combiprob{C}{\stepmax +1} \geq 0$, and constraints \eqref{small combi} are satisfied by the output $\sigma$ of the algorithm. In conclusion, the algorithm outputs a feasible solution of \OPOC.
\hfill \Halmos
\endproof

%
%
%

The output of Algorithm~\ref{ALG3}, by design, satisfies constraints \eqref{Equal_2}, and also constraints \eqref{small combi} for the maximal chains in $\allpaths{\stepmax+1}$. Recall that the remaining maximal chains were removed after an iteration $k$ in order to maintain the conservation law on the resulting set $\allpaths{k+1}$. This conservation law played an essential role in proving Proposition~\ref{Terminates}, i.e., in showing that constraints~\eqref{small combi} are also satisfied for the maximal chains that are not in $\allpaths{\stepmax+1}$ (see the proof of Lemma~\ref{Dominates}). \rem{Thus, Algorithm~\ref{ALG3}'s output is a feasible solution of \OPOC. Next, we show that this solution is \rev{in fact} optimal.}

\subsection*{Part 3: Optimality of Algorithm~\ref{ALG3}.} The final part of the proof of Theorem~\ref{optimal value} consists in showing that the total weight used by the algorithm is exactly $\max\{\max\{\rho_x, \ x \in \ground\},\max\{\pi_C, \ \chain \in \Chains\}\}$. This is done by considering the following quantity: \rev{for every} $k \in \llbracket 1,\stepmax + 1 \rrbracket$, $\maxweight{k} \coloneqq \max\{\max\{\edgeprob{x}{k}, \ x \in \ground\},\max\{\pathprob{\chain}{k}, \ \chain \in \Chains\}\}$. First,  we show that \rev{for every} $k \in \llbracket 1,\stepmax\rrbracket$, $\maxweight{k+1} = \maxweight{k} - \weight{k}$. Then, we show that $\maxweight{\stepmax +1} = 0$. Using a telescoping series, we obtain the desired result. Lemma~\ref{Dominates} \rev{is also used} to conclude that $\max\{\pathprob{C}{k}, \ C \in \Chains\}$ is attained by a maximal chain $C \in \allpaths{k+1}$.

\begin{proposition}\label{Termination}
The total weight used by Algorithm~\ref{ALG3} when it terminates is $\max\{\max\{\rho_x, \ x \in \ground\},\max\{\pi_C, \ \chain \in \Chains\}\}$.
\end{proposition}

\proof{Proof of Proposition~\ref{Termination}.}

For all $k \in \llbracket 1,\stepmax +1\rrbracket$, let $\maxweight{k} \coloneqq \max\{\max\{\edgeprob{x}{k}, \ x \in \ground\},\max\{\pathprob{\chain}{k}, \ \chain \in \Chains\}\}$.
%
%
First, we show that \rev{for every} $k \in \llbracket 1,\stepmax\rrbracket$, $\maxweight{k+1} = \maxweight{k} - \weight{k}$.
Consider $k \in \llbracket 1,\stepmax \rrbracket$, and let $\chain \in \Chains \backslash \allpaths{k+1}$. Then, there exists $k_1 \leq k$ such that $\chain \in \allpaths{k_1}\backslash\allpaths{k_1+1}$. If $\chain \cap \component{k_1} = \emptyset$, then $\pathprob{\chain}{k+1} \leq \pathprob{\chain}{k} \leq \pathprob{\chain}{k_1} \overset{\eqref{Relation_k}}{=}   -\combiprob{\chain}{k_1} \overset{\eqref{Inequality_k}}{\leq} 0$.
If $C \cap \component{k_1} \neq \emptyset$, then \rev{Lemma~\ref{Dominates} implies that} \rev{there exists} $C^{(2)} \in  \allpaths{k_1+1}$ such that\rem{  $\combiprob{C}{k_1} \geq \combiprob{C^{(2)}}{k_1}$ and $C^{(2)} \cap \component{k_1} \supseteq C \cap \component{k_1}$ (Lemma~\ref{Dominates}). This implies that} \rev{for all} $l \in \llbracket k_1, \stepmax +1\rrbracket, \ \combiprob{C}{l} \overset{\eqref{dominated}}{\geq} \combiprob{C^{(2)}}{l}$, and $C^{(2)} \cap \component{l} \supseteq C \cap \component{l}$. \rev{Consequently}, we obtain:
\begin{align*}
\forall l \in \llbracket k_1, \stepmax +1\rrbracket, \ \pathprob{C}{l} &\overset{\eqref{Relation_k}}{=} \sum_{x \in  C \cap \component{l}} \edgeprob{x}{l} - \combiprob{C}{l} + \pathprob{C^{(2)}}{l} + \combiprob{C^{(2)}}{l} - \sum_{x \in  C\cap \component{l}} \edgeprob{x}{l} - \sum_{x \in  (C^{(2)} \cap \component{l})\backslash(C\cap \component{l})} \edgeprob{x}{l}   \overset{\eqref{dominated}}{\leq} \pathprob{C^{(2)}}{l}.
\end{align*}


In particular, \rem{we deduce that }$\pathprob{C}{k} \leq \pathprob{C^{(2)}}{k}$ and $\pathprob{C}{k+1} \leq \pathprob{C^{(2)}}{k+1}$. As in Proposition~\ref{Terminates}, we construct a sequence of maximal chains $(C^{(s)})$, a sequence of increasing integers $(k_s)$, and a termination point $s^\prime \in \mathbb{N}$, such that $C^{(1)} = \chain$ \rev{and for all} $s \in \llbracket 1,s^\prime -1 \rrbracket, \ C^{(s)} \in \allpaths{k_s}\backslash\allpaths{k_s+1}$, $\pathprob{C^{(s)}}{k} \leq \pathprob{C^{(s+1)}}{k}$, and $\pathprob{C^{(s)}}{k+1} \leq \pathprob{C^{(s+1)}}{k+1}$. At termination, $C^{(s^\prime)} \in \allpaths{k_{s^\prime}}$, and either $k_{s^\prime} = k+1$, or $k_{s^\prime} < k+1$ and $C^{(s^\prime)} \cap \component{k_{s^\prime}} = \emptyset$. 
If $k_{s^\prime} = k+1$, then \rem{we conclude that }$\pathprob{C}{k} \leq \pathprob{C^{(s^\prime)}}{k}$ and $\pathprob{C}{k+1} \leq \pathprob{C^{(s^\prime)}}{k+1}$, with $C^{(s^\prime)} \in \allpaths{k +1}$. 
If $k_{s^\prime} < k+1$ and $C^{(s^\prime)} \cap \component{k_{s^\prime}} = \emptyset$, then $\pathprob{C}{k+1} \overset{\eqref{update_pi}}{\leq}  \pathprob{C}{k} \leq \pathprob{C^{(s^\prime)}}{k} \overset{\eqref{update_pi}}{\leq} \pathprob{C^{(s^\prime)}}{k_{s^\prime}} \overset{\eqref{Relation_k}}{=}   -\combiprob{C^{(s^\prime)}}{k_{s^\prime}}  \overset{\eqref{Inequality_k}}{\leq} 0\leq \edgeprob{x}{k+1}  \overset{\eqref{update_vectors}}{\leq} \edgeprob{x}{k}$ \rev{for all} $x \in \ground$. 
%
Thus, \rem{we deduce that }$\maxweight{k} = \max\{\max\{\edgeprob{x}{k}, \ x \in \ground\},\max\{\pathprob{C}{k}, \ C \in \allpaths{k+1}\}\}$, and $\maxweight{k+1}= \max\{\max\{\edgeprob{x}{k+1}, \ x \in \ground\},\max\{\pathprob{C}{k+1}, \ C \in \allpaths{k+1}\}\}$.



\rev{Since $\component{k} \neq \emptyset$, $\edgeprob{x}{k} \geq \edgeprob{x}{k+1} \geq 0$ for every $x \in \component{k}$, and $\edgeprob{x}{k} = \edgeprob{x}{k+1} = 0$ for every $x \in \ground \backslash \component{k}$, then $\max\{\edgeprob{x}{k}, \ x \in \ground\} = \max\{\edgeprob{x}{k}, \ x \in \component{k}\}$, and  $\max\{\edgeprob{x}{k+1}, \ x \in \ground\} = \max\{\edgeprob{x}{k+1}, \ x \in \component{k}\}$.}


Next, \rev{let} $x \in \component{k}\backslash \setmin{k}$.  Then, \rev{there exists} $y \in S^k$\rem{ \neq x \in \component{k}$} such that $y \preceq_{\pathtight{k}} x$\rem{, and $y \in \setmin{k}$ is a minimal element in $P^k$}. By definition, \rev{there exists} $C \in \pathtight{k}$ such that $y,x \in C$\rem{, and $y \prec x$}. In fact, $y$ is the minimal element of $C \cap \component{k}$ in $P^k$, and $C \in\allpaths{k+1}$. Since $C \in \pathtight{k}$, then  $\pathprob{C}{k} \overset{\eqref{Relation_k}}{=} \sum_{x^\prime \in C}\edgeprob{x^\prime}{k} \geq \edgeprob{x}{k} + \edgeprob{y}{k} \geq \edgeprob{x}{k} $.  Furthermore, since $y \in \setmin{k}$, then $\weight{k} \overset{\aref{max_weight}}{\leq} \edgeprob{y}{k}$. Thus, \rem{we obtain that }$\edgeprob{x}{k+1} = \edgeprob{x}{k} \leq \pathprob{C}{k} - \edgeprob{y}{k} \leq \pathprob{C}{k} - \weight{k} \overset{\eqref{update_pi}}{=} \pathprob{C}{k+1}$, from which we conclude that $\maxweight{k} = \max\{\max\{\edgeprob{x}{k}, \ x \in \setmin{k}\},\max\{\pathprob{\chain}{k}, \ \chain \in \allpaths{k+1}\}\}$, and $\maxweight{k+1} =\max\{\max\{\edgeprob{x}{k+1}, \ x \in \setmin{k}\},\max\{\pathprob{\chain}{k+1}, \ \chain \in \allpaths{k+1}\}\}$.

Finally, we note that \rev{for all} $\chain \in \allpaths{k+1}$, $\pathprob{\chain}{k+1} \overset{\eqref{update_pi}}{=} \pathprob{\chain}{k} - \weight{k}$ since $\setmin{k} \cap \chain \neq \emptyset$, and \rev{for all} $x \in \setmin{k}$, $\edgeprob{x}{k+1} \overset{\aref{update_vectors}}{=} \edgeprob{x}{k} - \weight{k}$. Putting everything together, we conclude:
%
%
%
\begin{align*}
\maxweight{k+1} 
& = \max\{\max\{\edgeprob{x}{k+1}, \ x \in \setmin{k}\},\max\{\pathprob{\chain}{k+1}, \ \chain \in \allpaths{k+1}\}\}\\
& = \max\{\max\{\edgeprob{x}{k}, \ x \in \setmin{k}\},\max\{\pathprob{\chain}{k}, \ \chain \in \allpaths{k+1}\}\} - \weight{k} = \maxweight{k} - \weight{k}.
\end{align*}


Next, we show that $\maxweight{\stepmax +1} =0$. First, $\edgeprob{x}{\stepmax+1} = 0$ \rev{for all} $x \in \ground$. Secondly, $\pathprob{C}{\stepmax +1} \overset{\eqref{Relation_k}}{=}- \combiprob{C}{\stepmax +1} \overset{\eqref{Inequality_k}}{\leq} 0$ \rev{for all} $C \in \allpaths{\stepmax+1}$. Thirdly, $\setmin{\stepmax} \neq \emptyset$ since $X^{\stepmax}\neq \emptyset$\rem{ is a nonempty poset}. This implies that $\maxweight{\stepmax +1} = \max\{\max\{\edgeprob{x}{\stepmax+1}, \ x \in \setmin{\stepmax}\},\max\{\pathprob{\chain}{\stepmax+1}, \ \chain \in \allpaths{\stepmax+1}\}\} = 0$. 
Finally, using a telescoping series, we obtain:
%
%
\begin{align*}
\sum_{\att \in \mathcal{P}} \sigma_{\att} \overset{\aref{max_weight}}{=}\sum_{k=1}^{\stepmax} \maxweight{k} - \maxweight{k+1} = \maxweight{1} - \underset{=0}{\underbrace{\maxweight{\stepmax+1}}} \overset{\aref{Initiate},\eqref{update_pi}}{=}  \max\{\max\{\rho_x, \ x \in \ground\},\max\{\pi_C, \ \chain \in \Chains\}\}.
\end{align*}
\vspace{-0.3cm}

\hfill
\Halmos
\endproof

In conclusion, Propositions~\ref{Never Negative}, \ref{Terminates}, and \ref{Termination} enable us to show that Algorithm~\ref{ALG3} outputs a feasible solution of \OPOC with objective value equal to $\max\{\max\{\rho_x, \ x \in \ground\},\max\{\pi_C, \ \chain \in \Chains\}\}$. Therefore $z^*_{\OPOCt} \leq  \max\{\max\{\rho_x, \ x \in \ground\},\max\{\pi_C, \ \chain \in \Chains\}\}$. Since we already established the reverse inequality at the end of Section~\ref{sec:equivalents}, we conclude that $z^*_{\text{\OPOC}} = \max\{\max\{\rho_x, \ x \in \ground\},\max\{\pi_C, \ \chain \in \Chains\}\}$, thus proving Theorem~\ref{optimal value}. 

Furthermore, since \rev{$\rho_x \leq 1$ \rev{for every} $x \in \ground$, and $\pi_C \leq 1$ for every $C \in \Chains$}, then $z^*_{\text{\OPOC}} \leq 1$. \rev{This implies that \FPOC is feasible: Given the output $\sigma$ of Algorithm~\ref{ALG3}, the vector $\widehat{\sigma} \in \revm{\mathbb{R}_{\geq0}^{\mathcal{P}}}$ obtained from $\sigma$ by additionally assigning $1 - z^*_\OPOCt$ to $\emptyset$ is a feasible solution of problem \FPOC, and proves Theorem~\ref{thm:finally}.}


\rev{In fact,} \OPOC is a generalization of \rev{the minimum-weighted fractional coloring problem on comparability graphs (Ho{\`a}ng \cite{HOANG1994133}}). The comparability graph of the poset $P = (X,\preceq)$ is an undirected graph whose set of \rev{nodes} is $X$ and whose edges are given by the pairs of comparable elements in $P$.
In the special case where \rev{for all} $C \in \Chains, \  \sum_{x \in C} \rho_{x} = \pi_C$ (i.e., inequality \eqref{Nec Cond} is tight), \OPOC is equivalent to the minimum-weighted fractional coloring problem on the comparability graph of $P$.  Algorithm~\ref{ALG3} can then be refined into Ho{\`a}ng's  $O(|X|^2)$-time algorithm. 

\rev{
Given $E_P$ the edge set of the cover graph of $P$ (as defined in Section~\ref{sec:order_theory}), the number of iterations of Algorithm~\ref{ALG3} is upper bounded by $|\ground| + |E_P|$. However, Algorithm~\ref{ALG3} requires at each iteration $k$ the storage of the possibly exponentially many chains $C$ in $\Chains^k$, along with their corresponding values $\delta^k_C$. Next, we develop an efficient implementation of Algorithm~\ref{ALG3} when $\pi$ is an affine function of the elements constituting each maximal chain of $P$.
}

\section{Affine case: a polynomial algorithm.} \label{sec:refinement}
Consider the problem \FPOC for a given poset $P = (X,\preceq)$, and vectors $\rho \in [0,1]^{X}$ and $\pi \in (-\infty,1]^{\mathcal{C}}$ satisfying~\eqref{Nec Cond}. In addition, we assume that the value of each maximal chain $\chain \in \Chains$ in $P$ is given by $\pi_\chain = \alpha - \sum_{x \in \chain} \beta_x$, with $\alpha \in \mathbb{R}$ and $\beta_x \in \mathbb{R}$ for every $x \in \ground$. We observe that $\pi$ satisfies the conservation law \eqref{Conservation}, and \FPOC is feasible.
In this section, we refine Algorithm~\ref{ALG3} for this special case and show that an optimal solution of \OPOC can be computed in polynomial time.

Our polynomial algorithm performs \aref{alg:Subposet} and \aref{max_weight} without enumerating all the maximal chains of $P$. Instead, it runs subroutines based on the shortest path algorithm in the directed cover graph of $P$ to construct the subposet $P^k$ and compute the maximum weight $w^k$ that can be assigned at each iteration $k \in \llbracket 1,\stepmax\rrbracket$. Let us discuss the execution of the first iteration of the algorithm.


Firstly, we augment the poset $P$ by adding an artificial source element $s$ and destination element $t$ that satisfy $s\preceq x \preceq t$, for every $x \in \ground$; let $P^*$ denote the augmented poset. Then, the algorithm stores the directed cover graph $H_{P^*} = (X\cup\{s,t\},E_{P^*})$ of $P^*$. An $s-t$ path of size $n$ is a sequence of edges $\{e_1 = (s_1,t_1),\dots,e_n = (s_n,t_n)\}$ such that $s_1 =s$, $t_n = t$, and for all $i \in \llbracket 1,n-1 \rrbracket$, $t_i = s_{i+1}$. Note that the set of maximal chains $\Chains$ of $P$ is equivalent to the set of $s-t$ paths of $H_{P^*}$: The set of nodes in $X$ visited by an $s-t$ path is a maximal chain $C \in \Chains$, and vice versa.

Next, the algorithm sets the length $\rho_y + \beta_y$ to every edge $(x,y) \in E_{P^*}$ with $y \neq t$, and sets the length $-\alpha$ to every edge $(x,t) \in E_{P^*}$ with $x \in X$. Thus, the length of every $s-t$ path in $H_{P^*}$, whose corresponding maximal chain is $C \in \Chains$, is $\sum_{x \in\chain} (\rho_x + \beta_x) - \alpha \overset{\eqref{delta_C}}{=} \delta_C$. We then compute the shortest distances between all pairs of nodes in $H_{P^*}$: We first topologically sort $H_{P^*}$, and then run the classical shortest path algorithm in directed acyclic graphs starting from each node of $H_{P^*}$. We store the shortest distances in a  matrix $M = (m_{xy})_{(x,y) \in (X\cup\{s,t\})^2}$.
By definition of $P^1 = (X^1,\preceq_{\pathtight{1}})$ in Algorithm~\ref{ALG3}, and since $\delta_C \overset{\eqref{Nec Cond}}{\geq} 0$ \revm{for every $C \in \mathcal{C}$}, we obtain that for every \revm{$(x,y) \in (X^1)^2$ with $x \neq y$}, $x \preceq_{\pathtight{1}}y$ if and only if the length $m_{sx}+m_{xy}+m_{yt}$ of a shortest $s-t$ path in $H_{P^*}$ that goes through $x$ and $y$ is 0. Thus, this shortest path subroutine replaces \aref{alg:Subposet}, and constructs $P^1$ in polynomial time. The algorithm then selects its set of minimal elements $S^1$.



Now, we compute the weight $w^1$ to assign to $S^1$ without enumerating all maximal chains $C \in \pathloose{1}$ such that $|S^1 \cap C| \geq 2$: Our algorithm constructs the subposet $\widehat{P}^1 \coloneqq ({S}^1\cup\{s,t\},\preceq_{\restrict{{S}^1\cup\{s,t\}}})$ of $P^*$, and stores its directed cover graph $H_{\widehat{P}^1} = ({S}^1\cup\{s,t\},E_{\widehat{P}^1})$. The length of each edge $(x,y) \in E_{\widehat{P}^1}$ is set to the shortest distance $m_{xy}$ from $x$ to $y$ in the graph $H_{P^*}$. Then, we extend the shortest path algorithm in directed acyclic graphs to obtain, for each $q \in \llbracket 1,|S^1|\rrbracket$, the distance $\widehat{\ell}^q$ of a shortest path from $s$ to $t$ that traverses $q$ elements of $S^1$. The maximum weight to assign to $S^1$ can be efficiently computed as $w^1 = \min\{\min\{{\rho}_{x}, \ x \in {S}^{1}\},\min\{\frac{{\widehat{\ell}}^{q}}{q - 1}, \ q \in \llbracket 2,|{S}^{1}| \rrbracket\}\}$, which replaces \aref{max_weight}.

%

Finally, the algorithm updates the vector $\rho$ and the set of elements with positive $\rho$. In addition, $w^1$ must be subtracted from the scalar $\alpha$ to capture the update \eqref{update_pi}. This in turn will change the lengths of the edges in $H_{P^*}$ for the next iteration. The key challenges for the analysis of the subsequent iterations $k$ are to account for the fact that some maximal chains are removed by Algorithm~\ref{ALG3}, and that the length of an $s-t$ path in $H_{P^*}$, whose corresponding maximal chain is $C \in \Chains$, is not necessarily $\delta_C^k$.
We now formally present Algorithm~\ref{PolyALG}.
\begin{algorithm}
\caption{\textbf{: Optimal solution of $(\mathcal{Q})$ in affine case}}\label{PolyALG}

  \rev{  \hspace*{\algorithmicindent} \textbf{Input}: Finite nonempty poset $P = (X,\preceq)$, scalar $\alpha \in \mathbb{R}$, and vectors $\rho \in \revm{\mathbb{R}_{\geq0}^{\ground}}$,  $\beta \in \revm{\mathbb{R}^{\ground}}$.\\
    \hspace*{\algorithmicindent} \textbf{Output}: Vector $\widetilde{\sigma} \in \revm{\mathbb{R}_{\geq0}^{\mathcal{P}}}$.}
\begin{algorithmic}[1]



\rev{
\StateNewtwo{Augment the poset into $P^* = (\ground \cup\{s,t\},\preceq)$, where $s\preceq x \preceq t$, $\forall x \in X$}
\StateNewtwo{Construct the directed cover graph $H_{P^*} = (\ground\cup\{s,t\},E_{P^*})$}

\State{$\widetilde{\rho}_{x}^{1} \gets \rho_{x}, \ \forall x \in X$, \quad \quad $\widetilde{\rho}^1_t \gets 0$, \quad \quad $\beta^1_x \gets \beta_x, \ \forall x \in X,$ \quad \quad $\beta_{t}^1 \gets -\alpha$, \quad \quad $\widetilde{X}^1 \gets  \{x \in \ground \ | \ \widetilde{\rho}_{x}^{1} >0\}$}



\State{$k \gets 1$}

\While{$\widetilde{X}^{k} \neq \emptyset$}

\State{Set the length of every edge $(x,y) \in E_{P^*}$ to $\widetilde{\rho}^k_y+\beta_y^k$}

\State{$\mat^k = (m^k_{xy})_{(x,y) \in (X\cup\{s,t\})^2} \gets$ all-pairs shortest distance matrix for the graph $H_{P^*}$} \label{all_pairs_m}

\State{Construct the poset $\widetilde{P}^k = (\widetilde{X}^k,\preceq_{k})$: $\forall \revm{x,y \in \widetilde{X}^k \text{ with } x \neq y}$, $x \prec_k y \Longleftrightarrow m^k_{sx} + m^k_{xy} +m^k_{yt}=0$}


%
%
%
%
%

\State{\revm{Select} $\widetilde{S}^{k}$ the set of minimal elements of $\widetilde{P}^k$}


\State{Construct the subposet $\widehat{P}^k = (\widetilde{S}^k\cup\{s,t\},\preceq_{\restrict{\widetilde{S}^k\cup\{s,t\}}})$ of $P^*$}

\State{Construct the directed cover graph $H_{\widehat{P}^k} = (\widetilde{S}^{k}\cup\{s,t\},E_{\widehat{P}^k})$, and topologically sort it}


\State{$\widehat{\ell}_{x}^{k,q} \gets +\infty$, $\forall x \in \ground\cup\{s,t\}$, $\forall q \in \llbracket -1,|\widetilde{S}^k|\rrbracket$, \quad \quad $\widehat{\ell}_{s}^{k,-1} \gets 0$}

\For{$x \in \widetilde{S}^{k}\cup\{s\}$ in topologically sorted order, and $y \in \widetilde{S}^{k}\cup\{t\}$ such that $(x,y) \in E_{\widehat{P}^k}$}\label{for_begin}
\For{$q \in \llbracket-1,|\widetilde{S}^k|-1\rrbracket$ such that $\widehat{\ell}_y^{k,q+1} > \widehat{\ell}_x^{k,q} + m_{xy}^k$}
\State{$\widehat{\ell}_y^{k,q+1} \gets \widehat{\ell}_x^{k,q} + m_{xy}^k$}
\EndFor
\EndFor\label{for_end}
\State{$\widetilde{w}^{k} \gets  \min\{\min\{\widetilde{\rho}_{x}^{k}, \ x \in \widetilde{S}^{k}\},\min\{\frac{\widehat{\ell}^{k,q}_t}{q - 1}, \ q \in \llbracket 2,|\widetilde{S}^{k}| \rrbracket\}\}$, \quad \quad $\widetilde{\sigma}_{\widetilde{S}^{k}} \gets \widetilde{w}^{k}$}\label{poly_max_weight}



\State{$\beta_x^{k+1} \gets \beta_x^k, \ \forall x \in X$, \quad \quad $\beta_t^{k+1} \gets \beta_t^k +  \widetilde{w}^k$, \quad  \quad $\widetilde{\rho}_{x}^{k+1} \gets \widetilde{\rho}_{x}^{k} - \widetilde{w}^{k} \mathds{1}_{\{x \in \widetilde{S}^{k}\}}, \ \forall x \in \ground\cup\{t\}$}\label{poly_update_vectors}
\State{$\widetilde{X}^{k+1} \gets \{x \in \widetilde{X}^{k} \ | \ \widetilde{\rho}_{x}^{k+1} >0\}$}



\State{$k \gets k+1$}

\EndWhile
}
\end{algorithmic}
\end{algorithm}

For every maximal chain $C \in \Chains$, we define the following sequence induced by Algorithm~\ref{PolyALG}, which represents the length of the corresponding $s-t$ path in  $H_{P^*}$ at the beginning of each iteration:
\begin{align}
\forall k \in \llbracket 1, \stepmax +1\rrbracket, \quad \ell_C^k \coloneqq \sum_{x \in \chain} (\widetilde{\rho}_{x}^{k} + \beta_x^k) + \beta^k_t. \label{Chain Length}
\end{align} 

We now proceed with proving by induction that Algorithm~\ref{PolyALG} is a refinement of Algorithm~\ref{ALG3}:
%
%

\begin{proposition}\label{Same}
Algorithm~\ref{ALG3}'s and Algorithm~\ref{PolyALG}'s outputs are identical. In particular, for every iteration $k \in \llbracket 1,\stepmax +1\rrbracket$, the following hold:
\begin{enumerate}
\item[(i)] The remaining \revm{values} for each element are identical: for every $x \in \ground$, $\widetilde{\rho}^k_x = \rho^k_x$, and $\widetilde{X}^k = X^k$.
\item[(ii)] For every maximal chain $C \in \Chains$, the length of its corresponding $s-t$ path in $H_{P^*}$ is at least $\delta_C^k$. Furthermore, this inequality is tight for every maximal chain in $\allpaths{k}$:
\begin{align}
&\forall C \in \Chains, \ \ell_C^k\geq \combiprob{\chain}{k}, \label{Length Inequal}\\
&\forall \chain \in \allpaths{k}, \ \ell_\chain^k =  \combiprob{\chain}{k}. \label{Length Equal}
\end{align}

\item[(iii)] $\widetilde{P}^k = (\widetilde{X}^k,\preceq_k)$ is a poset identical to $P^k = (X^k,\preceq_{\pathtight{k}})$, and $\widetilde{S}^k = S^k$.

\item[(iv)] The weights assigned by both algorithms are identical: $\widetilde{w}^k = w^k$.

\end{enumerate}

\end{proposition}

\proof{Proof of Proposition~\ref{Same}.} We show $(i)-(iv)$ by induction.

First, consider $k=1$. Since $\widetilde{\rho} = \rho = \rho^1$, then $\widetilde{X}^1 = X^1$. Furthermore, for all $C \in \Chains, \ \ell_C^1 \overset{\eqref{Chain Length}}{=} \sum_{x \in C} (\rho_x + \beta_x) - \alpha \overset{\eqref{delta_C}}{=} \delta_C^1 \overset{\eqref{Nec Cond}}{\geq}  0$.
%
%
%
Therefore, for every \revm{$(x,y) \in (X^1)^2$ with $x \neq y$}, $x \preceq_{\pathtight{1}} y$ if and only if \rev{there exists} $C \in \pathtight{1}$ such that $x,y \in C$, which in turn is equivalent to the length of a shortest path in $H_{P^*}$ traversing  $x$ and $y$ being 0.
Thus, $\widetilde{P}^1 = P^1$, and $\widetilde{S}^1 = S^1$. Next, since $\ell^1=\delta^1$, we obtain that:
\begin{align*}
\widetilde{w}^1 &\overset{\bref{poly_max_weight}}{=}  \min \Big\{\min\{\widetilde{\rho}_{x}^{1}, \ x \in \widetilde{S}^{1}\},\min\Big\{\frac{\widehat{\ell}^{1,q}_t}{q - 1}, \ q \in \llbracket 2,|\widetilde{S}^{1}| \rrbracket\Big\}\Big\} \\
&=  \min\Big\{\min\{{\rho}_{x}^{1}, \ x \in {S}^{1}\},\min\Big\{\frac{{\delta}^{1}_C}{|S^1\cap C| - 1}, \ C \in \Chains \ | \ |S^1 \cap C| \geq 2\Big\}\Big\}.
\end{align*}

Note that $|S^1 \cap C| \leq 1$ for every maximal chain $C \in \pathtight{1}$, by definition of $P^1$. Since $\allpaths{1} = \Chains$, we deduce that $\{C \in \Chains \ | \ |S^1 \cap C| \geq 2\} = \{C \in \pathloose{1} \ | \ |S^1 \cap C| \geq 2\} $. Therefore:
\begin{align*}
\widetilde{w}^1 = \min\Big\{\min\{{\rho}_{x}^{1}, \ x \in {S}^{1}\},\min\Big\{\frac{{\delta}^{1}_C}{|S^1\cap C| - 1}, \ C \in \pathloose{1} \ | \ |S^1 \cap C| \geq 2\Big\}\Big\} \overset{\aref{max_weight}}{=} w^1.
\end{align*}

%

We now assume that $(i)-(iv)$ hold for $k \in \llbracket 1,\stepmax\rrbracket$, and show that they also hold for $k+1$:
\begin{enumerate}
\item[$(i)$] Since $\widetilde{\rho}^k = \rho^k$, $\widetilde{S}^k = S^k$, and $\widetilde{w}^k = w^k$, then for every $x \in X$, $\widetilde{\rho}_x^{k+1} \overset{\bref{poly_update_vectors}}{=} \widetilde{\rho}_x^k - \widetilde{w}^k\mathds{1}_{\{x \in \widetilde{S}^k\}} = {\rho}_x^k - {w}^k\mathds{1}_{\{x \in {S}^k\}}  \overset{\aref{update_vectors}}{=}  \rho_x^{k+1}$. This also implies that $\widetilde{X}^{k+1} = X^{k+1}$.

\item[$(ii)$] For every maximal chain $C \in \Chains$, $\ell_C^k \geq \delta_C^k$ implies that:
\begin{align}
\ell_C^{k+1} \overset{\eqref{Chain Length},\bref{poly_update_vectors}}{=} \ell_C^{k} - \widetilde{w}^k |\widetilde{S}^k \cap C| + \widetilde{w}^k  \geq  \delta_C^{k} - w^k|\setmin{k} \cap C| + w^k\mathds{1}_{\{\setmin{k} \cap C \neq \emptyset\}} \overset{\aref{update_vectors}}{=}  \delta_C^{k+1}. \label{ell_delta}
\end{align}

If $C \in \allpaths{k+1}$, then $S^k \cap C \neq \emptyset$. Since $\allpaths{k+1} \subseteq \allpaths{k}$, then $\ell_C^k = \delta_C^k$ by inductive hypothesis. Therefore, \eqref{ell_delta} is tight for $C \in \allpaths{k+1}$.

\item[$(iii)$] Consider \revm{$(x,y) \in (X^{k+1})^2$ with $x \neq y$}.  If $x \preceq_{\pathtight{k+1}} y$, then \rev{there exists} $C^* \in \pathtight{k+1} \subseteq \allpaths{k+1}$ such that $x,y \in C^*$. Consequently, $\ell_{C^*}^{k+1} \overset{\eqref{Length Equal}}{=} \delta_{C^*}^{k+1} = 0 \leq \delta_C^{k+1}  \overset{\eqref{Length Inequal}}{\leq} \ell_C^{k+1}$ \rev{for every} $C \in \Chains$. Therefore, the $s-t$ path corresponding to $C^*$ is \revm{a} shortest path in $H_{P^*}$ that goes through $x$ and $y$, and has length 0. Thus, $x\preceq_{k+1}y$.

Now, assume that $x$ and $y$ are not comparable in $P^{k+1}$. Two cases arise:
\begin{itemize}
\item[Case 1:] $x$ and $y$ are not comparable in $P$. Then, there is no $s-t$ path in $H_{P^*}$ that goes through $x$ and $y$, which implies that $x$ and $y$ are not comparable in $\widetilde{P}^{k+1}$.

\item[Case 2:] $x \prec y$ in $P$. Then, $\delta_C^{k+1} > 0$ \rev{for all} $C \in \allpaths{k+1}$ \rev{such that} $x,y \in C$, by definition of $P^{k+1}$. Let $C^\prime \in \Chains$ be the maximal chain corresponding to a shortest path in $H_{P^*}$ that goes through $x$ and $y$. If $C^\prime \in \allpaths{k+1}$, then $\ell_{C^\prime}^{k+1} \overset{\eqref{Length Equal}}{=} \delta_{C^\prime}^{k+1} > 0$. If $C^\prime \notin \allpaths{k+1}$, then by applying Lemma~\ref{Dominates} as in Section~\ref{Big Proof}, we obtain that there exists a maximal chain $C^{(2)} \in \allpaths{k+1}$ such that $\delta^{k+1}_{C^\prime} \geq \delta^{k+1}_{C^{(2)}}$ and $C^{(2)} \cap X^{k+1} \supseteq C^{\prime} \cap \component{k+1}$. Consequently, $x,y \in C^{(2)}$, and $\ell_{C^\prime}^{k+1} \overset{\eqref{Length Inequal}}{\geq}  \delta_{C^\prime}^{k+1}  \geq \delta_{C^{(2)}}^{k+1} >0$. Thus, $x$ and $y$ are not comparable in $\widetilde{P}^{k+1}$.

\end{itemize}

In conclusion, $\widetilde{P}^{k+1} = P^{k+1}$, and $\widetilde{S}^{k+1}  = S^{k+1}$.

\item[$(iv)$] First, we note that: 
\begin{align}
\min\Big\{\frac{\widehat{\ell}^{k+1,q}_t}{q - 1}, \ q \in \llbracket 2,|\widetilde{S}^{k+1}| \rrbracket\Big\} = \min\Big\{\frac{{\ell}^{k+1}_C}{|{S}^{k+1} \cap C| - 1}, \ C \in \Chains \ | \ |S^{k+1} \cap C| \geq 2\Big\}.\label{Pb_Chains}
\end{align}

If the minimization problem in \eqref{Pb_Chains} is infeasible, that is, there is no maximal chain $C \in \Chains$ such that $|S^{k+1} \cap C| \geq 2$, then $\{C \in \pathloose{k+1} \ | \ |S^{k+1} \cap C| \geq 2\} = \emptyset$. In this case, we obtain $\widetilde{w}^{k+1} \overset{\bref{poly_max_weight}}{=} \min\{\widetilde{\rho}_x^{k+1}, \ x \in \widetilde{S}^k\} = \min\{{\rho}_x^{k+1}, \ x \in {S}^k\} \overset{\aref{max_weight}}{=} w^{k+1}$.

Next, consider the case where  \eqref{Pb_Chains} is feasible. We now show that the optimal value of \eqref{Pb_Chains} is achieved by a maximal chain in $\allpaths{k+1}$: Let $C^* \in \Chains$ be an optimal solution of \eqref{Pb_Chains}, and assume that $C^* \notin \allpaths{k+1}$. Since $C^* \cap X^{k+1} \supseteq C^* \cap S^{k+1}  \neq \emptyset$, then by applying Lemma~\ref{Dominates}, there exists a maximal chain $C^{(2)} \in \allpaths{k+1}$ such that $\delta_{C^*}^{k+1} \geq \delta_{C^{(2)}}^{k+1}$ and $C^{(2)} \cap S^{k+1} \supseteq C^{*} \cap S^{k+1}$.
Therefore, we obtain:
\begin{align*}
\frac{\ell_{C^*}^{k+1}}{|S^{k+1} \cap C^*| - 1} \overset{\eqref{Length Inequal}}{\geq} \frac{\delta_{C^*}^{k+1}}{|S^{k+1} \cap C^*| - 1} \geq \frac{\delta_{C^{(2)}}^{k+1}}{|S^{k+1} \cap C^{(2)}| - 1} \overset{\eqref{Length Equal}}{=} \frac{\ell_{C^{(2)}}^{k+1}}{|S^{k+1} \cap C^{(2)}| - 1}.
\end{align*}

Thus, $C^{(2)} \in \allpaths{k+1}$ is also an optimal solution of \eqref{Pb_Chains}. Then, we derive the following inequality:
\begin{align*}
\forall C \in \pathloose{k+1} \ | \ |S^{k+1} \cap C| \geq 2, \ \frac{\delta_{C^{(2)}}^{k+1}}{|S^{k+1} \cap C^{(2)}| - 1} &\overset{\eqref{Length Equal}}{=}  \frac{\ell_{C^{(2)}}^{k+1}}{|S^{k+1} \cap C^{(2)}| - 1} \\
&\leq  \frac{\ell_{C}^{k+1}}{|S^{k+1} \cap C| - 1} \overset{\eqref{Length Equal}}{=} \frac{\delta_{C}^{k+1}}{|S^{k+1} \cap C| - 1}.
\end{align*}

Therefore, $C^{(2)} \in \argmin\{\frac{\delta_{C}^{k+1}}{|S^{k+1} \cap C| - 1}, \ C \in \pathloose{k+1} \ | \ |S^{k+1} \cap C| \geq 2|\}$, and we obtain:
\begin{align*}
\widetilde{w}^{k+1} & \overset{\bref{poly_max_weight}}{=} \min\Big\{\min\{\widetilde{\rho}^{k+1}_x, \ x \in \widetilde{S}^{k+1}\},\min\Big\{\frac{{\ell}^{k+1}_C}{|\widetilde{S}^{k+1} \cap C| - 1}, \ C \in \Chains \ | \ |\widetilde{S}^{k+1} \cap C| \geq 2\Big\}\Big\}\\
& \overset{\eqref{Length Equal}}{=} \min\Big\{\min\{{\rho}^{k+1}_x, \ x \in {S}^{k+1}\},\min\Big\{\frac{{\delta}^{k+1}_C}{|{S}^{k+1} \cap C| - 1}, \ C \in \pathloose{k+1} \ | \ |S^{k+1} \cap C| \geq 2\Big\}\Big\} \overset{\aref{max_weight}}{=}{w}^{k+1}. 
\end{align*}
%

\end{enumerate}
We conclude by induction that $(i)-(iv)$ hold for every $k \in \llbracket 1,\revm{n^*+1}\rrbracket$. 
\hfill \Halmos
\endproof

In conclusion, Algorithm~\ref{PolyALG} computes an optimal solution of \OPOC when $\pi$ is an affine function of the elements constituting each maximal chain. Importantly, Algorithm~\ref{PolyALG} is a polynomial algorithm: Its running time is governed by \bref{all_pairs_m} and \bref{for_begin}-\bref{for_end}, which both require $O(|\ground|(|\ground|+|E_P|))$ operations since $H_{P^*}$ and $H_{\widehat{P}^k}$ are directed acyclic graphs (Ahuja et al. \cite{Ahuja:1993:NFT:137406}). Since the algorithm terminates after $\stepmax \leq |\ground|+|E_P|$ iterations, Algorithm~\ref{PolyALG} runs in $O(|\ground|(|\ground|+|E_P|)^2)$ time.


\renewcommand{\att}{I}

\section{\color{black}Applications to network interdiction.}\label{sec:games}

\rev{In this section, we introduce a strategic interdiction game involving a routing entity and an interdictor interacting on a flow network.
%
We use Theorems~\ref{thm:finally} and \ref{optimal value} on the existence of probability distributions over posets to characterize the equilibria of this game.
We also provide a solution approach for the equilibrium computation of this game, which involves solving a minimum-cost circulation problem and running Algorithm~\ref{PolyALG}.
%
}

\subsection{Game-theoretic model.}\label{ss:model}


{\color{black} \,} Consider a flow network, modeled as a \rev{simple} directed connected acyclic graph $\mathcal{G} = (\nodes,\edges)$, where $\nodes$ (resp. $\edges$) represents the set of nodes (resp. the set of edges) of the network. For each edge $(i,j) \in \mathcal{E}$, let $c_{ij} \in \revm{\mathbb{R}_{>0}}$ denote its capacity. \rev{Assume} that a single commodity can \rev{be sent through} $\mathcal{G}$ from a source node $s \in \mathcal{V}$ to a destination node $t\in \mathcal{V}$.
\rem{\rev{Recall that} an $s-t$ path $\lambda$ of size $n$ is a sequence of edges $\{e_1 = (s_1,t_1),\dots,e_n = (s_n,t_n)\}$ such that $s_1 =s$, $t_n = t$, and for all $k \in \llbracket 1,n-1 \rrbracket$, $t_k = s_{k+1}$.} \rev{Let $\Lambda$ denote} the set containing all $s-t$ paths of $\mathcal{G}$. 

A flow, \rev{denoted} by the vector $\flow \in \revm{\mathbb{R}_{\geq0}^{\Lambda}}$, enters the network from $s$ and leaves from $t$. A flow $\flow$ is said to be feasible if the flow through each edge does not exceed its capacity; that is, for all $(i,j) \in \mathcal{E},  \ \flow_{ij} \coloneqq \sum_{\{\lambda \in \Lambda \, | \, (i,j) \in \lambda\}} \flow_\lambda \leq c_{ij}$.
Let $\mathcal{F}$ denote the set of feasible flows of $\mathcal{G}$. 
Given a feasible flow $\flow \in \mathcal{F}$, let $\Value{\flow} \coloneqq \sum_{\lambda \in \Lambda} \flow_\lambda$ denote the amount of flow sent from the node $s$ to the node $t$. 
Each edge $(i,j)\in \edges$ is associated with a  marginal transportation cost, denoted $b_{ij} \in \revm{\mathbb{R}_{>0}}$. For each $s-t$ path $\lambda \in \Lambda$, $b_\lambda \coloneqq \sum_{(i,j) \in \lambda} b_{ij} $ represents the cost of transporting one unit of flow through $\lambda$. Given a feasible flow $\flow \in \mathcal{F}$, $\Costf{\flow} \coloneqq \sum_{\lambda \in \Lambda} b_\lambda \flow_\lambda$ denotes the total transportation cost of $\flow$. 
%
%
%
%

%
%
%


\rev{We define} a two-player game $\Gamma \coloneqq \langle\{1,2\},(\mathcal{F},\mathcal{I}),(u_1,u_2)\rangle$ on the flow network $\mathcal{G}$. Player 1 (\defender) is the routing entity that chooses to route a flow $\flow \in \mathcal{F}$ of goods through the network, and player 2 (\attacker) is the interdictor who simultaneously chooses a subset of edges $\att \in 2^\edges$ to interdict. 
%
%
The action set for \defender (resp. \attacker)  is $\mathcal{F}$  (resp. $\mathcal{I} \coloneqq 2^{\mathcal{E}}$). 
For every edge $(i,j) \in \edges$, $d_{ij} \in \revm{\mathbb{R}_{>0}}$ denotes the cost of interdicting $(i,j)$. Thus, the cost of \rev{an} interdiction $\att \in \mathcal{I}$ is given by $\Cost{\att} \coloneqq \sum_{(i,j) \in \att} d_{ij}$. In this model, \attacker (resp. \defender) gains (resp. looses) the flow that crosses the edges that are interdicted by \attacker. \rev{The model captures strategic routing situations when \defender cannot observe \attacker's actions before sending its flow and cannot re-route its flow after the interdiction. We do not consider partial edge interdictions for the sake of simplicity.} 
The \emph{effective flow} when a flow ${\flow}$ is chosen by \defender and an interdiction $\att$ is chosen by \attacker is ${\flow}^{\att}$, where $\flow^{\att}_\lambda  = \flow_\lambda \mathds{1}_{\{\lambda \cap \att = \emptyset\}}$ \rev{for all $\lambda \in \Lambda$}. We also suppose that the transportation cost incurred by \defender is for the initial flow $\flow$ and not for the effective flow $\flow^\att$. \rev{This modeling choice reflects \rev{an ex ante} monetary \rev{fee} paid by \defender to the network owner who provides \defender the access to send a quantity of flow through the network.}




The payoff of \defender is the value of effective flow assessed by \defender net the cost of transporting the initial flow: $u_1({\flow},{\att}) =p_1 \Eff{\flow}{\att} - \Costf{\flow}$, 
where $p_1\in \revm{\mathbb{R}_{>0}}$ is the marginal value of effective flow for \defender. 
Similarly, the payoff of \attacker is the value of  interdicted flow assessed by \attacker net the cost of interdiction: $u_2({\flow},\att)  = p_2 (\Value{\flow} - \Eff{\flow}{\att}) - \Cost{\att}$, 
where $p_2\in \revm{\mathbb{R}_{>0}}$ is the marginal value of interdicted flow for \attacker.



\rev{In playing the game $\Gamma$,} \defender can route goods in the network using a flow $\flow$ realized from a chosen probability distribution on the set $\mathcal{F}$, and  \attacker can interdict subsets of edges according to a  probability distribution on the set $\mathcal{I}$.  Specifically, \defender and \attacker respectively choose a mixed routing strategy $\sigma^1 \in \Delta(\mathcal{F})$ and a mixed interdiction strategy $\sigma^2 \in \Delta(\mathcal{I})$, where $\Delta(\mathcal{F}) = \{\sigma^1 \in \revm{\mathbb{R}_{\geq0}^{\mathcal{F}}} \ | \ \sum_{{\flow} \in \mathcal{F}} \sigma^1_\flow = 1 \}$,  and $\Delta(\mathcal{I}) = \{\sigma^2 \in \revm{\mathbb{R}_{\geq0}^{\mathcal{I}}}\ | \ \sum_{{\att} \in \mathcal{I}} \sigma^2_{\att} = 1\}$ denote the strategy sets.
%
%
Here, $\sigma^1_{\flow}$ (resp. $\sigma^2_\att$) represents the probability assigned to the flow $\flow$ (resp. interdiction $\att$) by \defender's routing strategy $\sigma^1$ (resp. \attacker's interdiction strategy $\sigma^2$).
The players' strategies are independent randomizations.
Given a strategy profile $\sigma = (\sigma^1,\sigma^2) \in \Delta(\mathcal{F}) \times \Delta(\mathcal{I})$, the expected payoffs are expressed as:
\begin{align}
U_1(\sigma^1,\sigma^2) &= p_1 \mathbb{E}_{\sigma}[\Eff{\flow}{\att}] - \mathbb{E}_{\sigma}[\Costf{\flow}],\label{payoff1}\\
U_2(\sigma^1,\sigma^2) & =p_2 \left(\mathbb{E}_{\sigma}[\Value{\flow}]-\mathbb{E}_{\sigma}[\Eff{\flow}{\att}]\right)
- \mathbb{E}_{\sigma}[\Cost{\att}].\label{payoff2}
\end{align}
\normalsize

\rev{We will also use the notations $U_i(\sigma^1,\att) = U_i(\sigma^1,\mathds{1}_{\{\att\}})$ and $U_i(\flow,\sigma^2) = U_i(\mathds{1}_{\{\flow\}},\sigma^2)$ for $i \in \{1,2\}$. We focus on characterizing the mixed strategy Nash equilibria of the game $\langle \{1,2\}, (\Delta(\mathcal{F}),\Delta(\mathcal{I})),(U_1,U_2)\rangle$.}
A strategy profile $({\sigma^1}^\ast,{\sigma^2}^\ast) \in \Delta(\mathcal{F}) \times \Delta(\mathcal{I})$ is a mixed strategy \emph{Nash Equilibrium} (NE)  of game $\Gamma$ if: \rev{for all} ${\sigma^1} \in \Delta(\mathcal{F}), \ U_1({\sigma^1}^*,\sigma^{2^*}) \geq U_1({\sigma^1},\sigma^{2^*})$, and \rev{for all}  ${\sigma^2} \in \Delta(\mathcal{I}), \ U_2({\sigma^1}^*,\sigma^{2^*}) \geq U_2({\sigma^1}^*,{\sigma^2})$. 
Equivalently, in a NE $(\sigma^{1^*},\sigma^{2^*})$, $\sigma^{1^*}$ (resp. $\sigma^{2^*}$) is a best response to $\sigma^{2^*}$ (resp. $\sigma^{1^*}$).
\rev{Let $\Sigma$ denote} the set of NE of $\Gamma$. 

We now proceed with the equilibrium analysis of the game $\Gamma$.

\subsection{Properties of Nash equilibria.}


$\Gamma$ is strategically equivalent to a zero-sum game; in particular, the following transformation preserves the set of NE:
%
%
\begin{align}
&\forall (\flow,\att) \in \mathcal{F} \times \mathcal{I}, \ \frac{1}{p_1}u_1(\flow,\att) + \frac{1}{p_2}\Cost{\att} = \Eff{\flow}{\att}   - \frac{1}{p_1}\Costf{\flow} + \frac{1}{p_2}\Cost{\att}\eqqcolon \widetilde{u}_1(\flow,\att), \label{transform1}\\
&\forall (\flow,\att) \in \mathcal{F} \times \mathcal{I}, \  \frac{1}{p_2}u_2(\flow,\att) - \Value{\flow} + \frac{1}{p_1}\Costf{\flow} =- \Eff{\flow}{\att} + \frac{1}{p_1}\Costf{\flow}  - \frac{1}{p_2}\Cost{\att}   = -\widetilde{u}_1(\flow,\att). \label{transform2}
\end{align}

Therefore, $\Gamma$ and $\widetilde{\Gamma} \coloneqq \langle \{1,2\}, (\mathcal{F}, \mathcal{I}), (\widetilde{u}_1,-\widetilde{u}_1) \rangle$ have the same equilibrium set. Additionally, NE of $\Gamma$ are interchangeable, i.e., if $({\sigma^1}^*,\sigma^{2^*}) \in \Sigma$ and $({\sigma^1}^{\prime},{\sigma^2}^{\prime})\in \Sigma$, then $({\sigma^1}^*,{\sigma^2}^{\prime})\in \Sigma$ and $({\sigma^1}^{\prime},{\sigma^2}^{*})\in \Sigma$. 
%
Also note that due to the splittable nature of the flow for any routing strategy $\sigma^1 \in\Delta(\mathcal{F})$ of \defender, one can consider an equivalent pure strategy \revm{$\bar{f} \in \mathcal{F}$ defined by $\bar{f}_\lambda = \mathbb{E}_{\sigma^1}[\flow_\lambda]$ for all $\lambda \in \Lambda$, which satisfies $U_i(\sigma^1,\sigma^2) = U_i(\bar{\flow},\sigma^2)$ for all $i \in \{1,2\}$ and $\sigma^2 \in \Delta(\mathcal{I})$, since $u_i(\cdot,\sigma^2)$ is an affine function.} 

\rev{The above-mentioned properties imply that} linear programming techniques \rev{can be used to obtain the NE of $\Gamma$}. However, this would entail solving a linear program \rev{of exponential size, containing $|\Lambda| +1$ variables and $2^{|\edges|} + |\edges|$ constraints.} 
%
%
%
 \rev{Instead, we derive an approach for efficiently solving $\Gamma$:  We show that by utilizing the primal and dual solutions of a minimum-cost circulation problem and applying our results on posets (Theorems~\ref{thm:finally} and \ref{optimal value}), we can obtain a complete equilibrium characterization for game $\Gamma$. Furthermore, using Algorithm~\ref{PolyALG}, we obtain a polynomial-time approach to compute NE of this game.}

\rev{We begin by considering the following ``natural'' network flow problem:}\hypertarget{MCCP}{} 
%
\begin{align*}
\begin{array}{lrll} (\mathcal{M})  & \quad\text{maximize} & \displaystyle  \Value{\flow} - \frac{1}{p_1}\Costf{\flow}&\\
& \text{subject to} & \displaystyle \sum_{\{\lambda \in \Lambda \, | \, (i,j) \in \lambda\}} \flow_\lambda \leq \min\left\{\frac{d_{ij}}{p_2},c_{ij}\right\}, & \forall (i,j) \in \edges\\[0.5cm]
& & \flow_\lambda \geq 0, & \forall \lambda \in \Lambda.\end{array}
\end{align*}

\rev{This problem} consists in finding a feasible flow $\flow$ in $\mathcal{F}$ that maximizes $u_1(\flow,\emptyset)$ with the \rev{constraint} that the flow through each edge $(i,j)$ is no more than $\frac{d_{ij}}{p_2}$. Game theoretically, this threshold captures \attacker's best response to \defender. \rev{Indeed,} if $\flow_{ij} > \frac{d_{ij}}{p_2}$ for some $(i,j) \in \edges$, then \attacker has an incentive to interdict $(i,j)$, resulting in an increase of \attacker's payoff (since $u_2(\flow,\{(i,j)\}) = p_2 \flow_{ij} - d_{ij} > 0$). Thus, \original can be viewed as the problem in which \defender maximizes its payoff while limiting \attacker's incentive to interdict any of the edges.
%
For each $s-t$ path $\lambda \in \Lambda$, let us denote $\pi^0_\lambda \coloneqq 1 - \frac{b_{\lambda}}{p_1}$. Then, the value $p_1\pi^0_\lambda$ represents the gain in \defender's payoff when one unit of flow traveling along $\lambda$ reaches the destination node. The primal and dual formulations of \original are given as follows:\hypertarget{LPs}{}
%
%
%
\begin{align*}
\begin{array}{lrll}  (\mathcal{M}_P): \ & \text{max} & \displaystyle\sum_{\lambda \in \Lambda}\pi^0_\lambda\flow_\lambda &\\
& \text{s.t.} & \displaystyle \sum_{\{\lambda \in \Lambda \, | \, (i,j) \in \lambda\}} \flow_\lambda \leq \frac{d_{ij}}{p_2}, & \forall (i,j) \in \edges\\[0.5cm]
&& \displaystyle \sum_{\{\lambda \in \Lambda \, | \, (i,j) \in \lambda\}} \flow_\lambda \leq c_{ij}, & \forall (i,j) \in \edges\\[0.5cm]
& & \flow_\lambda \geq 0, & \forall \lambda \in \Lambda\end{array} \quad \ \vrule \ \quad 
\begin{array}{lrll}(\mathcal{M}_D): \ & \text{min} & \displaystyle\sum_{(i,j) \in \edges}\left(\frac{d_{ij}}{p_2} \rho_{ij} + c_{ij}\mu_{ij}\right) &\mcr
& \text{s.t.} & \displaystyle \sum_{(i,j) \in \lambda} (\rho_{ij}+\mu_{ij}) \geq \pi^0_\lambda, & \forall \lambda \in \Lambda\mcr
& & \rho_{ij} \geq 0, & \forall (i,j) \in \edges\mcr
& & \mu_{ij} \geq 0, & \forall (i,j) \in \edges\end{array}
\end{align*}


\rev{Let $\mathcal{O}^*_{\primal}$ (resp. $\mathcal{O}^*_{\dual}$) denote the set of optimal solutions of \primal (resp. \dual).  By strong duality, the optimal value of  \primal  is identical to that of \dual; we denote it by $z^*_{\original}$. }
Note that \primal and \dual may have an exponential number of variables and constraints, respectively. However, equivalent \rev{polynomial-size} primal and dual formulations of \original can be derived; see Appendix~\ref{sec:MCCP}. 
Thus, $\flow^* \rev{\in \mathcal{O}^*_{\primal}}$ and $(\rho^*,\mu^*)\rev{\in\mathcal{O}^*_{\dual}}$ can be \rev{efficiently} computed by using an interior point method (Karmarkar \cite{Karmarkar1984}) or a dual network simplex algorithm  (Orlin et al. \cite{Orlin1993}). \rev{Alternatively, \original can be formulated as a minimum-cost circulation problem in a graph $\mathcal{G}^\prime = (\nodes^\prime,\edges^\prime)$ such that $\nodes^\prime = \nodes$, $\edges^\prime = \edges \cup \{(t,s)\}$. The capacity of each edge $(i,j) \in \edges$ is given by $\min\{\frac{d_{ij}}{p_2},c_{ij}\}$, and edge $(t,s)$ is uncapacitated. The transportation cost of  each edge $(i,j) \in \edges$ is given by $\frac{b_{ij}}{p_1}$, and the transportation cost of edge $(t,s)$ is $-1$. Thus, \rev{\primal and \dual can be solved using known combinatorial algorithms} (Ahuja et al. \cite{Ahuja:1993:NFT:137406}).}


%
%

\rev{From complementary slackness, we know that any pair of optimal primal and dual solutions $\flow^* \in \mathcal{O}^*_{\primal}$ and $(\rho^*,\mu^*) \in \mathcal{O}^*_{\dual}$ satisfies the following properties:} 
\begin{align}
\forall (i,j) \in \edges, \ \rho_{ij}^* > 0 \ &\Longrightarrow \ \flow_{ij}^* = \sum_{\{\lambda \in \Lambda \, | \, (i,j) \in \lambda\}}\flow_{\lambda}^* = \frac{d_{ij}}{p_2},\label{cs1_2}\\
\forall (i,j) \in \edges, \ \mu_{ij}^* > 0 \ &\Longrightarrow \ \flow_{ij}^* = \sum_{\{\lambda \in \Lambda \, | \, (i,j) \in \lambda\}}\flow_{\lambda}^* = c_{ij},\label{cs3_2}\\
\forall \lambda \in \Lambda, \ \flow_{\lambda}^* > 0 \ &\Longrightarrow \ \sum_{(i,j) \in \lambda}(\rho_{ij}^* + \mu_{ij}^*) = \pi^0_\lambda. \label{cs2_2}
\end{align}


These properties, along with \rev{Theorems~\ref{thm:finally} and \ref{optimal value}}, enable us to derive the following result:

\begin{theorem}\label{One_NE_general}
\rev{A strategy profile $(\sigma^{1^*},\sigma^{2^*}) \in \Delta(\mathcal{F}) \times \Delta(\mathcal{I})$ is a NE of the game $\Gamma$ if and only if there exists a pair of optimal primal and dual solutions $\flow^* \in \mathcal{O}^*_{\primal}$ and $(\rho^*,\mu^*) \in \mathcal{O}^*_{\dual}$ such that:
\begin{align}
\forall \lambda \in \Lambda, \ & \quad \  \; \sum_{\flow \in \mathcal{F}}\sigma^{1^*}_\flow \flow_\lambda = \flow^*_\lambda, \label{Expected_Flow}\\
\forall (i,j) \in \edges, \ & \sum_{\{\att \in \mathcal{I}\, | \, (i,j) \in \att\}}{\sigma}^{2^*}_{\att}  = \rho^*_{ij}, \label{eq_strat}\\
\forall \lambda \in \Lambda, \  &\sum_{\{\att \in \mathcal{I} \, | \,\att \cap \lambda \neq \emptyset\}} {\sigma}^{2^*}_{\att}\geq \pi^*_\lambda,\label{ineq_strat}
\end{align}
where $\pi_\lambda^* \coloneqq \pi^0_\lambda - \sum_{(i,j) \in \lambda}\mu_{ij}^*$ for all $\lambda \in \Lambda$.
The corresponding equilibrium payoffs are $U_1(\sigma^{1^*},\sigma^{2^*}) = p_1 \sum_{(i,j) \in \edges} c_{ij}\mu_{ij}^*$ and $U_2(\sigma^{1^*},\sigma^{2^*}) = 0$.
}

\rem{
Consider $\flow^*$ and $(\rho^*,\mu^*)$  optimal solutions of \primal and \dual, respectively. Theorem~\ref{thm:finally} guarantees the existence of an interdiction strategy $\widetilde{\sigma}^2 \in \Delta(\mathcal{I})$ satisfying: 
\begin{align}
\forall (i,j) \in \edges, \ &\sum_{\{\att \in \mathcal{I}\, | \, (i,j) \in \att\}}\widetilde{\sigma}^{2}_{\att}  = \rho^*_{ij}, \label{eq_strat}\\
\forall \lambda \in \Lambda, \  &\sum_{\{\att \in \mathcal{I} \, | \,\att \cap \lambda \neq \emptyset\}} \widetilde{\sigma}^{2}_{\att}\geq \pi^*_\lambda,\label{ineq_strat}
\end{align}
where $\forall \lambda \in \Lambda, \ \pi^*_\lambda \coloneqq\pi^0_\lambda - \sum_{(i,j) \in \lambda}\mu_{ij}^*$.

The strategy profile $(\flow^*,\widetilde{\sigma}^2) \in  \mathcal{F}\times \Delta(\mathcal{I})$ is a NE of  the game $\Gamma$. The corresponding equilibrium payoffs are $U_1(\flow^*,\widetilde{\sigma}^2) = p_1 \sum_{(i,j) \in \edges} c_{ij}\mu_{ij}^*$ and $U_2(\flow^*,\widetilde{\sigma}^2) = 0$.}
\end{theorem}


\rev{Thus, optimal primal and dual solutions of \original provide necessary and sufficient conditions for a strategy profile to be a NE.  In particular, optimal primal solutions represent the expected flows sent by \defender in equilibrium. Additionally, optimal dual solutions  characterize the marginal probabilities with which network components are interdicted in equilibrium; that is, 
\attacker's equilibrium strategy interdicts each edge $(i,j) \in \edges$ with probability $\rho^*_{ij}$, and interdicts each path $\lambda \in \Lambda$ with probability at least  $\pi^0_\lambda - \sum_{(i,j) \in \lambda}\mu_{ij}^*$.}


\rev{While showing that \eqref{Expected_Flow}-\eqref{ineq_strat} are sufficient conditions is relatively straightforward, the key challenge lies in proving that they are also necessary. In proving Theorem~\ref{One_NE_general}, we first show the existence of an interdiction strategy ${\sigma}^{2^*} \in \Delta(\mathcal{I})$ satisfying \eqref{eq_strat} and \eqref{ineq_strat} given an optimal dual solution of \original. In fact, this existence problem is an instantiation of problem \FPOC that we introduced in Section~\ref{sec:Setting} and positively answered in Theorem~\ref{thm:finally}.
Secondly, showing that \eqref{eq_strat} and \eqref{ineq_strat} are necessary conditions satisfied by \attacker's interdiction strategies in equilibrium involves exploiting strong duality in the strategically equivalent zero-sum game $\widetilde{\Gamma}$.}
Finally, the necessary condition \eqref{Expected_Flow} is a consequence of the $s-t$ paths having positive transportation costs. \revm{The proof exploits Theorem~\ref{optimal value},} which guarantees the existence of \attacker's strategy that, \revm{with positive probability $1 - \max\{\max\{\rho^*_{ij}, \ (i,j) \in \edges\},\max\{1 - \frac{b_\lambda}{p_1} - \sum_{(i,j) \in \lambda}\mu_{ij}^*, \ \lambda \in \Lambda\}\}$, does not interdict any edges at all in equilibrium.} 

\revm{We remark that for the case when the path transportation costs are assumed to be nonnegative (instead of strictly positive), conditions  \eqref{eq_strat} and \eqref{ineq_strat} are still necessary and sufficient for equilibrium interdiction strategies. However, \eqref{Expected_Flow} is only a sufficient condition for equilibrium routing strategies.
%
%
Indeed, if a path with low interdiction costs has zero cost of transportation, \attacker will interdict this path with probability 1. Any flow sent by \defender along this path will then always be interdicted, and in fact \defender can select an equilibrium strategy that saturates this path and violates constraints in \primal.} 

\rev{In fact, dual solutions of \original can be used to infer additional equilibrium properties:} Given an $s-t$ path $\lambda \in \Lambda$, $\pi^0_\lambda$ is the probability above which $\lambda$ \rev{must} be interdicted by \attacker \rev{\ to limit \defender's incentive to send any flow through the network} . However, when edges belonging to $\lambda$ have high interdiction costs, \attacker \rev{\ chooses} not to  interdict these edges, \rev{which may result in the interdiction probability of $\lambda$ being less than $\pi^0_\lambda$}. \rev{This} reduction of interdiction probability of $\lambda$ is captured by $\sum_{(i,j) \in \lambda}\mu^*_{ij}$. By complementary slackness \eqref{cs3_2},  $\mu_{ij}^* >0$ for $(i,j) \in \lambda$ only when $c_{ij} =\flow^*_{ij} \leq \frac{d_{ij}}{p_2}$\rem{, i.e., when the interdiction cost of $(i,j)$ is too high.}. \rev{Hence, the equilibrium} interdiction probability of $\lambda$ is given by $\pi^*_{\lambda} = \pi^0_\lambda - \sum_{(i,j) \in \lambda}\mu^*_{ij}$.

Consequently, if an $s-t$ path $\lambda \in \Lambda$ is such that $\sum_{(i,j) \in \lambda} \mu^*_{ij} >0$, then each unit of flow sent through $\lambda$ increases \defender's payoff by $p_1\sum_{(i,j) \in \lambda} \mu^*_{ij}$. This is captured by \defender's equilibrium strateg\rev{ies}\rev{, with expected flow $\flow^* \in \mathcal{O}^*_{\primal}$, that saturate} every edge $(i,j) \in \edges$ for which $\mu^*_{ij} >0$. Since $\flow^*$ only takes $s-t$ paths that are interdicted with probability exactly $\pi^*$, the resulting equilibrium payoff for \defender is \rev{given by $p_1 \sum_{(i,j) \in \edges} c_{ij}\mu_{ij}^*$}. \rev{Note also that $\flow^*$ does not take any $s-t$ path $\lambda$ for which $\pi_\lambda^0 <0$. This captures the fact that \defender has no incentive to send its flow through $s-t$  paths $\lambda$ for which $b_\lambda > p_1$.}
Recall \rev{from \original} that $\flow^*$ is such that interdicting any edge does not increase \attacker's payoff. Furthermore, \attacker only interdicts edges for which \rev{\revm{her} value from the} interdicted flow compensates the interdiction cost (from \eqref{cs1_2}). Thus, her payoff is 0 in equilibrium.
\rev{It is interesting to note that \defender's strategies and payoff in equilibrium can be expressed in terms of edge values, and are independent of the chosen path decomposition of $\flow^*$.}

\proof{Proof of \rev{Theorem}~\ref{One_NE_general}.} 

\rev{We prove this theorem by showing that conditions \eqref{Expected_Flow}-\eqref{ineq_strat} are sufficient for a strategy profile to be a NE (Step 1); satisfied by at least one strategy profile (Step 2); and satisfied by every NE (Step 3).}

\begin{itemize}
\item[\rev{\textbf{Step 1:}}] \rev{Let $\flow^* \in\mathcal{O}^*_{\primal}$ and $(\rho^*,\mu^*)  \in\mathcal{O}^*_{\dual}$\rem{ be optimal solutions of \primal and \dual, respectively}. First, we show that a strategy profile $(\sigma^{1^*},\sigma^{2^*}) \in \Delta(\mathcal{F}) \times \Delta(\mathcal{I})$ satisfying \eqref{Expected_Flow}-\eqref{ineq_strat} is a NE of $\Gamma$.} We write the following inequality for \defender's payoff:
\begin{align}
&\forall \flow \in \mathcal{F}, \ U_1(\flow,\sigma^{2^*}) \overset{\eqref{payoff1}}{=}  p_1 \sum_{\lambda \in \Lambda} \flow_\lambda \mathbb{E}_{\sigma^{2^*}}[1 -\mathds{1}_{\{\att \cap \lambda  \neq \emptyset\}}] -\sum_{\lambda \in \Lambda} b_\lambda \flow_\lambda  =  p_1 \sum_{\lambda \in \Lambda} \pi^0_\lambda \flow_\lambda - p_1 \sum_{\lambda \in \Lambda} \flow_\lambda \sum_{\{\att \in \mathcal{I} \, | \, \att \cap \lambda \neq \emptyset\}}\sigma^{2^*}_\att  \nonumber\\
&\overset{\eqref{ineq_strat}}{\leq}p_1 \sum_{\lambda \in \Lambda} \flow_\lambda \sum_{(i,j) \in \lambda} \mu_{ij}^*=  p_1 \sum_{(i,j) \in \edges} \flow_{ij}\mu_{ij}^*  \leq p_1 \sum_{(i,j) \in \edges} c_{ij}\mu_{ij}^*. \label{4th_(in)eq_2}
\end{align}


Now, given $\lambda \in \Lambda$ such that $\flow^*_\lambda >0$, we obtain: 
\begin{align}
\sum_{\{\att \in \mathcal{I} \, | \, \att \cap \lambda \neq \emptyset\}}\sigma^{2^*}_\att & \leq \sum_{\att \in \mathcal{I}}\sigma^{2^*}_\att  |\att \cap \lambda| = \sum_{(i,j) \in \lambda}\sum_{\att \in \mathcal{I}}\sigma^{2^*}_\att \mathds{1}_{\{(i,j) \in \att\}} \overset{\eqref{eq_strat}}{=}  \sum_{(i,j) \in \lambda} \rho_{ij}^*\overset{\eqref{cs2_2},\eqref{ineq_strat}}{\leq}\sum_{\{\att \in \mathcal{I} \, | \, \att \cap \lambda \neq \emptyset\}}\sigma^{2^*}_\att.  \label{For much later} 
\end{align}

Furthermore, \rev{for all} $(i,j) \in \edges$ such that $\mu_{ij}^* > 0$, $\flow^*_{ij} \overset{\eqref{cs3_2}}{=} c_{ij}$. Then, inequality \eqref{4th_(in)eq_2} is tight for $\flow^*$, and  $\rev{U_1(\sigma^{1^*},\sigma^{2^*}) \overset{\eqref{Expected_Flow}}{=} \ } U_1(\flow^*,\sigma^{2^*}) = p_1 \sum_{(i,j) \in \edges} c_{ij}\mu_{ij}^*$.

Similarly, regarding \attacker's payoff, we first derive the following inequality:
\begin{align}
\forall \att \in \mathcal{I}, \ &\sum_{(i,j) \in \att} \frac{d_{ij}}{p_2} \geq  \sum_{(i,j) \in \att} \sum_{\{\lambda \in \Lambda \, | \, (i,j) \in \lambda\}}\flow_{\lambda}^*  = \sum_{\lambda \in \Lambda}\flow_\lambda^* |\att \cap \lambda| \geq \sum_{\lambda \in \Lambda}\flow_\lambda^*\mathds{1}_{\{\att \cap \lambda \neq \emptyset\}}  =    \Value{\flow^*} - \Value{\flow^{*\att}}.\label{2nd_(in)eq_2}
\end{align}

Therefore, \rev{for all} $\att \in \mathcal{I}, \ \rev{U_2(\sigma^{1^*},I) \overset{\eqref{Expected_Flow}}{=} \  }U_2(\flow^*,\att) \overset{\eqref{payoff2}}{=} p_2( \Value{\flow^*} - \Value{\flow^{*\att}}) - \sum_{(i,j) \in \att} d_{ij}\overset{\eqref{2nd_(in)eq_2}}{\leq} 0$. 

Now, \revm{consider $I \in \supp(\sigma^{2^*})$. From \eqref{For much later}, we obtain that for every $\lambda \in \Lambda$ such that $f_\lambda^* >0$, $|I\cap \lambda| \leq 1$. Furthermore, for every $(i,j) \in I$,}
%
%
%
$\sum_{\{\lambda \in \Lambda \, | \, (i,j) \in \lambda\}} \flow^*_\lambda \overset{\eqref{cs1_2}}{=} \frac{d_{ij}}{p_2}$, since $\rho^*_{ij} >0$. 
Thus, \rev{for all} $\att \in \supp(\sigma^{2^*})$, inequality \eqref{2nd_(in)eq_2} is tight, and $U_2(\rev{\sigma^{1^*}},\att) = 0$. Therefore, $\rev{U_2(\sigma^{1^*},\sigma^{2^*}) = 0}$, and $(\sigma^{1^*},\sigma^{2^*})$ is a NE.

\item[\rev{\textbf{Step 2}:}] \rev{Let $\flow^*  \in\mathcal{O}^*_{\primal}$ and $(\rho^*,\mu^*)  \in\mathcal{O}^*_{\dual}$. Next, we show that there exists a strategy profile satisfying \eqref{Expected_Flow}-\eqref{ineq_strat}, and obtain the value of the zero-sum game $\widetilde{\Gamma}$. Trivially, if \defender chooses the pure strategy $\flow^*$, \eqref{Expected_Flow} is then satisfied. We now argue that there exists an interdiction strategy $\widetilde{\sigma}^2 \in \Delta(\mathcal{I})$ satisfying \eqref{eq_strat} and \eqref{ineq_strat}.} First, we define the following binary relation on $\edges$, denoted $\preceq_{\mathcal{G}}$: Given $(u,v) \in \mathcal{E}^2,$ $u\preceq_\mathcal{G} v$ if either $u =v$, or there exists an $s-t$ path $\lambda \in \Lambda$ that traverses $u$ and $v$ in this order. Since $\mathcal{G}$ is a directed acyclic connected graph, we \rev{obtain} the following lemma, which is proven in Appendix~\ref{sec:additional}:
\begin{lemma}\label{poset}
$P_{\mathcal{G}} = (\mathcal{E},\preceq_{\mathcal{G}})$ is a poset, whose set of maximal chains is the set of $s-t$ paths $\Lambda$.
\end{lemma}


Thus, showing that there exists $\widetilde{\sigma}^2 \in \Delta(\mathcal{I})$ \rev{satisfying} \eqref{eq_strat} and \eqref{ineq_strat} is an instantiation of problem \FPOC.
Since $(\rho^*,\mu^*) \in \mathcal{O}^*_{\dual}$, then condition~\eqref{Nec Cond} is satisfied, i.e., \rev{for all} $\lambda \in \Lambda, \ \sum_{(i,j) \in \lambda} \rho^*_{ij} \geq\pi^*_\lambda$. Additionally, for any $s-t$ path $\lambda \in \Lambda, \ \pi^*_\lambda = 1 - \sum_{(i,j) \in \lambda} (\frac{b_{ij}}{p_1} + \mu_{ij}^*)$, and $\pi^*$ is an affine function of the \rev{edges} constituting each $s-t$ path. Therefore, $\pi^*$ satisfies the conservation law described in \eqref{Conservation}. 
Finally, since $\rho^*_{ij} \in [0,1]$ \rev{for all} $(i,j) \in \edges$, and $\pi^*_\lambda \leq 1$ \rev{for all} $\lambda \in \Lambda$,  all conditions of Theorem~\ref{thm:finally} are satisfied, and \rev{there exists} an interdiction strategy $\widetilde{\sigma}^2 \in \Delta(\mathcal{I})$ satisfying \eqref{eq_strat} and \eqref{ineq_strat}. \rev{In particular, $\widetilde{\sigma}^2$ can be constructed from Algorithm~\ref{PolyALG}.}

\rev{From Step 1, $(\flow^*,\widetilde{\sigma}^2) \in \Sigma$. Then, \defender's equilibrium payoff in the zero-sum game $\widetilde{\Gamma}$ is:}
\begin{align}
\rev{\forall (\sigma^{1^*},\sigma^{2^*}) \in \nash, \ \widetilde{U}_1(\sigma^{1^*},\sigma^{2^*}) \overset{\eqref{Expected_Flow}}{=} \widetilde{U}_1(\flow^*,\widetilde{\sigma}^{2}) \overset{\eqref{transform1}}{=} \frac{1}{p_1} {U}_1(\flow^*,\widetilde{\sigma}^{2}) + \frac{1}{p_2}\mathbb{E}_{\widetilde{\sigma}^{2}}[\Cost{\att}]  \overset{\eqref{eq_strat}}{=} z^*_{\original}.\label{common}}
\end{align}

\item[\rev{\textbf{Step 3}:}]

\rev{Let $(\sigma^{1^\prime},\sigma^{2^\prime}) \in \Delta(\mathcal{F}) \times \Delta(\mathcal{I})$ be a NE of $\Gamma$. We now show that $(\sigma^{1^\prime},\sigma^{2^\prime})$ satisfies \eqref{Expected_Flow}-\eqref{ineq_strat} for some pair of optimal primal and dual solutions of \original. In particular, we first prove that $\flow^\prime \coloneqq \mathbb{E}_{\sigma^{1^\prime}}[\flow]$ is necessarily an optimal solution of \primal: Given $(\rho^*,\mu^*)  \in\mathcal{O}^*_{\dual}$, consider the equilibrium interdiction strategy $\widetilde{\sigma}^2$ described in Step 2 and constructed from Algorithm~\ref{PolyALG}. From Theorem~\ref{optimal value}, $\widetilde{\sigma}^2_{\emptyset} = 1 - \max\{\max\{\rho^*_{ij}, \ (i,j) \in \edges\},\max\{\pi^*_\lambda, \ \lambda \in \Lambda\}\} > 0$. Since $\emptyset \in \supp(\widetilde{\sigma}^2)$ and  $(\sigma^{1^\prime},\widetilde{\sigma}^2) \in \Sigma$, then for all $(i,j) \in \edges, \ 0 \leq U_2(\sigma^{1^\prime},\emptyset)  - U_2(\sigma^{1^\prime},\{(i,j)\}) \overset{\eqref{payoff2}}{=}  d_{ij}  - p_2 \flow^\prime_{ij}.$ Therefore, $\flow^\prime$ is a feasible solution of \dual. In addition,  $z^*_{\original} \overset{\eqref{common}}{=} \widetilde{U}_1(\sigma^{1^\prime},\emptyset) \overset{\eqref{transform1}}{=} \Value{\flow^\prime} - \frac{1}{p_1}\Costf{\flow^\prime}$\rem{, that is, the objective value of $\flow^\prime$ is equal to the optimal value of \primal}.
%
%
%
Thus, $\flow^\prime  \in\mathcal{O}^*_{\primal}$.


Secondly, we show that $\sigma^{2^\prime}$ necessarily satisfies \eqref{eq_strat} and \eqref{ineq_strat} for some optimal solution of \dual. For every $(i,j) \in \edges$, let $\rho^\prime_{ij} \coloneqq \sum_{\{\att \in \mathcal{I} \, | \, (i,j) \in \att\}} \sigma^{2^\prime}_I$. We can then derive the following inequalities:
\begin{align}
z^*_{\original} &\overset{\eqref{common}}{=}\max_{\flow \in \mathcal{F}} \widetilde{U}_1(\flow,\sigma^{2^\prime}) \overset{\eqref{transform1}}{=}   \sum_{(i,j) \in \edges} \frac{d_{ij}}{p_2}\rho_{ij}^\prime + \max_{\flow \in \mathcal{F}} \Big\{\sum_{\lambda \in \Lambda} \flow_\lambda (\pi_\lambda^0 - \sum_{\{\att \in \mathcal{I} \, | \,\att \cap \lambda \neq \emptyset\}} {\sigma}^{2^\prime}_{\att})\Big\} \nonumber\\
& = \sum_{(i,j) \in \edges} \frac{d_{ij}}{p_2}\rho_{ij}^\prime + \min_{\mu \in \revm{\mathbb{R}_{\geq0}^{\edges}}}\Big\{\sum_{(i,j) \in \edges} c_{ij}\mu_{ij} \ \Big| \  \forall \lambda \in \Lambda, \ \sum_{(i,j) \in \lambda} \mu_{ij} \geq \pi_\lambda^0 -\sum_{\{\att \in \mathcal{I} \, | \,\att \cap \lambda \neq \emptyset\}} {\sigma}^{2^\prime}_{\att}\Big\} \label{sd}\\
& \geq \sum_{(i,j) \in \edges} \frac{d_{ij}}{p_2}\rho_{ij}^\prime + \min_{\mu \in \revm{\mathbb{R}_{\geq0}^{\edges}}}\Big\{\sum_{(i,j) \in \edges} c_{ij}\mu_{ij} \ \Big| \ \forall \lambda \in \Lambda, \ \sum_{(i,j) \in \lambda} \mu_{ij} \geq \pi_\lambda^0 -\sum_{(i,j) \in \lambda} \rho_{ij}^\prime\Big\} \label{Bigger set}\\
& \geq \min_{\rho,\mu \in \revm{\mathbb{R}_{\geq0}^{\edges}}} \Big\{ \sum_{(i,j) \in \edges} (\frac{d_{ij}}{p_2}\rho_{ij} + c_{ij}\mu_{ij}) \ \Big| \  \forall \lambda \in \Lambda, \  \sum_{(i,j) \in \lambda} (\rho_{ij} + \mu_{ij}) \geq \pi_\lambda^0\Big\} = z^*_{\original}. \label{more variables}
\end{align}
Thus, inequalities \eqref{Bigger set}-\eqref{more variables} are tight. Note that \eqref{sd} is a consequence of the strong duality theorem\revm{, and \eqref{Bigger set} holds since $\sum_{\{\att \in \mathcal{I} \, | \,\att \cap \lambda \neq \emptyset\}} {\sigma}^{2^\prime}_{\att} \leq \sum_{(i,j) \in \lambda} \rho_{ij}^\prime$ for every $\lambda \in \Lambda$.} 

Let $\mu^\prime \coloneqq \argmin_{\mu \in \revm{\mathbb{R}_{\geq0}^{\edges}}}\big\{\sum_{(i,j) \in \edges} c_{ij}\mu_{ij} \ \big| \  \forall \lambda \in \Lambda, \ \sum_{(i,j) \in \lambda} \mu_{ij} \geq \pi_\lambda^0 -\sum_{\{\att \in \mathcal{I} \, | \,\att \cap \lambda \neq \emptyset\}} {\sigma}^{2^\prime}_{\att}\big\}$. Then, \eqref{Bigger set} and \eqref{more variables} imply that $(\rho^\prime,\mu^\prime)  \in\mathcal{O}^*_{\dual}$.
Furthermore, by definition of $\mu^\prime$, $\sum_{\{\att \in \mathcal{I} \, | \,\att \cap \lambda \neq \emptyset\}} {\sigma}^{2^\prime}_{\att} \geq \pi_\lambda^0 - \sum_{(i,j) \in \lambda} \mu_{ij}^\prime$ for every $\lambda \in \Lambda$.
Thus, $\sigma^{2^\prime}$ satisfies \eqref{eq_strat} and \eqref{ineq_strat} with $(\rho^\prime,\mu^\prime)  \in\mathcal{O}^*_{\dual}$. 
}
\hfill
\Halmos

\end{itemize}

\endproof

\rev{A direct consequence of Theorem~\ref{One_NE_general} is that some quantities related to the players' actions in equilibrium can be computed in closed form using the game parameters and the optimal primal and dual solutions of \original. They are summarized in the following corollary:} 

\rev{

\begin{corollary}\label{Corollary}
NE of the game $\Gamma$ satisfy the following properties:
\begin{enumerate}

\item[(i)] Expected amount (resp. cost) of initial flow sent by \defender is given by $\Value{\flow^*}$ (resp. $\Costf{\flow^*}$),


\item[(ii)] Expected cost of \attacker's interdiction strategy is given by $\sum_{(i,j) \in \edges} d_{ij} \rho^*_{ij}$,
 
\item[(iii)] Expected amount of interdicted flow is given by $\sum_{(i,j) \in \edges} \frac{d_{ij}}{p_2} \rho^*_{ij}$,

\item[(iv)] Expected amount of effective flow is given by $\Value{\flow^*} - \sum_{(i,j) \in \edges} \frac{d_{ij}}{p_2} \rho^*_{ij}$,

\end{enumerate}
where $\flow^* \in \mathcal{O}^*_{\primal}$ and $(\rho^*,\mu^*) \in \mathcal{O}^*_{\dual}$.
\end{corollary}

\rem{
\proof{Proof of Corollary~\ref{Corollary}.}
Let $(\sigma^{1^*},\sigma^{2^*})$ be a NE of the game $\Gamma$. From Theorem~\ref{One_NE_general}, there exist optimal solutions $\flow^*$ and $(\rho^*,\mu^*)$ of \primal and \dual respectively such that $(\sigma^{1^*},\sigma^{2^*})$ satisfies \eqref{Expected_Flow}-\eqref{ineq_strat}. Then, trivially, $\mathbb{E}_{\sigma^{1^*}}[\Value{\flow}] = \Value{\flow^*}$ and $\mathbb{E}_{\sigma^{1^*}}[\Costf{\flow}] = \Costf{\flow^*}$. Similarly, $\mathbb{E}_{\sigma^{2^*}}[\Cost{\att}] = \sum_{(i,j) \in \edges} d_{ij} \rho_{ij}^*$. Next, we obtain $\mathbb{E}_{\sigma^*}[\Value{\flow - \flow^\att}] = \frac{1}{p_2}U_2(\sigma^{1^*},\sigma^{2^*}) + \frac{1}{p_2}\mathbb{E}_{\sigma^{2^*}}[\Cost{\att}] = z^*_{\original} - \sum_{(i,j) \in \edges} c_{ij} \mu_{ij}^*$. Finally, we obtain that $\mathbb{E}_{\sigma^*}[\Value{\flow^\att}] = \Value{\flow^*}- z^*_{\original} + \sum_{(i,j) \in \edges} c_{ij} \mu_{ij}^* = \frac{1}{p_1}\Costf{\flow^*} + \sum_{(i,j) \in \edges} c_{ij} \mu_{ij}^*$.
\hfill \Halmos
\endproof}
}

\rev{Given $\flow^* \in \mathcal{O}^*_{\primal}$ and $(\rho^*,\mu^*) \in \mathcal{O}^*_{\dual}$, the expected amount of interdicted flow achievable by any interdiction strategy satisfying \eqref{eq_strat} is upper bounded by $\sum_{(i,j) \in \edges}
\frac{d_{ij}}{p_2} \rho^*_{ij}$. $(iii)$ in Corollary~\ref{Corollary} shows that this upper bound is achieved by \attacker's strategy in equilibrium. In other words, given the marginal edge interdiction probabilities $\rho^*$, \attacker randomizes its interdictions to maximize the amount of interdicted flow, while still limiting \defender's incentive to deviate from its strategy.}

\rev{Note that despite the exponential number of actions of both players, a NE can be computed in polynomial time. Indeed, we first solve the polynomial-size formulation of \original, and use the flow decomposition algorithm to obtain $\flow^* \in \mathcal{O}^*_{\primal}$ and $(\rho^*,\mu^*) \in \mathcal{O}^*_{\dual}$ (see Appendix~\ref{sec:MCCP}). Since $\pi^*$ is an affine function of the edges constituting each $s-t$ path, we run Algorithm~\ref{PolyALG} on the poset $P_{\mathcal{G}} = (\edges,\preceq_{\mathcal{G}})$ (Lemma~\ref{poset}) to compute an interdiction strategy $\widetilde{\sigma}^2 \in \Delta(\mathcal{I})$ satisfying \eqref{eq_strat} and \eqref{ineq_strat}. Given $H_{P_{\mathcal{G}}} = (\edges,E_{P_{\mathcal{G}}})$ the directed cover graph of $P_{\mathcal{G}}$, we deduce that $\widetilde{\sigma}^2$ can be obtained in $O(|\edges|(|\edges| + |E_{P_{\mathcal{G}}}|)^2)$ time. Since $\mathcal{G}$ is a simple directed acyclic graph, the degree of each $(i,j) \in \edges $ in $H_{P_{\mathcal{G}}}$ is at most $|\nodes|-2$, since $(i,j) \in \edges$ is adjacent to at most $|\nodes|-2$ edges $(i^\prime,j^\prime)$ in $\mathcal{G}$ such that $j=i^\prime$ or $j^\prime = i$. Therefore, the total number of edges in $H_{P_{\mathcal{G}}}$ is upper bounded by $|E_{P_{\mathcal{G}}}| \leq \frac{1}{2}|\edges|(|\nodes|-2)$. In conclusion, the NE $(\flow^*,\widetilde{\sigma}^2)$ is computed in $O(|\nodes|^2|\edges|^3)$ time. In this NE, \defender sends its flow along at most $|\edges|$ $s-t$ paths (from the flow decomposition theorem), and \attacker randomizes over at most $|\edges|+\frac{1}{2}|\edges|(|\nodes|-2) + 1$ interdictions (given the number of iterations of Algorithm~\ref{PolyALG}).}

We remark that in the simpler case where each $s-t$ path has an identical transportation cost, \original can be viewed as a maximum flow problem. \rev{Then, this solution} approach simply computes a NE of $\Gamma$ from a maximum flow for \defender, and a minimum-cut set for \attacker.

\rev{For the sake of completeness, we characterize the game instances for which pure NE exist. From Theorem~\ref{One_NE_general}, a pure NE exists if and only if there exists $(\rho^*,\mu^*) \in \mathcal{O}^*_{\dual}$ such that $\rho^* \in \{0,1\}^{\edges}$. Since \revm{$b_\lambda > 0$ for every $\lambda \in \Lambda$, then} $\rho^*_{ij} < 1$ for every $(i,j) \in \edges$ at optimality of \dual, \revm{and} a pure NE exists if and only if $\rho^*_{ij} = 0$ for every $(i,j) \in \edges$. The corresponding NE are $(\flow^*,\emptyset)$ with $\flow^* \in \mathcal{O}^*_{\primal}$. Note that this case occurs when the interdiction costs for \attacker or transportation costs for \defender are too high.}


\rem{
\begin{corollary}\label{pure_NE}
The game $\Gamma$ has a pure NE if and only if an optimal solution $(\rho^*,\mu^*)$ of \dual is such that $\rho_{ij}^* = 0$, for every $(i,j) \in \edges$. The corresponding NE is $(\flow^*,\emptyset)$ with $\flow^*  \in\mathcal{O}^*_{\primal}$.
\end{corollary}

\proof{Proof of Corollary~\ref{pure_NE}.}
Let $\flow^*\in \mathcal{O}^*_{\primal}$. If there exists $(\rho^*,\mu^*)\in \mathcal{O}^*_{\dual}$ such that $\rho^*_{ij} = 0$, for every $(i,j) \in \edges$, then $(\flow^*,\emptyset)$ satisfies \eqref{Expected_Flow}-\eqref{ineq_strat} and is a pure NE. Now, let $(\flow^*,\att^*)$ be a pure NE, and let $\rho^*_{ij} = \mathds{1}_{\{(i,j) \in \att^*\}} \in \{0,1\}$. From Theorem~\ref{One_NE_general}, we automatically deduce that $\flow^* \in \mathcal{O}^*_{\primal}$, and there exists $\mu^* \in \mathbb{R}^\edges_+$ such that $(\rho^*,\mu^*) \in \mathcal{O}^*_{\dual}$. By optimality of $(\rho^*,\mu^*)$, we know that $\rho^*_{ij} <1$ for every $(i,j) \in \edges$. Thus, $\rho^*_{ij} = 0$, for every $(i,j) \in \edges$, and $\att^* = \emptyset$.
\hfill \Halmos
\endproof

}

\rev{Finally}, we \rev{analyze} the set of $s-t$ paths (resp. set of edges) that are chosen (resp. interdicted) in at least one NE. \rev{
From Theorem~\ref{One_NE_general}, \rem{we know that} the set of $s-t$ paths chosen by \defender in at least one NE is given by $\bigcup_{\flow^* \in \mathcal{O}^*_{\primal}} \supp(\flow^*)$. Similarly, the set of edges interdicted by \attacker in at least one NE is given by $\bigcup_{(\rho^*,\mu^*) \in \mathcal{O}^*_{\dual}} \supp(\rho^*)$.
To efficiently compute these sets of critical components, we utilize the notion of \emph{strict complementary slackness}.} Specifically, optimal solutions $\flow^\dag$ and $(\rho^\dag$, $\mu^\dag)$ of \primal and \dual satisfy strict complementary slackness if:
\begin{align}
\forall (i,j) \in \edges,& \ \text{either } \rho_{ij}^\dag > 0 \ \text{ or } \ \flow_{ij}^\dag = \sum_{\{\lambda \in \Lambda \, | \, (i,j) \in \lambda\}}\flow_{\lambda}^\dag < \frac{d_{ij}}{p_2},\label{scs1_2}\\
\forall (i,j) \in \edges,& \ \text{either } \mu_{ij}^\dag > 0 \ \text{ or } \  \flow_{ij}^\dag = \sum_{\{\lambda \in \Lambda \, | \, (i,j) \in \lambda\}}\flow_{\lambda}^\dag < c_{ij},\label{scs3_2}\\
\forall \lambda \in \Lambda,& \ \text{either } \flow_{\lambda}^\dag > 0 \ \text{ or } \ \sum_{(i,j) \in \lambda}(\rho_{ij}^\dag + \mu_{ij}^\dag) > \pi^0_\lambda. \label{scs2_2}
\end{align}

We say that $\flow^\dag$ and $(\rho^\dag,\mu^\dag)$ form a \emph{strictly complementary primal-dual pair} of optimal solutions  of \original. \rev{Note that} such a pair is guaranteed to exist by the Goldman-Tucker theorem \cite{Goldman1957}. 
%
%
\rem{From Theorem~\ref{One_NE_general}, we already know that there exists a NE of $\Gamma$ where \defender's strategy is $\flow^\dag$ and \attacker's strategy is such that each edge $(i,j)$ is interdicted with probability $\rho^\dag_{ij}$. In fact, we can show that $\flow^\dag$ and $\rho^\dag$ characterize the $s-t$ paths and edges that are chosen by both players in equilibrium:}
\rev{We now show the following result:
}
%
\begin{proposition}\label{Nec_conds}
Let $\flow^\dag$ and $(\rho^\dag,\mu^\dag)$ be a strictly complementary primal-dual pair of optimal solutions of \original. The set of $s-t$ paths (resp. the set of edges) \rem{that are} chosen with positive probability by \defender's strategy (resp. \attacker's strategy) in at least one NE is given by $\supp(\flow^\dag)$ (resp. $\supp(\rho^\dag)$).
\end{proposition}

\proof{Proof of Proposition~\ref{Nec_conds}.} Let $\flow^\dag$ and $(\rho^\dag,\mu^\dag)$ be optimal solutions of \primal and \dual that satisfy strict complementary slackness. We denote $\widetilde{\sigma}^2 \in \Delta(\mathcal{I})$ the interdiction strategy, constructed from Algorithm~\ref{PolyALG}, \rev{that} interdicts every edge $(i,j) \in \edges$ with probability $\rho^\dag_{ij}$, and interdicts every $s-t$ path $\lambda \in \Lambda$ with probability at least $\pi^0_\lambda - \sum_{(i,j) \in \lambda}\mu^\dag_{ij}$. 
%
%
Given $\Sigma$ the set of NE of the game $\Gamma$, let $\mathcal{H}_1 \coloneqq \bigcup_{(\sigma^{1^*},\sigma^{2^*}) \in \Sigma} \bigcup_{\flow \in \supp(\sigma^{1^*})} \rev{\supp(\flow)}$ and $\mathcal{H}_2 \coloneqq \bigcup_{(\sigma^{1^*},\sigma^{2^*}) \in \Sigma} \bigcup_{\att \in \supp(\sigma^{2^*})}\att$.

\rem{To show the reversed inclusions, we exploit properties of zero-sum games: Recall that the game $\Gamma$ is strategically equivalent to the game $\widetilde{\Gamma} = \langle \{1,2\}, (\mathcal{F}, \mathcal{I}), (\widetilde{u}_1,-\widetilde{u}_1) \rangle$ where $\widetilde{u}_1$ is given by \eqref{transform1}. Therefore, each player's payoff in $\widetilde{\Gamma}$ is identical in any NE. 
We note the following equality:
\begin{align}
 \mathbb{E}_{\widetilde{\sigma}^2}[\Value{\flow^\dag} - \Value{\flow^{\dag\att}}] & \overset{\eqref{<=1}}{=} \sum_{\lambda \in \Lambda}\flow_{\lambda}^\dag\mathbb{E}_{\widetilde{\sigma}^2}[|\att \cap \lambda|] \overset{\eqref{eq_strat}}{=}\sum_{\lambda \in \Lambda}\flow_{\lambda}^\dag \sum_{(i,j) \in \lambda} \rho_{ij}^\dag  \overset{\eqref{scs3_2},\eqref{scs2_2}}{=}  z^*_{\original} - \sum_{(i,j) \in \edges} c_{ij}\mu_{ij}^\dag , \label{Eq_lost}
\end{align}
where $z^*_{\original}$ is the optimal value of \original.
This enables us to obtain \defender's equilibrium payoff in the zero-sum game $\widetilde{\Gamma}$: For every $(\sigma^{1^*},\sigma^{2^*}) \in \Sigma,$
\begin{align}
\widetilde{U}_1(\sigma^{1^*},\sigma^{2^*}) & =  \widetilde{U}_1(\flow^\dag,\widetilde{\sigma}^2) \overset{\eqref{transform1}}{=}  \mathbb{E}_{\widetilde{\sigma}^2}[\Value{\flow^{\dag\att}}] - \Value{\flow^\dag}+ \Value{\flow^\dag}- \frac{1}{p_1}\Costf{\flow^\dag} + \frac{1}{p_2}\sum_{\att \in \mathcal{I}}\widetilde{\sigma}^2_{\att}\sum_{(i,j) \in \att}d_{ij}\nonumber\\
& \overset{\eqref{eq_strat}}{=}  -\mathbb{E}_{\widetilde{\sigma}^2}[\Value{\flow^\dag} - \Value{\flow^{\dag\att}}] +z_{\original}^* + \frac{1}{p_2} \sum_{(i,j) \in \edges} d_{ij} \rho^\dag_{ij}  \overset{\eqref{Eq_lost}}{=}  z^*_{\original}. \label{common}
\end{align}}


From Theorem~\ref{One_NE_general}, we know that $(\flow^\dag,\widetilde{\sigma}^2) \rev{\in \Sigma}$\rem{ is a NE}. Consequently, $\mathcal{H}_1 \supseteq  \supp(\flow^\dag)$, and $\mathcal{H}_2 \supseteq \supp(\rho^\dag)$. To show the reverse inclusions, consider $(\flow^*,\sigma^{2^*}) \in \Sigma$. \rev{Theorem~\ref{One_NE_general} implies that there exists $(\rho^*,\mu^*) \in \mathcal{O}^*_{\dual}$ such that $\sum_{\{I \in \mathcal{I} \, | \, (i,j) \in I\}}\sigma^{2^*}_I = \rho^*_{ij}$. 
Consider $(i,j) \in \edges$ such that $\rho^*_{ij} > 0$. By complementary slackness between $(\rho^*,\mu^*)$ and $\flow^\dag$, $\frac{d_{ij}}{p_2} \overset{\eqref{cs1_2}}{=} \sum_{\{\lambda \in \Lambda \, | \, (i,j) \in \lambda \}} \flow^\dag_\lambda$. Then, from strict complementary slackness \eqref{scs1_2}, $\rho^\dag_{ij} >0$. Therefore, $\mathcal{H}_2 \subseteq \supp(\rho^\dag)$, which implies that $\mathcal{H}_2 = \supp(\rho^\dag)$.

}


\rev{Similarly, given $(\flow^*,\sigma^{2^*}) \in \Sigma$, Theorem~\ref{One_NE_general} implies that $\flow^* \in \mathcal{O}^*_{\primal}$. Then, by complementary slackness between $\flow^*$ and $(\rho^\dag,\mu^\dag)$, $\sum_{(i,j) \in \lambda} (\rho^\dag_{ij} + \mu^\dag_{ij}) \overset{\eqref{cs2_2} }{=} \pi_\lambda^0$ for every $\lambda \in \Lambda$ such that $\flow_\lambda^* > 0$. From \eqref{scs2_2}, $\flow^\dag_\lambda >0$. Therefore, $\mathcal{H}_1 \subseteq \supp(\flow^\dag)$, and we conclude that $\mathcal{H}_1 = \supp(\flow^\dag)$.}
\hfill \Halmos
\endproof

\rev{Thus, by computing a strictly complementary primal-dual pair of optimal solutions $\flow^\dag$ and $(\rho^\dag,\mu^\dag)$ of \original, Proposition~\ref{Nec_conds} shows that the set of critical $s-t$ paths of the network is given by $\supp(\flow^\dag)$, and the set of critical network edges is given by $\supp(\rho^\dag)$. Such a pair can be efficiently computed using any of the existing methods in the literature (see Balinski and Tucker \cite{doi:10.1137/1011060}, Adler et al. \cite{Adler1992}, Jansen et al. \cite{doi:10.1080/02331939408843952}).}
%
%
%

\begin{figure}[htbp]
\centering
\begin{tikzpicture}[auto,x=2.2cm, y=1.4cm,
  thick,main node/.style={circle,draw},flow_a/.style ={blue!100}]
\tikzstyle{edge} = [draw,thick,->]
\tikzstyle{cut} = [draw,very thick,-]
\tikzstyle{flow} = [draw,line width = 1pt,->,blue!100]
\small
	\node[main node] (s) at (1,0) {$s$};
	\node[main node] (1) at (2,0) {1};
	\node[main node] (2) at (1.5,-1) {2};
	\node[main node] (t) at (3,0) {$t$};

	\path[edge,->,>=stealth',shorten >=1pt,blue]
	(s) edge node[below= 0.15cm]{\tb{1},\textcolor{black}{2},\tr{1}} node[above = 0.15cm]{\tr{$\widetilde{\sigma}^2_{s1} = 0.1$}} (1)	
	(2) edge node[below right]{\tb{1},\textcolor{black}{2},\tr{2}}(1)
	(s) edge node[below left]{\tb{1},\textcolor{black}{2},\tr{2}}(2)
	(1) edge node[below= 0.15cm]{\tb{2},\textcolor{black}{3},\tr{2}} node[above=0.15cm]{\tr{$\widetilde{\sigma}^2_{1t} = 0.7$}} (t);

\node (15) at (1.5,0) {\large\Cross};
\node (15) at (2.5,0) {\large\Cross};

\normalsize
\end{tikzpicture}
\caption{NE when $p_1 = 10$, $p_2 =1$. $b_{ij} = 1$ \rev{for all} $(i,j) \in \edges$. The label of each edge $(i,j)$ represents $(\flow^\dag_{ij},c_{ij},d_{ij})$. Edge $(s,1)$ is interdicted by the equilibrium interdiction strategy $\widetilde{\sigma}^2$, but is not part of the  minimum-cut set.}
\label{not a min cut-set}
\end{figure}
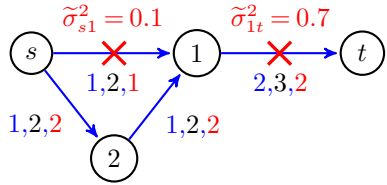

We note that in the setting that we consider, \attacker may need to interdict edges that are not part of any minimum-cut set, and can even belong to different cut sets; Figure~\ref{not a min cut-set} illustrates an example. In this example, the equilibrium interdiction strategy targets edges $(s,1)$ and $(1,t)$ that do not belong to a same cut set. Thus, Proposition~\ref{Nec_conds} generalizes the previously studied max-flow min-cut-based metrics of network criticality (see Assadi et al. \cite{Assadi:2014:MVP:2685231.2685250}, Dwivedi and Yu \cite{6062676}, Gueye et al. \cite{10.1007/978-3-642-35582-0_20}).

%
%
%

%
%
%



%
%
%





\rem{
Finally, we can derive additional equilibrium properties for the setting where each edge is potentially worth interdicting by \attacker, i.e., when $\frac{d_{ij}}{p_2} < c_{ij}, \ \forall (i,j) \in \edges$. Recall that $\frac{d_{ij}}{p_2}$ is the threshold on the flow $\flow_{ij}$ that determines \attacker's incentive to interdict edge $(i,j)$ or not.  If edge $(i,j)$ is such that $\frac{d_{ij}}{p_2} \geq c_{ij}$, then for any feasible flow $\flow \in \mathcal{F}$, $\flow_{ij} \leq \frac{d_{ij}}{p_2}$, and interdicting that edge does not increase \attacker's payoff. On the other hand, if $\frac{d_{ij}}{p_2} < c_{ij}$, then \attacker has an incentive to interdict $(i,j)$ if \defender routes more than $\frac{d_{ij}}{p_2}$ units of flow through that edge.  Next, we exploit the strategic equivalence to the zero-sum game $\widetilde{\Gamma}$, as well as  Theorems~\ref{thm:finally} and \ref{optimal value}, to derive additional results for this special case. 


\begin{proposition}\label{thm:mainresult}
If $\ \forall (i,j) \in \edges, \ \frac{d_{ij}}{p_2} < c_{ij}$, then any NE $\sigma^* = (\sigma^{1^*},\sigma^{2^*}) \in \Sigma$ satisfies the following properties:
\begin{enumerate}
\item[(i)] Both players' equilibrium payoffs are constant and given by: $U_1({\sigma^1}^*,\sigma^{2^*}) = U_2({\sigma^1}^*,\sigma^{2^*}) = 0$.

\item[(ii)] \defender's routing strategy satisfies: $\mathbb{E}_{\sigma^{1^*}}[p_1\Value{\flow} - \Costf{\flow}] =p_1z^*_{\original}$.

\item[(iii)] The expected cost of \attacker's interdiction strategy is given by: $\mathbb{E}_{\sigma^{2^*}}[\Cost{\att}]= p_2 z^*_{\original}$.
 
\item[(iv)] The expected amount of interdicted flow is given by: $\mathbb{E}_{\sigma^*}[\Value{\flow} - \Eff{\flow}{\att}]  =z^*_{\original}.$

\end{enumerate}
%
%
%
%
%
%
\end{proposition}

\proof{Proof of Proposition~\ref{thm:mainresult}.}

In \eqref{common}, we established that $\forall (\sigma^{1^*},\sigma^{2^*}) \in \Sigma$, $\widetilde{U}_1(\sigma^{1^*},\sigma^{2^*}) = z^*_{\original}$. Let $\flow^*$ and $(\rho^*,\mu^*)$ denote optimal solutions of \primal and \dual, respectively. Since $\forall (i,j) \in \edges$, $\frac{d_{ij}}{p_2} < c_{ij}$, then $\forall (i,j) \in \edges, \ \flow^*_{ij} \leq \frac{d_{ij}}{p_2} < c_{ij}$. Therefore,  from \eqref{cs3_2}, we deduce that $\forall (i,j) \in \edges, \ \mu^*_{ij} = 0$. Let $\widetilde{\sigma}^2 \in \Delta(\mathcal{I})$ denote the interdiction strategy constructed from Algorithm~\ref{ALG3} that satisfies~\eqref{eq_strat} and \eqref{ineq_strat}. We denote $\flow^0 \in \mathcal{F}$ the action of not sending any flow in the network, i.e., $\flow^0_\lambda = 0$, $\forall \lambda \in \Lambda$, and we denote $\flow^\prime \coloneqq (1+\epsilon) \flow^*$, with $\epsilon = \min\{p_2\frac{c_{ij}}{d_{ij}} - 1, \  (i,j) \in \edges\} > 0$. Then, $\flow^\prime \in \mathcal{F}$.

Let us consider $\widetilde{\sigma}^1 \in \Delta(\mathcal{F})$ defined by: $\widetilde{\sigma}^1_{\flow^\prime} = \frac{1}{1+\epsilon}$, and $\widetilde{\sigma}^1_{\flow^0} = \frac{\epsilon}{1+\epsilon}$. Then, we  show that $(\widetilde{\sigma}^1,\widetilde{\sigma}^2)$ is a NE.
Regarding \defender's payoff, since $\mu_{ij}^* = 0, \ \forall (i,j) \in \edges$, we can rewrite \eqref{4th_(in)eq_2} as follows:
\begin{align*}
\forall \flow \in \mathcal{F}, \ U_1(\flow,\widetilde{\sigma}^2) &\overset{\eqref{payoff1}}{=}  p_1 \sum_{\lambda \in \Lambda} \pi^0_\lambda \flow_\lambda - p_1 \sum_{\lambda \in \Lambda} \flow_\lambda \sum_{\{\att \in \mathcal{I} \, | \, \att \cap \lambda \neq \emptyset\}}\widetilde{\sigma}^2_\att  \overset{\eqref{ineq_strat}}{\leq} p_1  \sum_{\lambda \in \Lambda} \pi^0_\lambda \flow_\lambda - p_1 \sum_{\lambda \in \Lambda} \flow_\lambda \pi^0_\lambda  = 0.
\end{align*}

Trivially, we obtain that $U_1(\flow^0,\widetilde{\sigma}^2) =  0$. Furthermore, we know from \eqref{eq_prob_lambda} that $\forall \lambda \in \Lambda$ such that $\flow^*_\lambda > 0, \ \sum_{\{\att \in \mathcal{I} \, | \, \att \cap \lambda \neq \emptyset\}}\widetilde{\sigma}^2_\att = \pi^0_\lambda$. Since $\flow^*_\lambda > 0 \Longleftrightarrow \flow^\prime_\lambda >0$, we deduce that $U_1(\flow^\prime,\widetilde{\sigma}^2) = 0$. 
Therefore $U_1(\widetilde{\sigma}^1,\widetilde{\sigma}^2) = 0$.

Regarding \attacker's payoff, we know that $\forall \sigma^2 \in \Delta(\mathcal{I}), \ U_2(\widetilde{\sigma}^1,\sigma^2) = U_2(\mathbb{E}_{\widetilde{\sigma}^1}[\flow],\sigma^2) = U_2(\flow^*,\sigma^2)$. 
Therefore, $U_2(\widetilde{\sigma}^1,\widetilde{\sigma}^2) = U_2(\flow^*,\widetilde{\sigma}^2)  \geq U_2(\flow^*,\sigma^2)  = U_2(\widetilde{\sigma}^1,\sigma^2), \ \forall \sigma^2 \in \Delta(\mathcal{I})$.
 Thus, $(\widetilde{\sigma}^1,\widetilde{\sigma}^2)$ is a NE.

We now consider $(\sigma^{1^*},\sigma^{2^*}) \in \Sigma$. Then, we know that $(\sigma^{1^*},\widetilde{\sigma}^2) \in \Sigma$ and $(\widetilde{\sigma}^1,\sigma^{2^*}) \in \Sigma$.
Since $\flow^0 \in \supp(\widetilde{\sigma}^1)$, we obtain that $p_2\widetilde{U}_1(\flow^0,\sigma^{2^*}) \overset{\eqref{transform1}}{=}  \mathbb{E}_{\sigma^{2^*}}[\Cost{\att}] \overset{\eqref{common}}{=} p_2z^*_{\original}$. 
Similarly, since $\emptyset \in \supp(\widetilde{\sigma}^2)$, then $p_1\widetilde{U}_1(\sigma^{1^*},\emptyset)  \overset{\eqref{transform1}}{=} \mathbb{E}_{\sigma^{1^*}}[p_1\Value{\flow} -\Costf{\flow}] \overset{\eqref{common}}{=} p_1z^*_{\original}$.
We deduce the players' equilibrium payoffs:
\begin{align*}
& U_1(\sigma^{1^*},\sigma^{2^*}) \overset{\eqref{transform1}}{=} p_1 \widetilde{U}_1(\sigma^{1^*},\sigma^{2^*}) - \frac{p_1}{p_2} \mathbb{E}_{\sigma^{2^*}}[\Cost{\att}] \overset{\eqref{common}}{=} p_1 z^*_{\original} -  p_1z^*_{\original} = 0,\\
&U_2(\sigma^{1^*},\sigma^{2^*}) \overset{\eqref{transform2}}{=} p_2 (-\widetilde{U}_1(\sigma^{1^*},\sigma^{2^*})) + p_2 \mathbb{E}_{\sigma^{2^*}}[\Value{\flow} - \frac{1}{p_1}\Costf{\flow}]  \overset{\eqref{common}}{=} -p_2 z^*_{\original} + p_2z^*_{\original} = 0.
\end{align*}

Finally, we characterize the expected amount of flow that is interdicted in any equilibrium: $\mathbb{E}_{\sigma^{*}}[\Value{\flow} - \Eff{\flow}{\att}] = \frac{1}{p_2} U_2(\sigma^{1^*},\sigma^{2^*}) + \frac{1}{p_2} \mathbb{E}_{\sigma^{2^*}}[\Cost{\att}] = z^*_{\original}.$
\hfill \Halmos
\endproof}

\rev{In summary}, our results in Section~\ref{sec:games} provide a new approach to \rev{solve} the \rev{strategic} \rev{interdiction} game $\Gamma$, and derive equilibrium properties for settings involving \rev{multiple interdictions,} heterogeneous cost parameters, and general network topology.

%
%
%


\section{Concluding remarks.}\label{sec:conclusion}

In this article, we studied an existence problem of probability distributions over partially ordered sets, and showed its \rev{applications} to a class of \rev{interdiction} games on flow networks. In the existence problem, we considered a poset, where each element and each maximal chain is associated with a value. Under two relevant conditions on these values, we showed that there exists a probability distribution over the subsets of this poset, with the following properties: the probability that each element (resp. maximal chain) is contained in a subset  (resp. intersects with a subset) is equal to (resp. as large as) the corresponding value. We provided a constructive proof of this result by designing a combinatorial algorithm that exploits structural properties of the problem. \rev{In the special case where the maximal chain values depend affinely on their constituting elements, we refined our algorithm to compute a probability distribution that satisfies the desired properties in polynomial time.}

By applying this existence result, we \rev{solved} a generic formulation of \rev{strategic} network \rev{interdiction} game between a routing entity and an interdictor. To overcome the computational and analytical challenges of the formulation, we proposed a new approach for \rev{characterizing} equilibria of the game. This approach relies on our existence result on posets, as well as optimal primal and dual solutions of a minimum-cost circulation problem.  \rev{In addition, we showed that Nash equilibria of the game can be efficiently computed with our refined algorithm on posets. Finally, we demonstrated that the critical network components that are chosen in equilibrium by both players can be computed from a strictly complementary primal-dual pair of optimal solutions of the circulation problem.}


%
%
%




%
%
%
 \begin{APPENDICES}

\ifarXiv
\OneAndAHalfSpacedXI

\fi


\section{Remaining proofs.}\label{sec:additional}

\proof{Proof of Lemma~\ref{Minimal Elements}.}

Let $P$ be a finite nonempty poset, and let $S$ be the set of minimal elements of $P$. If $|S| =1$, then $S$ is an antichain of $P$. Now, assume that $|S| \geq 2$, and consider \revm{$(x,y) \in S^2$ with $x \neq y$}. Since $x$ (resp. $y$) is a minimal element of $P$, then $y \nprec x$ (resp. $x \nprec y$). Therefore, $x$ and $y$ are incomparable, and $S$ is an antichain of $P$.

Now, consider a maximal chain $C \in \Chains$, and assume that $C$ does not contain any minimal element of $P$. Let $x$ be the minimal element of $(C,\preceq_{\restrict{C}})$. Since $x$ is not a minimal element of $P$, there exists $y \in \ground \backslash C$ such that $y \prec x$. By transitivity of $\preceq$, \rem{we deduce that }$y\prec x^\prime$ \rev{for every} $x^\prime \in C$. Therefore, $C\cup \{y\}$ is a chain containing $C$, which contradicts the maximality of $C$. Thus, every maximal chain of $P$ intersects with the set of minimal elements of $P$.
\hfill
\Halmos
\endproof

\proof{Proof of Lemma~\ref{new poset general}.}

Consider $\ground^\prime \subseteq \ground$, and $\Chains^\prime  \subseteq \Chains$ that preserves the decomposition of maximal chains intersecting in $\ground^\prime$. Let us show that $\preceq_{\Chains^\prime}$ defined in Section~\ref{sec:order_theory} is a partial order on $X^\prime$:
\begin{itemize}
\item[--] Reflexivity: For every $x \in \ground^\prime$, $x \preceq_{\Chains^\prime}x$ by definition.
\item[--] Antisymmetry:  Consider $(x,y) \in (\ground^\prime)^2$ such that $x \preceq_{\Chains^\prime} y$ and $y \preceq_{\Chains^\prime} x$. If $x \neq y$, then we would have $x \prec y$ and $y \prec x$, which contradicts $\preceq$ being a partial order. Therefore, $x=y$.

\item[--] Transitivity: Consider $(x,y,z) \in (\ground^\prime)^3$, and assume that $x \preceq_{\Chains^\prime} y$ and $y \preceq_{\Chains^\prime} z$. If $x = y$ or $y=z$, then we trivially obtain that $x \preceq_{\Chains^\prime} z$. Now, let us assume that $x \neq y$ and $y \neq z$. By definition of $\preceq_{\Chains^\prime}$, \rev{there exists} $C^1 \in\Chains^\prime$ \rev{such that} $(x,y) \in (C^1)^2$ and $x \prec y$. Similarly, \rev{there exists} $C^2 \in \Chains^\prime$ \rev{such that} $(y,z) \in (C^2)^2$ and $y \prec z$.  
We can rewrite $C^1$ and $C^2$ as follows: $C^1 = \{x_0,\dots,x_l=x,x_{l+1},\dots,x_{l+m}=y,x_{l+m+1},\dots,x_{l+m+n}\}$ and $C^2 = \{y_0,\dots,y_q=y,y_{q+1},\dots,y_{q+r}=z,y_{q+r+1},\dots,y_{q+r+s}\}$. Now, consider the maximal chain $C_1^2 = \{x_0,\dots,x_l=x,x_{l+1},\dots,x_{l+m}=y,y_{q+1},\dots,y_{q+r}=z,y_{q+r+1},\dots,y_{q+r+s}\}$, as illustrated in Figure~\ref{Proof1}.

%
%
%

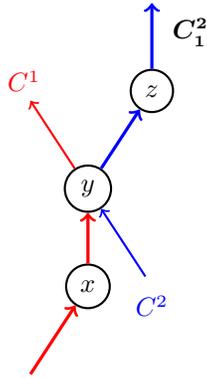
\begin{figure}[htbp]
    \centering
  \begin{tikzpicture}[auto,x=1.7cm, y=1.3cm,
  thick,main node/.style={circle,draw},flow_a/.style ={blue!100}]
  \small
  
\tikzstyle{edge} = [draw,thick,->]
\tikzstyle{cut} = [draw,very thick,-]
\tikzstyle{flow} = [draw,line width = 1pt,->,blue!100]
	\node[main node] (1) at (0,0) {$x$};
	\node[main node] (2) at (0,1) {$y$};
	\node[main node] (3) at (0.5,2) {$z$};
	\node (4) at (-0.5,-1) {};
	\node (5) at (-0.5,2) {};
	\node (6) at (0.5,0) {};
	\node (7) at (0.5,3) {};
	
%
	
%
	
	\path[edge]
	(1) edge[red,very thick] (2)
	(2) edge[blue,very thick] (3)
	(4) edge[red,very thick] (1)
	(2) edge[red] (5)
	(6) edge[blue] (2)
	(3) edge[blue,very thick] (7);

%
\node (11) at (-0.5,2.1) {\color{red} $C^1$};
\node (12) at (0.5,-0.2) {\color{blue} $C^2$};
\node (13) at (0.8,2.6) {\color{black}\textbf{$\boldsymbol{C_1^2}$}};

\normalsize
\end{tikzpicture}
    \caption{Illustration of the transitivity of $\preceq_{\Chains^\prime}$. $C_1^2$ is represented by the thick chain.}
    \label{Proof1}
\end{figure}

%
%
%

Since $C^1$ and $C^2$ intersect in $y \in \ground^\prime$, and $\Chains^\prime$ preserves the decomposition of maximal chains intersecting in $\ground^\prime$, we deduce that $C_1^2 \in \Chains^\prime$ as well. Furthermore, $(x,z) \in (C_1^2)^2$, and the transitivity of $\preceq$ implies that $x \prec z$. Therefore, $x \preceq_{\Chains^\prime} z$.

\end{itemize}

Thus, $\preceq_{\Chains^\prime}$ is a partial order on $\ground^\prime$, and $P^\prime = (\ground^\prime,\preceq_{\Chains^\prime})$ is a poset.

Let $C \subseteq \ground^\prime$ be a maximal chain of $P^\prime$ of size at least two. Let us rewrite $C = \{x_1,\dots,x_n\}$ with $n \geq 2$, where \rev{for all} $k \in \llbracket 1,n-1\rrbracket, \ x_k \prec:_{\Chains^\prime} x_{k+1}$. We show by induction on $k \in \llbracket 2,n \rrbracket$ that \rev{there exists} $C^\prime \in \Chains^\prime$ such that $\{x_1,\dots,x_k\} \subseteq C^\prime$. If $k=2$, then by definition, \rev{there exists} $C^\prime \in \Chains^\prime$ such that $\{x_1,x_2\} \subseteq \Chains^\prime$.  Now, assume that the result \rev{holds} for $k \in \llbracket 2,n-1\rrbracket$. Consider $C^1 \in \Chains^\prime$ such that $\{x_1,\dots,x_k\} \subseteq C^1$. Since $x_k \prec_{\Chains^\prime} x_{k+1}$, then  \rev{there exists} $C^2 \subseteq \Chains^\prime$ such that $(x_k,x_{k+1}) \in (C^2)^2$. Analogously, we can show that $C_1^2$ (illustrated in Figure~\ref{Proof1}), which is in $\Chains^\prime$, contains $\{x_1,\dots,x_{k+1}\}$.
%
Therefore, by induction, \rem{we obtain that }\rev{there exists} $C^\prime \in \Chains^\prime$ such that $C = \{x_1,\dots,x_n\} \subseteq C^\prime$. Since $C \subseteq \ground^\prime$, then we have  $C = C \cap \ground^\prime \subseteq C^\prime \cap \ground^\prime$.

Now, assume that \rev{there exists} $x^\prime \in C^\prime \cap \ground^\prime \backslash C$. For every $ k \in \llbracket 1,n\rrbracket$, $(x_k,x^\prime) \in (C^\prime)^2$. Therefore, $x^\prime$ is comparable in $P^\prime$ with every element of the chain $C$. This implies that $C \cup\{x^\prime\}$ is a chain in $P^\prime$, which contradicts the maximality of $C$ in $P^\prime$. Therefore, $C = C^\prime \cap \ground^\prime$.
\hfill \Halmos
\endproof

\proof{Proof of Lemma~\ref{poset}.} Let us show that $\preceq_{\mathcal{G}}$ is a partial order on $\edges$.
\begin{itemize}
\item[--] Reflexivity: For every $u \in \edges$, $u \preceq_{\mathcal{G}} u$ by definition.
\item[--] Antisymmetry:  Consider $(u,v) \in \edges^2$ such that $u \preceq_{\mathcal{G}} v$ and $v \preceq_{\mathcal{G}} u$. If $u \neq v$, then there exist $\lambda^1$ and $\lambda^2$ in $\Lambda$ such that $\lambda^1$ traverses $u$ and $v$ in this order, and $\lambda^2$ traverses $v$ and $u$ in this order. They can be written as follows: $\lambda^1 = \{u_1,\dots,u_n,u,u_{n+1},\dots,u_{n+m},v,u_{n+m+1},\dots,u_{n+m+p}\}$ and $\lambda^2 = \{v_1,\dots,v_q,v,v_{q+1},\dots,v_{q+r},u,v_{q+r+1},\dots,v_{q+r+s}\}$. Then, $\{u,u_{n+1},\dots,u_{n+m},v,v_{q+1},\dots,v_{q+r}\}$ is a cycle (see Figure~\ref{Antisymmetry}), which contradicts $\mathcal{G}$ being acyclic. Therefore $u=v$.

%
%
%

\begin{figure}[htbp]
    \centering
   \begin{tikzpicture}[->,>=stealth',shorten >=1pt,auto,x=2.0cm, y=1.1cm,
  thick,main node/.style={circle,draw},flow_a/.style ={blue!100}]
\tikzstyle{edge} = [draw,thick,->]
\tikzstyle{cut} = [draw,very thick,-]
\tikzstyle{flow} = [draw,line width = 1pt,->,blue!100]
\small
	\node[main node] (s) at (0,1) {s};
	\node[main node] (1) at (1,2) {1};
	\node[main node] (2) at (2,2) {2};
	\node[main node] (3) at (1,0) {3};
	\node[main node] (4) at (2,0) {4};
	\node[main node] (t) at (3,1) {t};

	\path[edge, violet, very thick]
	(1) edge node{$u$}(2)
	(3) edge node {$v$} (4);
	
	\path[edge, blue]
	(s) edge node{$\lambda^1$}  (1)	
	(2) edge[very thick] (3)	
	(4) edge (t);
	
	\path[edge, red]
	(s) edge node[below left]{$\lambda^2$} (3)
	(4) edge[very thick] (1)
	(2) edge (t);

\normalsize
\end{tikzpicture}

    \caption{Proof of antisymmetry of $\preceq_{\mathcal{G}}$ by contradiction: if $u \preceq_\mathcal{G} v$, $v \preceq_\mathcal{G} u$, and $u \neq v$, then one can see that $u$ and $v$ necessarily belong to a cycle (shown in thick edges), although $\mathcal{G}$ is acyclic.}
    \label{Antisymmetry}
\end{figure}
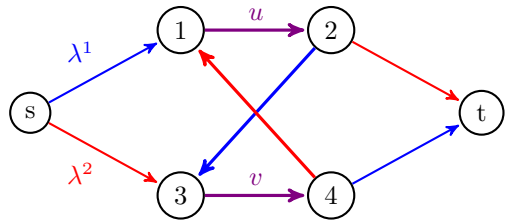

%
%
%


\item[--] Transitivity: Consider $(u,v,w) \in \edges^3$, and assume that $u \preceq_{\mathcal{G}} v$ and $v \preceq_{\mathcal{G}} w$. If $u = v$ or $v=w$, then we trivially obtain that $u \preceq_{\mathcal{G}} w$. Now, let us assume that $u \neq v$ and $v \neq w$. Then, there exist $\lambda^1$ and $\lambda^2$ in $\Lambda$ such that $\lambda^1$ traverses $u$ and $v$ in this order, and $\lambda^2$ traverses $v$ and $w$ in this order. They can be written as $\lambda^1 = \{u_1,\dots,u_n,u,u_{n+1},\dots,u_{n+m},v,u_{n+m+1},\dots,u_{n+m+p}\}$ and $\lambda^2 = \{v_1,\dots,v_q,v,v_{q+1},\dots,v_{q+r},w,v_{q+r+1},\dots,v_{q+r+s}\}$. Then, $\lambda_1^2 = \{u_1,\dots,u_n,u,u_{n+1},\dots,u_{n+m},v,v_{q+1},\dots,v_{q+r},w,v_{q+r+1},\dots,v_{q+r+s}\}$ is an $s-t$ path (see Figure~\ref{Transitivity paths}), and traverses $u$ and $w$ in this order. Therefore, $u \preceq_\mathcal{G} w$.

%
%
%

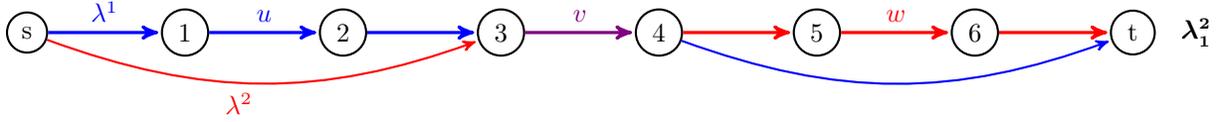
\begin{figure}[htbp]
    \centering
        \begin{tikzpicture}[->,>=stealth',shorten >=1pt,auto,x=2.1cm, y=1.3cm,
  thick,main node/.style={circle,draw},flow_a/.style ={blue!100}]
\tikzstyle{edge} = [draw,thick,->]
\tikzstyle{cut} = [draw,very thick,-]
\tikzstyle{flow} = [draw,line width = 1pt,->,blue!100]
\small
	\node[main node] (s) at (0,2) {s};
	\node[main node] (1) at (1,2) {1};
	\node[main node] (2) at (2,2) {2};
	\node[main node] (3) at (3,2) {3};
	\node[main node] (4) at (4,2) {4};
	\node[main node] (5) at (5,2) {5};
	\node[main node] (6) at (6,2) {6};
	\node[main node] (t) at (7,2) {t};

	\path[edge, very thick]
	(3) edge[violet] node {$v$} (4);
	
	\path[edge, blue]
	(1) edge[very thick] node{$u$}(2)
	(s) edge[very thick] node{$\lambda^1$}  (1)	
	(2) edge[very thick] (3)	
	(4) edge[bend right=20] (t);
	
	\path[edge, red]
	(s) edge[bend right=20] node[below left]{$\lambda^2$} (3)
	(4) edge[very thick] (5)
	(5) edge[very thick] node {$w$} (6)
	(6) edge[very thick] (t);

\node at (7.4,2) {$\boldsymbol{\lambda_1^2}$};

\normalsize
\end{tikzpicture}
   \caption{Proof of transitivity of $\preceq_\mathcal{G}$: if $u \preceq_\mathcal{G} v$, and $v \preceq_\mathcal{G} w$, then one can construct an $s-t$ path $\lambda_1^2$ (in thick line) that traverses $u$ and $w$ in this order.}
    \label{Transitivity paths}
\end{figure}

%
%
%

\end{itemize}

In conclusion, $P_{\mathcal{G}} = (\edges,\preceq_\mathcal{G})$ is a poset. 

Next, we prove that the set of maximal chains $\Chains $ of $P_{\mathcal{G}}$ is $\Lambda$. \rem{First, we show that $\Chains \subseteq \Lambda$. }Consider a maximal chain $C \in \Chains$ of $P_{\mathcal{G}}$. If $C = \{u\}$ is of size 1, then necessarily $u = (s,t)$, because $\mathcal{G}$ is connected. Therefore, $C = \{u\}$ is an $s-t$ path. Now, assume that $|C| \geq2$. Let us write $C = \{u_1,\dots,u_n\}$, where \rev{for every} $k \in \llbracket 1,n-1\rrbracket, \ u_k \prec:_{\mathcal{G}} u_{k+1}$. Since $u_1 \prec_\mathcal{G} u_2$ and $u_2 \prec_\mathcal{G} u_3$, then there exist $\lambda^1$ and $\lambda^2$ in $\Lambda$ such that $\lambda^1$ traverses $u_1$ and $u_2$ in this order, and $\lambda^2$ traverses $u_2$ and $u_3$ in this order. When \rev{proving} the transitivity of $\preceq_\mathcal{G}$ in the proof of Lemma~\ref{poset}, we \rev{showed} that there exists $\lambda_1^2 \in \Lambda$ that traverses $u_1$, $u_2$, and $u_3$ in this order. \rev{By repeating} this process, we obtain an $s-t$ path $\lambda \in \Lambda$ such that $C \subseteq \lambda$.

Now, assume that \rev{there exists} $u \in \lambda \backslash C$. Since $C \subseteq \lambda$, and $u \in \lambda$, then  $u$ is comparable with every element of $C$ (by definition of $\preceq_\mathcal{G}$). Therefore $C \cup \{u\}$ is a chain in $P_{\mathcal{G}}$, which contradicts the maximality of $C$. Therefore $C = \lambda$ and $\mathcal{C} \subseteq \Lambda$.

To show the reverse inclusion, consider an $s-t$ path $\lambda \in \Lambda$. \rev{By} definition of $\preceq_\mathcal{G}$, $\lambda$ is a chain in $P_{\mathcal{G}}$. Let us assume that \rem{$\lambda$ is not a maximal chain of $P_{\mathcal{G}}$, i.e., }there exists a maximal chain $C \in \Chains$ such that $\lambda \subsetneq C$. Let us write $\lambda = \{u_1,\dots,u_n\}$ where  \rev{for every} $k \in \llbracket 1,n - 1\rrbracket$, $u_k \prec_{\mathcal{G}} u_{k+1}$, and let $v \in C \backslash \lambda$.  Since $\lambda \subset C$ and $v \in C$, then $v$ is comparable with every element of $\lambda$. By transitivity of $\preceq_\mathcal{G}$, if \rev{there exists} $k \in \llbracket 1,n\rrbracket$ such that $v \prec_{\mathcal{G}} u_k$, then \rev{for every} $l \in \llbracket k,n \rrbracket$, $v \prec_{\mathcal{G}} u_l$. Similarly, if \rev{there exists} $k \in \llbracket 1,n\rrbracket$ such that $u_k \prec_{\mathcal{G}} v$, then \rev{for every} $l \in \llbracket 1,k \rrbracket$, $u_l \prec_{\mathcal{G}} v$. Therefore, three cases can arise:
\begin{itemize}
\item[--] $v \prec_{\mathcal{G}} u_1$. In this case, \rev{there exists} $\lambda^1 = \{w_1,\dots,w_n,v,w_{n+1},\dots,w_{n+m},u_{1},w_{n+m+1},\dots,w_{n+m+p}\}\in \Lambda$. However, since $\lambda$ is an $s-t$ path, then the start node of $u_1$ is $s$, which is also the start node of $w_1$. Therefore, $\{w_1,\dots,w_n,v,w_{n+1},\dots, w_{n+m}\}$ is a cycle, which is a contradiction.
\item[--] $u_n  \prec_{\mathcal{G}} v$. In this case, \rev{there exists} $\lambda^1 = \{v_1,\dots,v_q,u_n,v_{q+1},\dots,v_{q+r},v,v_{q+r+1},\dots,v_{q+r+s}\} \in \Lambda$. Analogously, \rem{we deduce that the end nodes of $u_n$ and $v_{q+r+s}$ are the destination node $t$, which implies that }$\{v_{q+1},\dots,v_{q+r},v,v_{q+r+1},\dots,v_{q+r+s}\}$ is a cycle in the acyclic graph $\mathcal{G}$.
\item[--] $u_k \prec_\mathcal{G} v \prec_\mathcal{G} u_{k+1}$ for some $k \in \llbracket 1,n-1 \rrbracket$. In this case, there exist  two $s-t$ paths $\lambda^1 = \{v_1,\dots,v_q,u_k,v_{q+1},\dots,v_{q+r},v,v_{q+r+1},\dots,v_{q+r+s}\} \in \Lambda$ and $\lambda^2 = \{w_1,\dots,w_n,v,w_{n+1},\dots,w_{n+m},u_{k+1},w_{n+m+1},\dots,w_{n+m+p}\} \in \Lambda $. One can verify that $\{v_{q+1},\dots,v_{q+r},v,w_{n+1},\dots,w_{n+m}\}$ is a cycle in $\mathcal{G}$.
 \rem{since the start node of $v_{q+1}$ is the end node of $w_{n+m}$. This is in fact the end node of $u_k$, which is also the start node of $u_{k+1}$ since $\lambda$ is a path.} This contradicts $\mathcal{G}$ being acyclic.
\end{itemize}

Thus, $\lambda = C$, and $\Lambda \subseteq \mathcal{C}$. In conclusion, $\mathcal{C} = \Lambda$.
\hfill
\Halmos
\endproof
 

\newpage
\section{Illustration of Algorithm~\ref{ALG3}.}\label{app:example}
 Consider the poset $P$ represented by the Hasse diagram given in Figure~\ref{ExPoset}. 
 
 \ifarXiv
 \fi
 
\begin{figure}[htbp]
    \centering
    \begin{tikzpicture}[->,>=stealth',shorten >=0pt,auto,x=1.8cm, y=2.0cm,
  thick,main node/.style={circle,draw},flow_a/.style ={blue!100}]
\tikzstyle{edge} = [draw,thick,->]
\tikzstyle{cut} = [draw,very thick,-]
\tikzstyle{flow} = [draw,line width = 1pt,->,blue!100]
\small
	\node[main node] (1) at (0,0) {1};
	\node[main node] (2) at (1,0) {2};
	\node[main node] (3) at (0.5,0.5) {3};
	\node[main node] (4) at (0,1) {4};
	\node[main node] (5) at (1,1) {5};

%
	
	\path[edge]
	(1) edge (3)
	(2) edge  (3)
	(3) edge  (4)
	(3) edge (5);

%

\normalsize
\end{tikzpicture}
    \caption{Hasse diagram of a poset $P$.}
    \label{ExPoset}
\end{figure}
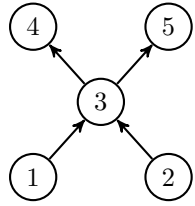

 
In this poset $P$, the set of maximal chains is given by $\Chains = \{\{1,3,4\},\{2,3,5\},\{1,3,5\},\{2,3,4\}\}$. We assume that the values assigned to each maximal chain are $\pi_{134} = \pi_{135} = 0.8$ and $\pi_{234} = \pi_{235} = 0.6$, and the values assigned to each element are $\rho_1 = 0.4$, $\rho_2 = 0.3$, $\rho_3 = 0.5$, $\rho_4 = 0.5$, $\rho_5 = 0.7$.

First, we can see that \rev{for all} $C \in \Chains, \ \sum_{x \in C} \rho_x \geq \pi_C$, and $\pi_{134} + \pi_{235}= \pi_{135} +\pi_{234}$. Therefore, conditions \eqref{Nec Cond} and \eqref{Conservation} are satisfied, and we can run Algorithm~\ref{ALG3} to optimally solve \OPOC (and construct a feasible solution of \FPOC). Figure~\ref{Steps Algo_bis} (resp. Figure~\ref{Steps Algo}), illustrates each iteration of the algorithm using the poset $P$ (resp. the posets $P^k$, for $k \in \llbracket 1,\stepmax\rrbracket$).
%
\begin{itemize}
\item $\boldsymbol{k=1:}$ $\component{1} = \ground = \llbracket 1,5\rrbracket$, $\allpaths{1} = \Chains$, $\edgeprob{x}{1} = \rho_x$ \rev{for all} $x \in \ground$. Note that $\delta_{134} = 0.6,\  \delta_{235} = 0.9, \ \delta_{135} = 0.8,$ and $\delta_{234} = 0.7$.
Since \rev{for all} $C \in \Chains, \ \combiprob{C}{1} = \delta_C >0$, then $\pathtight{1} =\emptyset$, and $\pathloose{1} = \Chains$. 
Therefore, each pair of elements in $P^1 = (\component{1},\preceq_{\pathtight{1}})$ is incomparable, and $\setmin{1} = \{1,2,3,4,5\}$. Then one can check that $\min_{x \in \setmin{1}} \edgeprob{x}{1} = 0.3$ and $\min_{\{C \in \pathloose{1} \, | \, |\setmin{1} \cap C| \geq 2\}} \frac{\combiprob{C}{1}}{|\setmin{1} \cap C|-1} = 0.3$. Therefore, $\sigma_{\setmin{1}} = \weight{1} = 0.3 = \edgeprob{2}{1} =  \frac{\combiprob{134}{1}}{|\setmin{1}\cap \{1,3,4\}| -1}$.

Next, the values are updated as follows: $\edgeprob{1}{2} = 0.1, \ \edgeprob{2}{2} = 0, \ \edgeprob{3}{2} = 0.2, \ \edgeprob{4}{2} = 0.2, \ \edgeprob{5}{2} = 0.4$, and $\combiprob{134}{2} = 0, \ \combiprob{235}{2} = 0.3, \ \combiprob{135}{2} = 0.2, \ \combiprob{234}{2} = 0.1$. Since each maximal chain's minimal element is in $\setmin{1}$, then $\allpaths{2} = \Chains$.
We conclude the first iteration of the algorithm by letting $\component{2} = \{1,3,4,5\}$, $\pathtight{2} = \{\{1,3,4\}\}$, and $\pathloose{2} = \{\{2,3,5\},\{1,3,5\},\{2,3,4\}\}$. 

\item $\boldsymbol{k=2:}$ The set of minimal elements of the new poset $P^2 = (\component{2},\preceq_{\pathtight{2}})$ is given by $\setmin{2} = \{1,5\}$\rem{ (see Figure~\ref{Steps Algo})}. Furthermore, $\min_{x \in \setmin{2}} \edgeprob{x}{2} = 0.1$ and $\min_{\{C \in \pathloose{2} \, | \, |\setmin{2} \cap C| \geq 2\}} \frac{\combiprob{C}{2}}{|\setmin{2} \cap C|-1} = 0.2$, which imply that $\sigma_{\setmin{2}} = \weight{2} = 0.1 = \edgeprob{1}{2}$.
Then, the values are updated as follows: $\edgeprob{1}{3} = 0, \ \edgeprob{2}{3} = 0, \ \edgeprob{3}{3} = 0.2, \ \edgeprob{4}{3} = 0.2, \ \edgeprob{5}{3} = 0.3$, and $\combiprob{134}{3} = 0, \ \combiprob{235}{3} = 0.3, \ \combiprob{135}{3} = 0.1, \ \combiprob{234}{3} = 0.1$.

Now, one can see that the minimal element of $\{2,3,5\} \cap \component{2}$ and $\{2,3,4\} \cap \component{2}$ in $P$ is $3\notin S^2$. Therefore, $\allpaths{3} = \{\{1,3,4\},\{1,3,5\}\}$,  $\component{3} = \{3,4,5\}$, $\pathtight{3} = \{\{1,3,4\}\}$, and $\pathloose{3} = \{\{1,3,5\}\}$. 

\item $\boldsymbol{k=3:}$ The set of minimal elements of $P^3  = (\component{3},\preceq_{\pathtight{3}})$ is given by $\setmin{3} = \{3,5\}$\rem{(see Figure~\ref{Steps Algo})}. Since $\min_{x \in \setmin{3}} \edgeprob{x}{3} = 0.2$, and $\min_{\{C \in \pathloose{3} \, | \, |\setmin{3} \cap C| \geq 2\}} \frac{\combiprob{C}{3}}{|\setmin{3} \cap C|-1} = 0.1$, then $\sigma_{\setmin{3}} = \weight{3} = 0.1 = \frac{\combiprob{135}{3}}{|\setmin{3} \cap \{1,3,5\}|-1} $. The values are updated as follows: $\edgeprob{1}{4} = 0, \ \edgeprob{2}{4} = 0, \ \edgeprob{3}{4} = 0.1, \ \edgeprob{4}{4} = 0.2, \ \edgeprob{5}{4} = 0.2$, and $\combiprob{134}{4} = 0, \ \combiprob{235}{4} = 0.2, \ \combiprob{135}{4} = 0, \ \combiprob{234}{4} = 0.1$.
Then, $\component{4} = \{3,4,5\}$,  $\allpaths{4} = \allpaths{3}$, $\pathtight{4} = \{\{1,3,4\},\{1,3,5\}\}$, and $\pathloose{4} = \emptyset$. 

\item $\boldsymbol{k=4:}$ The set of minimal elements of $P^4= (\component{4},\preceq_{\pathtight{4}})$ is $\setmin{4} = \{3\}$\rem{ (see Figure~\ref{Steps Algo})}. Then, $\sigma_{\setmin{4}} = \weight{4} = \min_{x \in \setmin{4}} \edgeprob{x}{4} = \edgeprob{3}{4} = 0.1$, and the new values are:  $\edgeprob{1}{5} = 0, \ \edgeprob{2}{5} = 0, \ \edgeprob{3}{5} = 0, \ \edgeprob{4}{5} = 0.2, \ \edgeprob{5}{5} = 0.2$, and $\combiprob{C}{5} = \combiprob{C}{4}$ \rev{for all} $C \in \Chains$.
The new sets are $\component{5} = \{4,5\}$, $\allpaths{5} = \allpaths{4}$, $\pathtight{5} = \{\{1,3,4\},\{1,3,5\}\}$, and $\pathloose{5} = \emptyset$.

\item $\boldsymbol{k=5:}$ The set of minimal elements of $P^5 = (\component{5},\preceq_{\pathtight{5}})$ is given by $\setmin{5} = \{4,5\}$\rem{ (Figure~\ref{Steps Algo})}, and the weight associated with it is $\sigma_{\setmin{5}} = \weight{5} = \edgeprob{4}{5} = \edgeprob{5}{5} = 0.2$. \rev{Then,} $\edgeprob{x}{6} = 0$ \rev{for all} $x \in \ground$, and $\combiprob{C}{6} = \combiprob{C}{5}$ \rev{for all} $C \in \Chains$.

\end{itemize}

Since $\component{6} = \emptyset$, the algorithm terminates and outputs $\sigma$ \rev{which} satisfies constraints \eqref{Equal_2} and \eqref{small combi}. The total weight utilized is $\sum_{S \in \mathcal{P}} \sigma_S = 0.8 = \max\{\max\{\rho_x, \ x \in \ground\},\max\{\pi_C, \ C \in \Chains\}\}$. 
Therefore, from Theorem~\ref{optimal value},  $\sigma$ is an optimal solution of \OPOC. Since $0.8 \leq 1$, then $\widehat{\sigma} \in \revm{\mathbb{R}_{\geq0}^{\mathcal{P}}}$ given by $\widehat{\sigma}_{S} = \sigma_S$ \rev{for every} $S \in \mathcal{P} \backslash\emptyset$, and $\widehat{\sigma}_\emptyset = 0.2$, is a feasible solution of \FPOC.

%
%
%

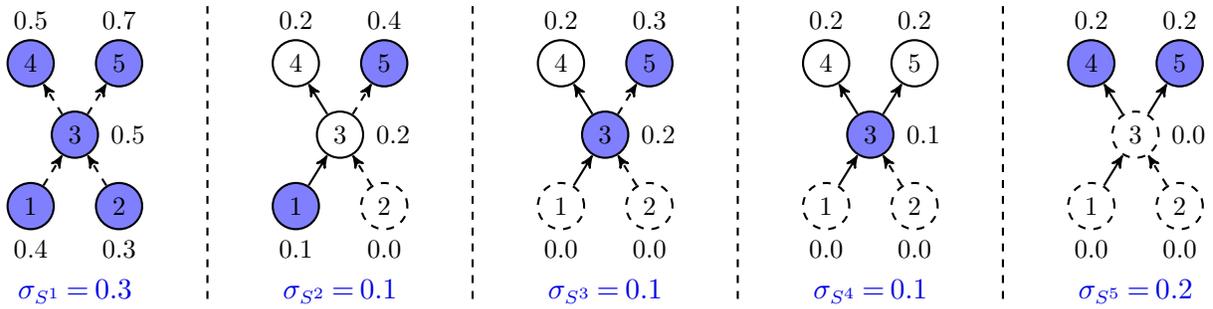
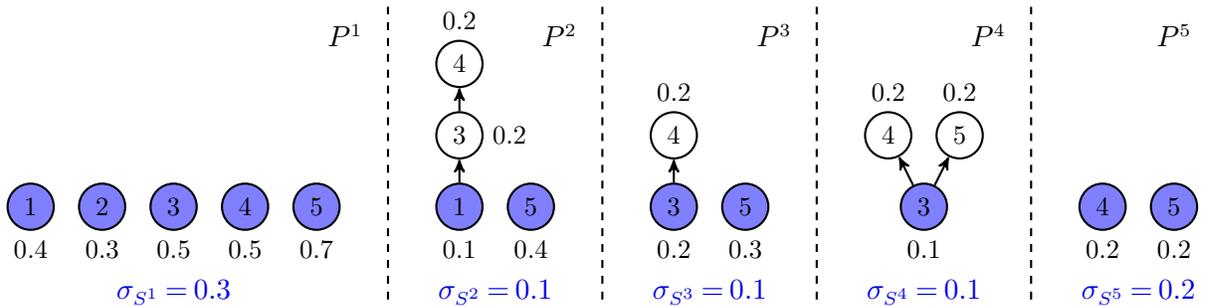
\begin{figure}[htbp]
    \centering
    \begin{subfigure}{0.95\textwidth}
    \centering
   \begin{tikzpicture}[->,>=stealth',shorten >=0pt,auto,x=2.35cm, y=1.9cm,
  thick,main node/.style={circle,draw},flow_a/.style ={blue!100}]
\tikzstyle{edge} = [draw,thick,->]
\tikzstyle{cut} = [draw,very thick,-]
\tikzstyle{flow} = [draw,line width = 1pt,->,blue!100]
\small
	\node[main node,fill=blue!50] (1) at (2.5,0) {1};
	\node[main node,fill=blue!50] (2) at (3,0) {2};
	\node[main node,fill=blue!50] (3) at (2.75,0.5) {3};
	\node[main node,fill=blue!50] (4) at (2.5,1) {4};
	\node[main node,fill=blue!50] (5) at (3,1) {5};
	
	\node[main node,fill=blue!50] (11) at (4,0) {1};
	\node[main node,dashed] (12) at (4.5,0) {2};
	\node[main node] (13) at (4.25,0.5) {3};
	\node[main node] (14) at (4,1) {4};
	\node[main node,fill=blue!50] (15) at (4.5,1) {5};
	
	\node[main node,dashed] (21) at (5.5,0) {1};
	\node[main node,dashed] (22) at (6,0) {2};
	\node[main node,fill=blue!50] (23) at (5.75,0.5) {3};
	\node[main node] (24) at (5.5,1) {4};
	\node[main node,fill=blue!50] (25) at (6,1) {5};
	
	\node[main node,dashed] (31) at (7.0,0) {1};
	\node[main node,dashed] (32) at (7.5,0) {2};
	\node[main node,fill=blue!50] (33) at (7.25,0.5) {3};
	\node[main node] (34) at (7,1) {4};
	\node[main node] (35) at (7.5,1) {5};
	
	\node[main node,dashed] (41) at (8.5,0) {1};
	\node[main node,dashed] (42) at (9.0,0) {2};
	\node[main node,dashed] (43) at (8.75,0.5) {3};
	\node[main node,fill=blue!50] (44) at (8.5,1) {4};
	\node[main node,fill=blue!50] (45) at (9,1) {5};
%
	
	\path[edge,dashed]
	(1) edge (3)
	(2) edge (3)
	(3) edge (4)
	(3) edge (5)
	(12) edge (13)
	(13) edge (15)
	(22) edge (23)
	(23) edge (25)
	(32) edge (33)
	(42) edge (43);
	
	\path[edge]
	(11) edge (13)
	(13) edge  (14)
	(23) edge (24)
	(33) edge (34)
	(33) edge (35)
	(21) edge (23)
	(31) edge (33)
	(41) edge (43)
	(43) edge (44)
	(43) edge (45);

%
\node (101) at (2.5,-0.3) {0.4};
\node (102) at (3,-0.3) {0.3};
\node (103) at (3.05,0.5) {0.5};
\node (104) at (2.5,1.3) {0.5};
\node (105) at (3,1.3) {0.7};

\node (111) at (4,-0.3) {0.1};
\node (112) at (4.5,-0.3) {0.0};
\node (113) at (4.55,0.5) {0.2};
\node (114) at (4,1.3) {0.2};
\node (115) at (4.5,1.3) {0.4};

\node (121) at (5.5,-0.3) {0.0};
\node (122) at (6,-0.3) {0.0};
\node (123) at (6.05,0.5) {0.2};
\node (124) at (5.5,1.3) {0.2};
\node (125) at (6,1.3) {0.3};

\node (131) at (7,-0.3) {0.0};
\node (132) at (7.5,-0.3) {0.0};
\node (133) at (7.55,0.5) {0.1};
\node (134) at (7.0,1.3) {0.2};
\node (135) at (7.5,1.3) {0.2};

\node (131) at (8.5,-0.3) {0.0};
\node (132) at (9.0,-0.3) {0.0};
\node (133) at (9.05,0.5) {0.0};
\node (134) at (8.5,1.3) {0.2};
\node (135) at (9,1.3) {0.2};

\draw[dashed,-] (3.5,-0.65) -- (3.5,1.4); 
\draw[dashed,-] (5,-0.65) -- (5,1.4); 
\draw[dashed,-] (6.5,-0.65) -- (6.5,1.4); 
\draw[dashed,-] (8,-0.65) -- (8,1.4); 


\node (02) at (2.75,-0.6) {\normalsize\textcolor{blue!90!black}{$\sigma_{\setmin{1}} = 0.3$}};
\node (02) at (4.25,-0.6) {\normalsize\textcolor{blue!90!black}{$\sigma_{\setmin{2}} = 0.1$}};
\node (02) at (5.75,-0.6) {\normalsize\textcolor{blue!90!black}{$\sigma_{\setmin{3}} = 0.1$}};
\node (02) at (7.25,-0.6) {\normalsize\textcolor{blue!90!black}{$\sigma_{\setmin{4}} = 0.1$}};
\node (02) at (8.75,-0.6) {\normalsize\textcolor{blue!90!black}{$\sigma_{\setmin{5}} = 0.2$}};

\normalsize
\end{tikzpicture}
    \caption{{\footnotesize Poset $P$ at the beginning of each iteration of the algorithm. The solid nodes are in $\component{k}$, the dashed nodes are in $\ground \backslash \component{k}$, and the blue nodes are in $\setmin{k}$. An edge is solid if there exists a maximal chain in $\pathtight{k}$ that contains both end nodes of the edge. 
 The values $\edgeprob{x}{k}$ are given  next to each element.}}
    \label{Steps Algo_bis}
\end{subfigure}

\begin{subfigure}{0.95\textwidth}
    \centering
    \ifarXiv
    \vspace{0.3cm}
  \else
    \fi
  \begin{tikzpicture}[->,>=stealth',shorten >=0pt,auto,x=1.9cm, y=1.9cm,
  thick,main node/.style={circle,draw},flow_a/.style ={blue!100}]

\tikzstyle{edge} = [draw,thick,->]
\tikzstyle{cut} = [draw,very thick,-]
\tikzstyle{flow} = [draw,line width = 1pt,->,blue!100]
\small
	\node[main node,fill=blue!50] (1) at (1,0) {1};
	\node[main node,fill=blue!50] (2) at (1.5,0) {2};
	\node[main node,fill=blue!50] (3) at (2,0) {3};
	\node[main node,fill=blue!50] (4) at (2.5,0) {4};
	\node[main node,fill=blue!50] (5) at (3,0) {5};

	\node[main node,fill=blue!50] (11) at (4,0) {1};
	\node[main node] (13) at (4,0.5) {3};
	\node[main node] (14) at (4,1) {4};
	\node[main node,fill=blue!50] (15) at (4.5,0) {5};
	
	\node[main node,fill=blue!50] (23) at (5.5,0) {3};
	\node[main node] (24) at (5.5,0.5) {4};
	\node[main node,fill=blue!50] (25) at (6,0) {5};
	
	\node[main node,fill=blue!50] (33) at (7.25,0) {3};
	\node[main node] (34) at (7,0.5) {4};
	\node[main node] (35) at (7.5,0.5) {5};
	
	\node[main node,fill=blue!50] (44) at (8.5,0) {4};
	\node[main node,fill=blue!50] (45) at (9,0) {5};
%
	
	\path[edge]
	(11) edge (13)
	(13) edge  (14)
	(23) edge (24)
	(33) edge (34)
	(33) edge (35);

%
\node (101) at (1,-0.3) {0.4};
\node (102) at (1.5,-0.3) {0.3};
\node (103) at (2,-0.3) {0.5};
\node (104) at (2.5,-0.3) {0.5};
\node (105) at (3,-0.3) {0.7};

\node (111) at (4,-0.3) {0.1};
\node (113) at (4.35,0.5) {0.2};
\node (114) at (4,1.3) {0.2};
\node (115) at (4.5,-0.3) {0.4};

\node (123) at (5.5,-0.3) {0.2};
\node (124) at (5.5,0.8) {0.2};
\node (125) at (6,-0.3) {0.3};

\node (133) at (7.25,-0.3) {0.1};
\node (134) at (7.0,0.8) {0.2};
\node (135) at (7.5,0.8) {0.2};

\node (134) at (8.5,-0.3) {0.2};
\node (135) at (9,-0.3) {0.2};

\draw[dashed,-] (3.5,-0.65) -- (3.5,1.4); 
\draw[dashed,-] (5,-0.65) -- (5,1.4); 
\draw[dashed,-] (6.5,-0.65) -- (6.5,1.4); 
\draw[dashed,-] (8,-0.65) -- (8,1.4); 

\node (01) at (3.2,1.2) {\normalsize$P^1$};
\node (01) at (4.7,1.2) {\normalsize$P^2$};
\node (01) at (6.2,1.2) {\normalsize$P^3$};
\node (01) at (7.7,1.2) {\normalsize$P^4$};
\node (01) at (9.0,1.2) {\normalsize$P^5$};

\node (02) at (2,-0.6) {\normalsize\textcolor{blue!90!black}{$\sigma_{\setmin{1}} = 0.3$}};
\node (02) at (4.25,-0.6) {\normalsize\textcolor{blue!90!black}{$\sigma_{\setmin{2}} = 0.1$}};
\node (02) at (5.75,-0.6) {\normalsize\textcolor{blue!90!black}{$\sigma_{\setmin{3}} = 0.1$}};
\node (02) at (7.25,-0.6) {\normalsize\textcolor{blue!90!black}{$\sigma_{\setmin{4}} = 0.1$}};
\node (02) at (8.75,-0.6) {\normalsize\textcolor{blue!90!black}{$\sigma_{\setmin{5}} = 0.2$}};

\normalsize
\end{tikzpicture}
    \caption{{\footnotesize$P^k$, for $k \in \llbracket 1,5 \rrbracket$. The values $\edgeprob{x}{k}$ are given  next to each element. $\setmin{k}$ is given by the blue nodes.}}
    \label{Steps Algo}
\end{subfigure}
\caption{Illustration of Algorithm~\ref{ALG3} for the poset $P$ given in Figure~\ref{ExPoset}.}
\end{figure}

\section{Minimum-cost circulation problem.}\label{sec:MCCP}

Primal and dual linear formulations of \original of polynomial size are given as follows:\hypertarget{Polyp}{}
%
%
\ifarXiv
\begin{align*}
\begin{array}{lrll}(\mathcal{M}_P^\prime) \ & \text{maximize} & \displaystyle\sum_{\{i \in \nodes \, | \, (i,t) \in \edges\}}\flow_{it} - \sum_{(i,j) \in \edges}\frac{b_{ij}}{p_1}\flow_{ij} &\mcr
& \text{subject to} &  \displaystyle\sum_{\{ j \in \nodes \, | \, (j,i) \in \mathcal{E}\}} \flow_{ji} = \sum_{\{j \in \nodes \, | \, (i,j) \in \mathcal{E}\}} \flow_{ij}, &\forall i \in \mathcal{V}\backslash \{s,t\} \mcr
& & 0 \leq \flow_{ij} \leq c_{ij}, & \forall (i,j) \in \mathcal{E}\mcr
& & 0 \leq \flow_{ij} \leq\displaystyle \frac{d_{ij}}{p_2}, & \forall (i,j) \in \mathcal{E}.\end{array}
\end{align*}

\fi
%
\hypertarget{Polyd}{}
\ifarXiv
\begin{align*}
\begin{array}{lrll}(\mathcal{M}_D^\prime) \ & \text{minimize} & \displaystyle\sum_{(i,j) \in \edges }c_{ij} \rho_{ij} + \frac{d_{ij}}{p_2} \mu_{ij} &\\
& \text{subject to} &  \displaystyle y_i - y_j + \rho_{ij} + \mu_{ij} \geq -\frac{b_{ij}}{p_1}, &\forall (i,j) \in \edges \ | \ i \neq s \text{ and }  j \neq t\\[0.3cm]
& &  \displaystyle - y_j + \rho_{sj} + \mu_{sj}\geq -\frac{b_{sj}}{p_1}, &\forall  j \in \nodes \ | \ (s,j) \in \edges\\[0.3cm]
& &  \displaystyle y_i + \rho_{it} + \mu_{it}\geq 1 -\frac{b_{it}}{p_1}, &\forall  i \in \nodes \ | \ (i,t) \in \edges\\
& & \rho_{ij} \geq 0, & \forall (i,j) \in \mathcal{E}\mcr
& & \mu_{ij} \geq 0, & \forall (i,j) \in \mathcal{E}.
\end{array}
\end{align*}

\fi

Let $z^*_{\polyo}$ denote the optimal value of \polyp and \polyd. 
We show the following result:
\begin{lemma}\label{Poly_size}
Any $s-t$ path decomposition of any optimal solution $\flow^\prime$ of \polyp is an optimal solution of \primal. Furthermore, given any optimal solution $(\rho^\prime,\mu^\prime,y^\prime)$ of \polyd, $(\rho^\prime,\mu^\prime)$ is an optimal solution of \dual. 
\end{lemma}

\proof{Proof of Lemma~\ref{Poly_size}.} Let $\flow^* \in \revm{\mathbb{R}_{\geq0}^{\Lambda}}$ be an optimal solution of \primal. Then, $\flow^\prime \in \revm{\mathbb{R}_{\geq0}^{\edges}}$ defined by $\flow^\prime_{ij} = \sum_{\{\lambda \in \Lambda \, | \, (i,j) \in \lambda\}} \flow^*_\lambda$ is a feasible solution of \polyp. Therefore, $z^*_{\polyo} \geq z^*_{\original}$.
Now, let $\flow^\prime \in \revm{\mathbb{R}_{\geq0}^{\edges}}$ be an optimal solution of \polyp. From the flow decomposition theorem, there exists a vector $\flow^* \in \revm{\mathbb{R}_{\geq0}^{\Lambda}}$ such that \rev{for all} $(i,j) \in \edges, \ \flow^\prime_{ij} = \sum_{\{\lambda \in \Lambda \, | \, (i,j) \in \lambda\}} \flow^*_\lambda$. Since $\flow^*$ is a feasible solution of \primal, \rev{then} $z^*_{\original} \geq z^*_{\polyo}$. In conclusion, $z^*_{\original} = z^*_{\polyo}$, and an optimal solution of \primal can be obtained by decomposing an optimal solution of \polyp into $s-t$ paths.

Now, consider an optimal solution $(\rho^\prime,\mu^\prime,y^\prime)$ of \polyd.
Then, one can verify that for every $s-t$ path $\lambda \in \Lambda$, $\sum_{(i,j) \in \lambda} (\rho^\prime_{ij} + \mu^\prime_{ij}) \geq 1 - \frac{1}{p_1}\sum_{(i,j) \in \lambda} b_{ij} = \pi^0_\lambda$ (the $y^\prime$ cancel in a telescopic manner along each $s-t$ path). Therefore, $(\rho^\prime,\mu^\prime)$ is a feasible solution of \dual. Since $z^*_{\polyo} = z^*_{\original}$, we conclude that  $(\rho^\prime,\mu^\prime)$ is an optimal solution of \dual. 
\hfill
\Halmos
\endproof



%
%


 \end{APPENDICES}

\section*{Acknowledgments}

This work was supported in part by the Singapore National Research Foundation through the Singapore MIT Alliance for Research
and Technology (SMART), the DoD Science of Security Research Lablet (SOS), FORCES (Foundations
Of Resilient CybEr-Physical Systems), which
receives support from the National Science Foundation
(NSF award numbers CNS-1238959, CNS-1238962,
CNS-1239054, CNS-1239166), NSF CAREER award CNS-1453126, and the AFRL LABLET - Science of Secure and Resilient Cyber-Physical Systems (Contract ID: FA8750-14-2-0180, SUB 2784-018400). \revm{We are grateful to Bernhard von Stengel, the Associate
Editor, and the reviewers who handled the submission, for useful suggestions.}



\bibliographystyle{informs2014} 
\bibliography{routing_games.bib}


\end{document}